\documentclass[aps,prl, twocolumn,amsmath,amssymb,longbibliography,superscriptaddress]{revtex4-1}% APS journal style
\usepackage{amsmath,amssymb,amsfonts,bm}
\usepackage{graphicx}
\usepackage{xcolor}
\usepackage{bbold}
\usepackage{graphicx}% Include figure files
\usepackage{dcolumn}% Align table columns on decimal point
\usepackage{epstopdf}
\usepackage{epsfig}
\usepackage{url}
\usepackage{hyperref}
\usepackage{verbatim}
\usepackage[export]{adjustbox}
\usepackage{scalerel}
\usepackage{braket}
\usepackage[normalem]{ulem}

\usepackage{lipsum}
\usepackage{enumitem}

\usepackage{float}

\newcommand{\be}{\begin{equation}}
\newcommand{\ee}{\end{equation}}
\newcommand{\ba}{\begin{aligned}}
	\newcommand{\ea}{\end{aligned}}

\newcommand{\thetasphi}{{\theta^*_{\phi}}}
\newcommand{\thetasLB}{{\theta^*_{\mathrm{LB}}}}

\newcommand{\rr}{\mathrm{Re}}
\newcommand{\ii}{\mathrm{Im}}

\newcommand{\thei}{t_{\mathrm{Hei}}}

\newcommand{\thetas}{\theta^*}

\newcommand{\tthei}{\tau_{\mathrm{Hei}}}

\newcommand{\tte}{\tau_{\mathrm{edge}}}
\newcommand{\kerr}{\mathcal{K}}

\newcommand{\Kc}{\mathcal{K}}

\newcommand{\thetasone}{{\theta^*_0}}
\newcommand{\thetastwo}{{\theta^*_{\pi/2}}}
\newcommand{\tauhei}{{\tau_{\mathrm{Hei}}}}

\newcommand{\tildeD}{{\tilde{\Delta}}}

\newcommand{\titleinfo}{Spectral statistics of non-Hermitian matrices and dissipative quantum chaos}

\graphicspath{{Figures/}}

\begin{document}
	
	\title{Spectral Statistics of Non-Hermitian Matrices and Dissipative Quantum Chaos}
	\author{Jiachen Li}
	\affiliation{Department  of  Physics,  Princeton  University,  Princeton,  New  Jersey  08544,  USA}
	\author{Toma\v{z} Prosen}
	\affiliation{Department of Physics, Faculty of Mathematics and Physics,University of Ljubljana, Jadranska 19, SI-1000 Ljubljana, Slovenia}
	\author{Amos Chan}
	\affiliation{Princeton Center for Theoretical Science, Princeton University, Princeton New Jersey 08544, USA}
	\date{\today}
	
	\begin{abstract}
		We propose a measure, which we call the \textit{dissipative spectral form factor} (DSFF), to characterize the spectral statistics of non-Hermitian (and non-Unitary) matrices. %
		We show that DSFF successfully diagnoses dissipative quantum chaos, and reveals correlations between real and imaginary parts of the complex eigenvalues up to arbitrary energy (and time) scale. 
		Specifically, we provide the exact solution of DSFF for the complex Ginibre ensemble (GinUE) and for a Poissonian random spectrum (Poisson)
		as minimal models of dissipative quantum chaotic and integrable systems respectively. 
		For dissipative quantum chaotic systems, we show that DSFF exhibits an exact rotational symmetry in its complex time argument $\tau$.
		Analogous to the spectral form factor (SFF) behaviour for Gaussian unitary ensemble,  DSFF for GinUE shows a ``dip-ramp-plateau'' behavior in $|\tau|$: 
		DSFF initially decreases, 
		increases at intermediate time scales, 
		and saturates after a generalized Heisenberg time which scales as the inverse mean level spacing.
		Remarkably, for large matrix size, the ``ramp'' of DSFF for GinUE increases \textit{quadratically} in $|\tau|$,
		in contrast to the \textit{linear} ramp in SFF for Hermitian ensembles. 
		For dissipative quantum integrable systems, we show that DSFF takes a constant value except for a region in complex time whose size and behavior depends on the eigenvalue density.  
		Numerically, we verify the above claims and additionally show that DSFF for real and quaternion real Ginibre ensembles coincides with the GinUE behaviour except for a region in complex time plane of measure zero in the limit of large matrix size.
		As a physical example, we consider the quantum kicked top model with dissipation, and show that it falls under the Ginibre universality class and Poisson as the `kick' is switched on or off. 
		Lastly, we study spectral statistics of ensembles of random classical stochastic matrices or Markov chains, and show that these models again fall under the Ginibre universality class.  
	\end{abstract}

	\maketitle
	
	\textit{Introduction.} 
	The study of spectral statistics is of fundamental importance in theoretical physics due to its universality 
	and utility as a robust diagnosis of quantum chaos~\cite{mehta,bohigas1984characterization}. 
	While Bohigas, Giannoni and Schmidt conjectured that chaotic quantum systems exhibit spectral correlation as those found in random matrix theory (RMT) in the same symmetry class~\cite{bohigas1984characterization},
	Berry and Tabor observed that integrable systems follow Poisson statistics of uncorrelated random variables~\cite{BerryTabor}.
	Both claims have withstood the test of time, and in particular, the signature of level repulsion have been found in a wide range of disciplines including nuclear resonance spectra \cite{nuclear1981}, mesoscopic physics \cite{meso1997, meso2015}, quantum chaos\cite{Rigol}, black hole physics \cite{gargar2016, Cotler_2017, complexity2017}, quantum chromodynamics \cite{qcd1995, qcd2018}, number theory \cite{number2008}, information theory \cite{informationtheory2004} and more.

	Non-Hermitian physics has advanced significantly in recent years in the study of optics \cite{Feng2017optics, carmichael1993optics, Elganainy2018optics}, acoustics~\cite{Ma2016acoustics, Cummer2016acoustics}, parity-time-symmetric systems~\cite{Bender2007PT, Ozdemir2019PT}, mesoscopic physics~\cite{quantumresonance1998, Muga2004mesoscopic, cao2015mesoscopic, rotter2017mesoscopic}, cold atoms~\cite{daley2014coldatom, kuhr2016coldatom}, driven dissipative systems~\cite{Deng2010, muller2012engineered, ritsch2013, Sieberer2016, Chou_2011}, biological systems~\cite{Marchetti2013}, disordered systems~\cite{meso1997}.
	Recent studies on spectral properties have focused on the shape of eigenvalue density, the spectral gap, and the spacing between nearest-neighbour eigenvalues \cite{akemann2019, can2019a, can2019b, sa2019a, sa2019b, sa2020a, sa2020b, rubio2021, luitz2020, denisov2019, Shklovskii2020, ueda2019nonhermMBL, Tzortzakakis2020, Peron2020, Borgonovi1991}. 
	%
	%However, due to the complication in dealing with complex-valued spectrum an analytically tractable measure that would characterize level correlation at all scales is still lacking.
	%
	%The goal of this letter is to introduce and analyze a simple indicator that characterizes the level statistics of non-Hermitian matrices up to arbitrary energy (and, equivalently, time) scale. 
	The goal of this letter is to introduce and analyze a simple indicator that characterizes the level statistics of non-Hermitian matrices up to an arbitrary energy (and, equivalently, time) scale, and show that it captures universal signatures of dissipative quantum chaos.
	We treat the complex-valued spectrum as a two-dimensional (2D) gas, and introduce the DSFF as the 2D Fourier transform of the density-density correlator of complex eigenvalues, which depends on a complex time parameter, $\tau$.
	We exactly compute DSFF for the GinUE and for a Poissonian random spectrum as minimal models of dissipative quantum chaotic and integrable systems respectively. 
	In particular, we show that DSFF for GinUE exhibits a dip-ramp-plateau behavior as a function of $|\tau|$, with an asymptotically \textit{quadratic} ramp, as opposed to the \textit{linear} ramp in SFF for Gaussian ensembles of Hermitian matrices. 
	We demonstrate the universality of these results by showing that they capture the level statistics of quantum kicked top model with dissipation and random classical stochastic matrices.
	We conjecture that for large enough complex time $|\tau|$, dissipative chaotic systems have DSFF behaviour that coincides with GinUE's \cite{upcoming}. 
	As such, the DSFF solution for GinUE  provides an important benchmark of dissipative quantum chaotic system, and can be treated as a complex analogue of SFF solution for GUE.

	\begin{figure}[t]
		\centering
		\includegraphics[width=0.475 \textwidth  ]{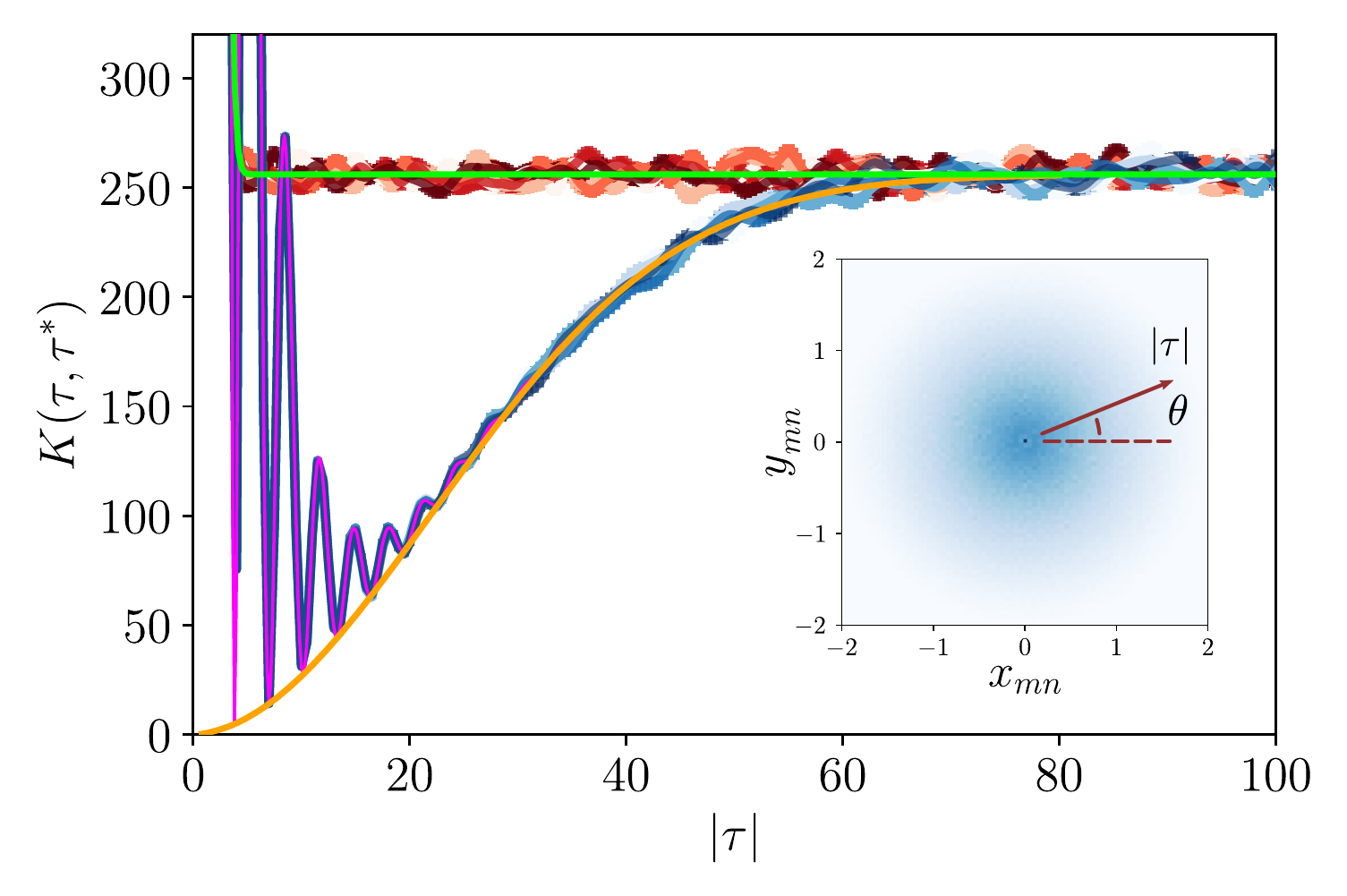}
		\caption{
			$K(\tau,\tau^*)$ vs. $|\tau|$ for the GinUE and a Poissonian-random spectrum (defined above \eqref{eq:dsff_poi}) with matrix size $N=256$. 
			Writing $\tau= |\tau|e^{i\theta}$, numerical simulations of GinUE (Poisson) for fixed $\theta = 0$ to $\theta = \pi/2$ in steps of $\pi/4$ are plotted in blue colors (red colors) in multiple shades.
			The analytical solutions of $K_{\mathrm{GinUE}}$ (Eq.~\eqref{eq:dsff_ginue}) and $K_{\mathrm{Poi}}$ (Eq. \eqref{eq:dsff_poi}) are plotted as the purple and green lines on top of the numerical data.  The connected part of $K_{\mathrm{GinUE}}$ (1st and 3rd terms in Eq.~\eqref{eq:dsff_ginue}) is plotted as the orange line.
			Inset: Two-dimensional histogram of $\{ z_{mn} \equiv z_m - z_n \}$ of the GinUE for $N=1024$ where increasing values are plotted with deeper blue colors.
			Note that the bin at the origin is occupied by $N$ (diagonal) contributions of $\{ z_{nn} =0 \}$, and there is a dip around the origin due to level repulsion.
			The computation of DSFF for fixed $\theta$ as a function of $|\tau|$ is equivalent to the computation of SFF as a function of $|\tau|$ of $\{ z_{m} \}$ projected onto the axis defined by $\theta$.
			The sample sizes are 5000 and 2000 for Poisson and GinUE respectively.
		}
		\label{fig:overall}
	\end{figure}
	
	%
	%
	%
	%
	%%%%%%%%%%%%%%%%%%%%%%%%%%%%

	\textit{Spectral form factor.} It is instructive to review the behaviour of SFF for closed quantum systems \cite{Haake}, with which DSFF shares several analogous features.
	Consider a closed quantum system described by a $N\times N$ Hermitian (or unitary) matrix with the density of states (DOS) $\rho(E) = \sum_n \delta(E- E_n)$ where $E_n$ is the $n$-th eigenlevel (or eigenphase).
	Note that in our convention, $\int dE \, \rho(E) = N$.
	The correlation between eigenlevels can be quantified by the (so-called ``2$\alpha$-point'') SFF, which is the ($\alpha$-th power of the) Fourier transform of two-level correlation function $\langle \rho(E) \rho(E+ \omega)\rangle$, and can be directly defined as
	\begin{equation}\label{eq:old_K}
	K_\alpha(t):= 
	\left\langle 
	\,
	\left[ 
	\sum_{n,m}  
	e^{-i(E_n-E_m)t}
	\right]^{\alpha}
	%e^{i(E_n - E_m)t} 
	\right\rangle
	%=
	% \left\langle
	% \Big| \Tr[U^t ] \Big|^{2\alpha} 
	% \right\rangle
	\;,
	\end{equation}
	where $\langle \cdot \rangle$ denotes the average over an ensemble of statistically similar systems. 
	Importantly, SFF captures correlations between eigenlevels at {\em all} scales, including the level repulsion and spectral rigidity, while at the same time
	it is one of the simplest non-trivial and analytically tractable diagnostic of quantum chaos
	\footnote{SFF involves only two sums over eigenlevels as supposed to four sums in observables like the out-of-time-order correlator or the quantum purity. Note also that the finer correlations between eigenlevels can be captured by the so-called higher-point SFF\cite{complexity2017,liu2018}}.
	Furthermore, SFF has recently been shown to capture novel signatures of quantum many-body physics like the deviation from RMT behaviour at early time (and consequently, the onset of chaos) %in many-body quantum chaotic systems
	~\cite{kos_sff_prx_2018, bertini2018exact, Roy2020random, flack2020statistics, cdc1, cdc2, friedman2019, moudgalya2020spectral, garratt2020manybody, Gharibyan_2018, chan2020lyap, chan2021transinv, Cotler_2017, complexity2017, saad2019semiclassical, Gharibyan_2018, garratt2020manybody, bertini2021random, swingle2020hydrodynamic, mblexistence}. 
	For integrable systems with Poisson statistics, we have $K_1(t) =  N$ for $t$ larger than a timescale set by the inverse width of DOS.  
	For quantum chaotic systems without symmetries, the generic behaviour of SFF can be understood by studying the Gaussian unitary ensemble (GUE). SFF for GUE initially decays and then grows linearly until the Heisenberg time, $\thei$, after which $K(t)=N$ reaches a plateau. 
	Qualitatively, this is referred to as the ``dip-ramp-plateau'' behaviour.
	The form of the early time decay is due to the (non-universal) form of the DOS, while
	the linear ramp reflects the phenomenon of spectral rigidity.
	$\thei$ is proportional to the inverse of the mean level spacing, and it encodes the largest physically relevant time scale of the system.

	\textit{Dissipative spectral form factor.} For a $N\times N$ non-Hermitian matrix with complex spectra, the SFF is exponentially growing or decaying in time due to the imaginary parts of the complex eigenvalues. 
	Moreover, traditional methods like the Green's function approach fail due to non-analyticity of the Green's function.
	To circumvent these problems, we consider 2D DOS, $\rho(z) =  \sum_n \delta(x - x_n)  \delta(y - y_n)$, where $z=x+iy$, $x_n = \rr \, z_n$, $y_n = \ii \, z_n$, and $z_n = x_n + i y_n$ is the $n$-th complex eigenvalue. 
	We introduce the (``$2\alpha$-point'') DSFF as the ensemble average of the ($\alpha$-th power of the) 2D Fourier transform of the two-level correlation function $\langle \rho  (x,y )\rho(x + \omega , y +\omega')\rangle$.
	%\footnote{Note that the $\alpha$-th power and the Fourier transform do not commute.}. 
	%
	We directly define DSFF as 
	\begin{equation} \label{eq:dsff_def_a}
	K_\alpha(t,s):= 
	\left\langle
	\left[ 
	\sum_{m,n} e^{i(x_n - x_m)t + i(y_n - y_m)s}
	\right]^\alpha
	\right\rangle
	\;,
	\end{equation}
	where $t$ and $s$ are two ``time'' variables conjugate to the $x_n - x_m$ and $y_n - y_m$ respectively.
	Importantly, the correlation between \textit{both} the real and imaginary parts of two given complex eigenvalues now contributes to DSFF as phases.
	Notice that if the spectrum is real, DSFF is effectively reduced to SFF as a function of $t$ for all $s$. 
	To obtain an intuition of how DSFF behaves, we write 
	$\vec{z}_{mn}\equiv (x_{mn} ,y_{mn} ) \equiv (x_m-x_n, y_m-y_n)$, 
	and $\vec{\tau} \equiv (t,s) = (|\tau| \cos \theta , |\tau| \sin \theta )$. The DSFF can now be written as 
	$
	K_\alpha (t,s)= 
	\left\langle
	\left[ 
	\sum_{m,n} e^{i \vec{z}_{mn} \cdot  \vec{\tau} }
	\right]^\alpha
	\right\rangle
	\;,
	$
	which allows a natural interpretation in the complex plane: 
	At fixed $\theta$ and as a function of $|\tau|$, ($2\alpha$-point) DSFF is the ($\alpha$-point) SFF of the projection of $\{z_m\}$ onto the radial axis specified by angle $\theta$ (illustrated in Fig.1 inset). 
	Reverting back to the notation with complex numbers, we define complex time, $\tau = t + is$ and write DSFF as
	$
	{K}_\alpha(\tau, \tau^*)
	=
	\left \langle
	\left| \sum_{n} e^{i(z_n \tau^* + z_n^* \tau ) /2 } \right|^{2\alpha}
	\right \rangle
	$,
	For ensembles where the two-point correlation function 
	$\langle 
	\rho(z_1) \rho(z_2) 
	\rangle $ 
	is known, DSFF at $\alpha=1$ can be written as an integral
	$
	K_1 %(\tau, \tau^*) 
	= 
	\int 
	d^2z_1 d^2 z_2 \,
	\langle 
	\rho(z_1) \rho(z_2) 
	\rangle 
	\,
	e^{i(z_1 \tau^* + z_1^* \tau - z_2^* \tau - z_2 \tau^* ) /2 }
	\;.
	$
	For the rest of the paper, we will drop the subscript and focus on the simplest and most relevant case $\alpha=1$.
	\textit{Dissipative quantum chaotic systems.} We use the GinUE as a minimal model of the dissipative quantum chaotic systems. The joint probability distribution function of eigenvalues of GinUE is known exactly, and the correlation function of eigenvalues can be expressed in terms of the kernel \cite{ginibre_1965},
	$
	\kerr(z_1, z_2) = 
	\frac{N}{\pi}
	e^{-\frac{N}{2}(|z_1|^2 + |z_2|^2)}
	\sum_{\ell= 0}^{N-1} \frac{(N z_1 z_2^*)^\ell}{\ell !}
	\;.
	$
	The 1-point correlation function, i.e. the DOS,
	is given by $\langle  \rho(z) \rangle = \Kc(z,z)$, and the kernel is normalized such that $\int d^2 z \, \langle  \rho(z) \rangle = \int d^2 z \, \Kc(z,z)= N$. Note that the DOS is isotropic, 
	and is asymptotically, as $N\to\infty$, flat on a unit disc $|z|<1$ and vanishing outside.
	The 2-point correlation function is
	$
	\langle 
	\rho(z_1) \rho(z_2) 
	\rangle 
	=
	\Kc (z_1, z_1) \delta(z_1 - z_2) 
	+ \Kc (z_1,z_1)\Kc (z_2,z_2)
	-
	\left| \Kc (z_1,z_2)\right|^2
	$.
	We will refer to the above three terms as the contact, disconnected and connected term respectively. 
	Using  $\langle \rho(z_1) \rho(z_2)\rangle$, we compute \eqref{eq:dsff_def_a} by expanding the exponential factors, and using the fact that the integrals over the phases of $z_1$ and $z_2$ kill all terms in the sum except the ones depending on $|z_1|$ and $|z_2|$. This gives %one of our main results,
	\begin{widetext}
		\be
		\label{eq:dsff_ginue}
		K_{\mathrm{GinUE}}(\tau, \tau^*) 
		= 
		N + 
		N^2
		\, _1F_1\left(N+1;2;-\frac{\left| \tau \right| ^2}{4 N}\right)^2
		-
		\sum_{n, m=0}^{N-1}
		\frac{(\max (m,n)!)^2 
		}{
			n! m! 
			(\left| m-n\right| !)^2}
		\, _1F_1\left(\max (m,n)+1;\left| m-n\right| +1;-\frac{\left| \tau\right| ^2}{4 N}\right)^2
		,
		\ee
	\end{widetext}
	where we have listed the three terms in the same ordering as in the 2-point correlation function.
	$_1F_1(a,b;z) = \sum_{n=0}^\infty a^{(n)} z^n / b^{(n)} n!$ is the Kummer confluent hypergeometric function, where  $a^{(n)}$ is the rising factorial. 
	Keeping the leading contributions in $N$, Eq.~\eqref{eq:dsff_ginue} becomes
	\be \label{eq:dsff_ginue_largeN}
	K_{\mathrm{GinUE}}(\tau , \tau^*) = N 
	+ N^2  \frac{ 4 J_1(\left| \tau \right| )^2  }{ \left| \tau\right| ^2}
	- N \exp\left( -\frac{|\tau|^2}{ 4N}  \right) \,,
	\ee
	where $J_{\mu}(x)$ is the Bessel function of the first kind.
	Along with Eq.~\eqref{eq:dsff_ginue}, Eq.~\eqref{eq:dsff_ginue_largeN} forms our main result, as they capture the universal spectral correlations of dissipaitve quantum chaotic systems. 
	%
	%%%%%%%%%%%%%% (1)
	Firstly, note that $K_{\mathrm{GinUE}}(\tau, \tau^*)  $ only depends on the absolute value of $\tau$, i.e. DSFF is manifestly rotational symmetric in complex time (see Fig.~\ref{fig:overall}).
	%%%%%%%%%%%%%% (2)
	Secondly, the qualitative behaviour of DSFF as a function of $|\tau|$ for dissipative quantum systems shows a dip-ramp-plateau structure, analogous to SFF for closed quantum systems: 
	At early time $|\tau | \lesssim \tte$, 
	DSFF dips from $K(0,0)= N^2$ with a form dominated by the disconnected piece \eqref{eq:dsff_ginue}; 
	At intermediate time $\tte \lesssim |\tau| \lesssim \tthei $,
	DSFF increases \textit{quadratically} $K_{\rm GinUE} \simeq |\tau|^2/4$ with precise form given by the sum of hypergeometric functions, or the Gaussian function for large $N$,
	until it reaches late time $ \tthei  \lesssim |\tau|$ with $ \tthei \sim \sqrt{N}$, where DSFF reaches a plateau at $N$. 
	Note that the DSFF GinUE ramp behaviour is drastically different from the corresponding SFF GUE behaviour, which is \textit{linear} in time.
	%%
	%
	%%%%%%%%%%%%%% (3)
	Thirdly, in analogy to SFF, the connected term in DSFF captures the spectral rigidity in the complex spectral plane.
	As apparent from the functional form of the connected term, we see that the Heisenberg time scales as the inverse of mean level spacing (in the complex plane), $\tthei = O(\Delta^{-1}) = O(\sqrt{N})$. \cite{supplementary}
	Again, this is in contrast to the corresponding Heisenberg time scaling for SFF, which scales as $N$. 
	%
	%
	%%%%%%%%%%%%%% (4)
	Fourthly, the non-oscillatory part of the disconnected term asymptotically scales as 
	$N^2 |\tau|^{-3}$. Setting the time $\tte$ where disconnected and connected contributions are of the same order, we find $\tte = O(N^{2/5})$.
	Note that as a function of $|\tau|$, the GinUE disconnected term of DSFF coincides with the GUE disconnected term of SFF, due to the fact that the projection of the DOS along the axis defined by $\theta$  exhibits a semi-circle shape, like the DOS of GUE. 
	%
	%%%%%%%%%%%%%%%%%%%(5)
	Fifthly, according to the interpretation described above, % and in Fig.~\ref{fig:overall} inset,
	DSFF at fixed $\theta$ is equivalent to the SFF of the projected spectrum $\{z_{m} \}$ along the axis defined by $\theta$. 
	While there is level repulsion in the 2D complex plane, there are accidental degeneracies between distanced pairs of eigenvalues in the set of projected $\{ z_{m} \}$, and remarkably, the lack of level repulsion along the projection axis  makes up for the difference between SFF of GUE, which has linear ramp in $t$ with $\thei = O(N)$, and DSFF of GinUE which has a quadratic ramp in $|\tau|$ with $\tthei  = O(\sqrt{N})$. 

	Lastly, the GinUE DSFF result can be interpreted via the mapping between GinUE and 2D log-potential Coulomb gas.
	DSFF is equivalent to the 2D static structure factor (SSF), defined as the Fourier transform of density-density fluctuation, 
	where the complex energy and the complex time in \eqref{eq:dsff_def_a} take the roles of position and wavevector $\vec{k}$. 
	For the Coulomb gas, with the assumption of ``perfect screening'', SSF is argued to have an asymptotic behaviour of $|\vec{k}|^2$ for small $|\vec{k}|$~\cite{BAUS19801}, which is consistent with the quadratic increase $|\tau|^2$ for large $N$ in \eqref{eq:dsff_ginue_largeN}. 
	Furthermore, a system is \textit{hyperuniform} if its SSF vanishes as $|\vec{k}|$ tends to zero. This implies that density fluctuation is suppressed at very large length scales~\cite{hyperuniformity2003,hyperuniformity2018}.
	This leads us to interpret that the spectrum of 
	GinUE
	is a 2D gas that displays hyperuniformity with a quadratic power-law form.

	The numerical data and analytical solutions are plotted in Fig.~\ref{fig:overall} for $N=256$ with excellent agreement. 
	We further computed DSFF for the real (GinOE) and quaternion real (GinSE) Ginibre ensembles~\cite{supplementary}.
	The rotational symmetry of $K$ in $\tau$ is broken, since eigenvalues of quaternion real (real) ensemble (are either real or) come in complex conjugate pairs, which leads to special behaviour of $z_{mn}$ near $\theta =0, \pi/2$~\cite{supplementary}.
	We define critical angles $\thetasone$ and $\thetastwo$ such that DSFF of GinOE and GinSE coincide with the one of GinUE for $\theta \in [\thetasone, \pi/2 - \thetastwo]$. 
	We numerically show and heuristically argue that $\thetasphi \propto N^{-1/2}$ with the only exception of $\thetastwo \propto N^{-0.56}$  for GinSE due to the lack of ``projected degeneracies'' \cite{supplementary}.
	We therefore conclude that, in large $N$, GinOE and GinSE coincide with the GinUE behaviour except for the angle $\theta=0$ and $\pi/2$.
	This is consistent with the fact that spectral correlations of these ensembles coincide with GinUE for eigenvalues away from the real axis~\cite{Nils1991, forrester2007, Borodin_2009, Akemann_2007, Sommers_2008}.

	\textit{Dissipative integrable systems.} We model the spectrum of dissipative quantum integrable systems with a set of uncorrelated normally distributed complex eigenvalues. The DOS is $\langle\rho (z)  \rangle  = N (2\pi)^{-1} e^{-|z|^2 /2}$ with $\int d^2 z \, \langle\rho (z)  \rangle  = N$. 
	The 2-point correlation function is
	$\langle\rho (z_1) \rho(z_2) \rangle 
	= \langle\rho(z_2) \rangle\delta(z_1 - z_2)
	+ \frac{N(N-1)}{N^2}\langle\rho (z_1) \rangle \langle \rho(z_2) \rangle$.
	The DSFF can be evaluated as
	\be \label{eq:dsff_poi}
	K_{\mathrm{Poi}} (\tau, \tau^*)= N + N(N-1) e^{-|\tau|^2 } \;.
	\ee
	We see that DSFF for Poissonian random spectrum takes a constant value of $N$ except for a small region of complex time near the origin. % depending on DOS. 
	The constant value $N$ is due to the diagonal contribution from DSFF, and the deviation from $N$ is due to the disconnected part and depends on the details of the DOS.
	The numerical data and analytical solution are plotted in Fig.~\ref{fig:overall} for $N=256$ with excellent agreement. 

\begin{figure}[t]
		\centering
		\includegraphics[width=0.475 \textwidth  ]{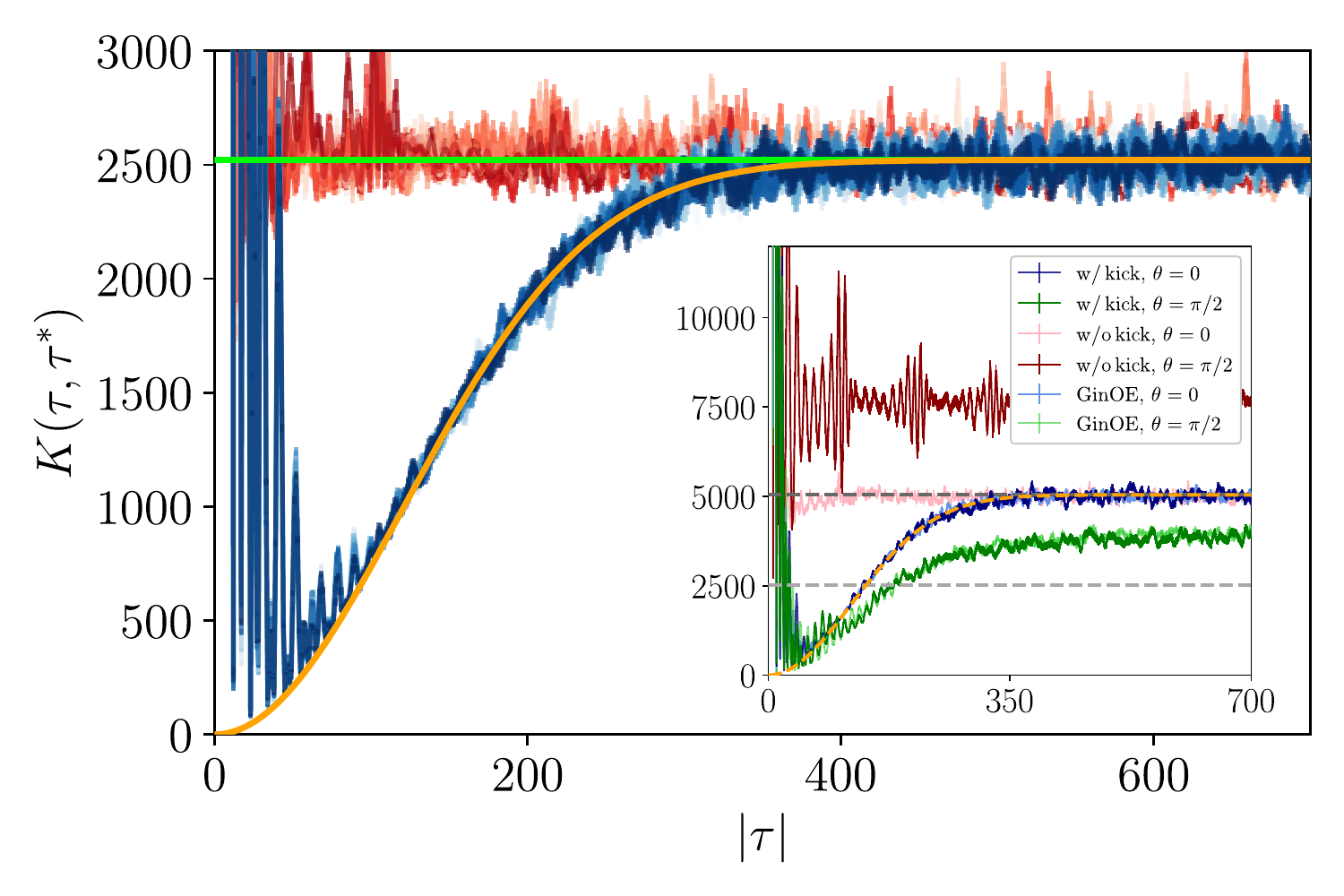}
		\caption{Main: DSFF of QKT with dissipation for $j=35$
			and Gaussianly-distributed $p \in \mathcal{N}(2,2/3)$ and $k_0 \in \mathcal{N}(10,3)$. 
			The blue (red) lines in different shades are for DSFF of QKT with the kick with $k_1=8$ (without the kick with $k_1=0$) for fixed $\theta \in [ \pi/16 , 7\pi/16]$ in steps of $\pi/16$. 
			The two cases fit the connected part of GinUE (orange line) and Poisson (green line) predictions as expected.
			Inset: DSFF of QKT and GinOE for angles $\theta = 0, \pi/2$ \cite{supplementary}.
			The sample sizes for QKT with and without the kick are 2500 and 4300 respectively.
		} \label{fig:qkt}
	\end{figure}

	\textit{Dissipative quantum kicked top model.} A simple but rich example of quantum systems that exhibit chaotic and integrable behaviours is the quantum kicked top (QKT) model \cite{sommers1988, Haake1987, loschmidt2006} which has been experimentally realized in \cite{Smith2004}. 
	The unitary evolution of QKT is governed by the Hamiltonian \cite{sommers1988},
	\be
	H(t) = p J_z +  \frac{k_0}{2j} J^2_z + \frac{k_1}{2j} J_{y}^2 \sum_{ n=- \infty}^{ \infty} \delta(t- n) \;,
	\ee
	where $\vec{J}= (J_x, J_y,J_z)$ are angular momentum operators that act on a single spin-$j$ particle and obey $[J_\alpha,J_\beta] = i \epsilon_{\alpha\beta\gamma} J_\gamma$, $\alpha,\beta,\gamma\in\{x,y,z\}$. 
	The first two terms describe the precession of the spin. 
	The third term describes a periodic kick 
	at integer time $n$. 
	We introduce the dissipation by considering the action of the quantum map in the Kraus form $\Phi (\rho) = \sum_{a} K_{a} e^{-iH} \rho e^{iH} K^\dagger_a$, 
	where
	$K_{1,2} = (J_x\pm i J_y)/ \sqrt{j(j+1)}$ and $K_3 = \sqrt{2} J_z/ \sqrt{j(j+1)}$ such that the constraint $\sum_{a} K^\dagger_a K_a = \mathbb{1}$ is satisfied to ensure trace preservation and complete positivity \cite{nielsen_chuang_2010}. 
	The time evolved density matrix is obtained by a successive action of $\Phi$, i.e. $\rho(t)=\Phi^t(\rho)$.
	%
	%We represent $\Phi$ as a superoperator, i.e. $\Phi = \sum_{a} (K_{a}\otimes K^*_{a})(e^{-iH}\otimes e^{iH})$.
	%
	We represent $\Phi$ as a superoperator, i.e. $\Phi = \sum_{a} (K_{a}\otimes K^*_{a})(U\otimes U^*)$, where $U = \mathcal{T}\exp[ -i \int_0^1 dt\,  H(t)]$ and $\mathcal{T}$ is the time-ordering.
	Note that the total angular momentum $\vec{J}^2$ and the parity are conserved and we will therefore study the restricted Hilbert space of size roughly half of $(2j+1)^2$.
	We analyse the DSFF of the spectrum of $\Phi$ in the symmetric subspace.
	Note that the spectrum is symmetric across the real axis~\cite{supplementary}.
	As shown in Fig.~\ref{fig:qkt}, as we turn on and off the kick parameter $k_1$, the DSFF coincides with the GinOE (for all angles~\cite{supplementary}) and Poisson behaviour with excellent agreement. Note that the DSFF of the corresponding Liouvillian operators of the Lindblad form of the QKT also converges to GinOE DSFF behaviour. 
	%
	%There are deviation from GinUE behaviour 
	%\red{We find that the DSFF behaviour converges to the one of GinOE} for angles close to $\theta=0, \pi/2$ due to the fact that eigenvalues of quantum channel are either real or come in complex conjugate pairs \cite{supplementary}. 

	\begin{figure}[t]
		\centering
		\includegraphics[width=0.475 \textwidth  ]{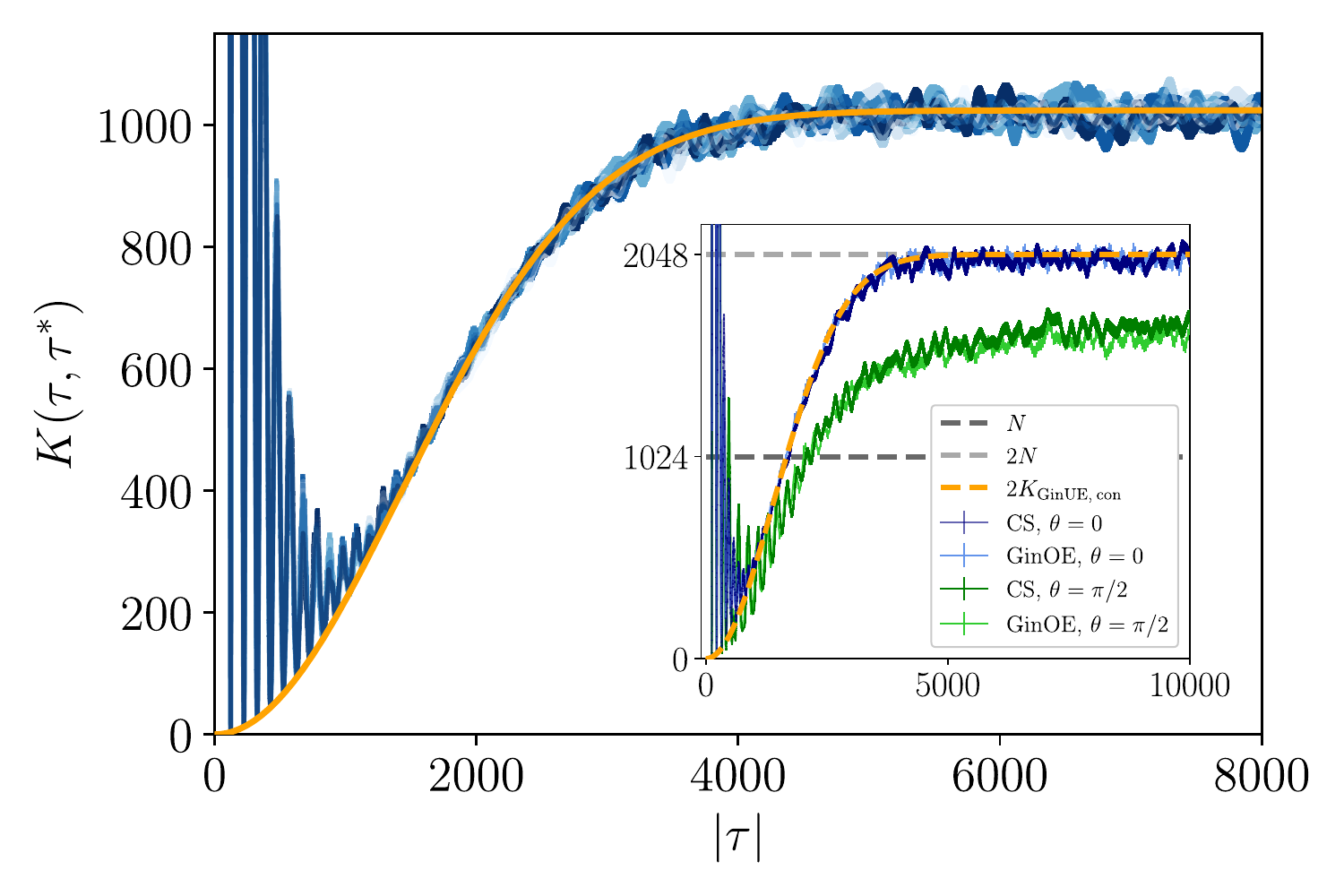}
		\caption{Main: DSFF of the random classical stochastic ensemble induced by CUE for $N= 1024$ with sample size 5000. The blue colors are for fixed $\theta \in [ \pi/16 , 7\pi/16]$ in steps of $\pi/16$. The data can be fitted with the connected part of GinUE (orange line).
			Inset: DSFF for CS and GinOE at angles $\theta=0, \pi/2$ \cite{supplementary}.
		} \label{fig:cs}
	\end{figure}

	\textit{Classical stochastic systems.}
	Another interesting class of non-Hermitian matrices are classical stochastic  matrices or Markov chains, which are matrices with real positive entries and with each column summing to unity. 
	A particular way to generate an ensemble of random stochastic matrices is to consider matrix $S$ with entries $S_{ij} = {|M_{ij}|^2}/{\sum_{i} |M_{ij}|^2}$,
	where $M_{ij}$ is a matrix chosen from a certain matrix ensemble.
	We consider classical stochastic (CS) matrices induced from two random ensembles: the circular unitary ensemble (CUE) (whose induced stochastic ensemble is called the uni-stochastic ensemble) and the GinUE.   
	Unistochastic matrices arise in the context of quantum graphs~\cite{Tanner_2001, Tanner_2000, Smilansky2006} and in the theory of majorization and characterization of quantum maps~\cite{marshall11, nielsen1999, nielsen2000a, cirac2002}.
	By the Perron-Frobenius theorem, the stochastic matrices have leading eigenvalues of unity, and the spectra are again symmetric across the real axis~\cite{Zyczkowski_2003, supplementary}.
	We plot the DSFF for the uni-stochastic ensemble in Fig.~\ref{fig:cs} and the other GinUE-induced ensemble in the supplementary material~\cite{supplementary}. 
	For both ensembles, the DSFF behaviour coincides with the GinOE behaviour (for all angles~\cite{supplementary}) with excellent agreement.
	%, except near the special angles $\theta=0$ and $\pi/2$ \red{where DSFF behaviour is GinOE-like}~\cite{supplementary}.
	
	%
	%

	\textit{Discussion.} 
	We have proposed and exactly computed DSFF for GinUE and a Poissonian-random spectrum as minimal models of dissipative quantum chaotic and integrable systems. In particular, we show that DSFF for GinUE has a dip-ramp-plateau behaviour with a quadratic ramp, and numerically demonstrated the universality of the result with the example of QKT and random classical stochastic ensembles.
	This work open up many exciting directions: DSFF can be used to classify dissipative quantum chaotic systems in different universality and symmetry classes, beyond the nearest-neighbour spacing distribution studied previously~\cite{Magnea_2008, ueda2019, Bernard2002, ueda2020univclass},
	and to unveil deviation of open quantum many-body systems from RMT behaviours at early time (cf. \cite{kos_sff_prx_2018, bertini2018exact, Roy2020random, flack2020statistics, cdc1, cdc2, friedman2019, moudgalya2020spectral, garratt2020manybody, Gharibyan_2018, chan2020lyap}). 
	In particular, it can be used to investigate the spectral properties across the measurement-induced phase transition \cite{Skinner_2019, Li_2018, Li_2019, chan2019, altman2019, Jian_2020, Zabalo_2020}. 
	These directions will be discussed in an upcoming work~\cite{upcoming}. 
	
	Lastly, note that DSFF contrasts with a related observable called \textit{dissipative form factor}~\cite{can2019a} (DFF) in several ways: DFF is a one-parameter function defined for the (Lindblad) superoperators, and the correlation between the imaginary parts of eigenvalues contribute to the DFF as an exponential factor (as supposed to a phase in DSFF). This makes DFF useful in capturing the scaling of the spectral gap, while DSFF is beneficial in unraveling the correlation between eigenvalues in the bulk of the spectrum.
	
	\textit{Acknowledgement.}
	We are thankful to Fiona Burnell, David Huse, Abhinev Prem, Shinsei Ryu, Lucas S\'{a}, Shivaji Sondhi and Salvatore Torquato for helpful discussions.
	AC is particularly grateful to Tankut Can for pointing out the connections to static structure factor and hyperuniformity, and to John Chalker and Andrea De Luca for collaborations on related projects.
	AC is supported by fellowships from the Croucher foundation and the PCTS at Princeton University. 
	TP acknowledges ERC Advanced grant 694544-OMNES and ARRS research program P1-0402.

	\bibliography{biblio}

	\onecolumngrid
	\newpage 
	
	\appendix
	\setcounter{equation}{0}
	\setcounter{figure}{0}
	\renewcommand{\thetable}{S\arabic{table}}
	\renewcommand{\theequation}{S\thesection.\arabic{equation}}
	\renewcommand{\thefigure}{S\arabic{figure}}
	\setcounter{secnumdepth}{2}
	
	\begin{center}
		{\Large Supplementary Material \\ 
			\vspace{0.2cm}
			\titleinfo
		}
	\end{center}
	In this supplementary material we provide additional details about:
	\begin{enumerate}[label=\Alph*]
		\item Dissipative spectral form factor (DSFF) for additional ensembles and discussion of DSFF at $\theta=0, \pi/2$
		\item Spectra of single realizations
		\item Density of states
		\item Scaling of Heisenberg time $\tthei$
		\item DSFF for modified spectra with ``projected degeneracies'' 
		\item DSFF for $\theta$ near $0$ and $\pi/2$
		\item Scaling of the critical angle $\thetas$
		\begin{itemize}
			\item[1.] Definition of $\thetas$
			\item[2.] Scaling of $\thetas$
			\item[3.] Arguments for the scaling of $\thetas$
		\end{itemize}
	\end{enumerate}

	\section{DSFF for additional ensembles and discussion of DSFF at $\theta = 0, \pi /2$}\label{app:dsff_ginoe_ginse}
\begin{figure}[H]
		\begin{minipage}[t]{0.469\textwidth}
			\includegraphics[width=\linewidth,keepaspectratio=true]{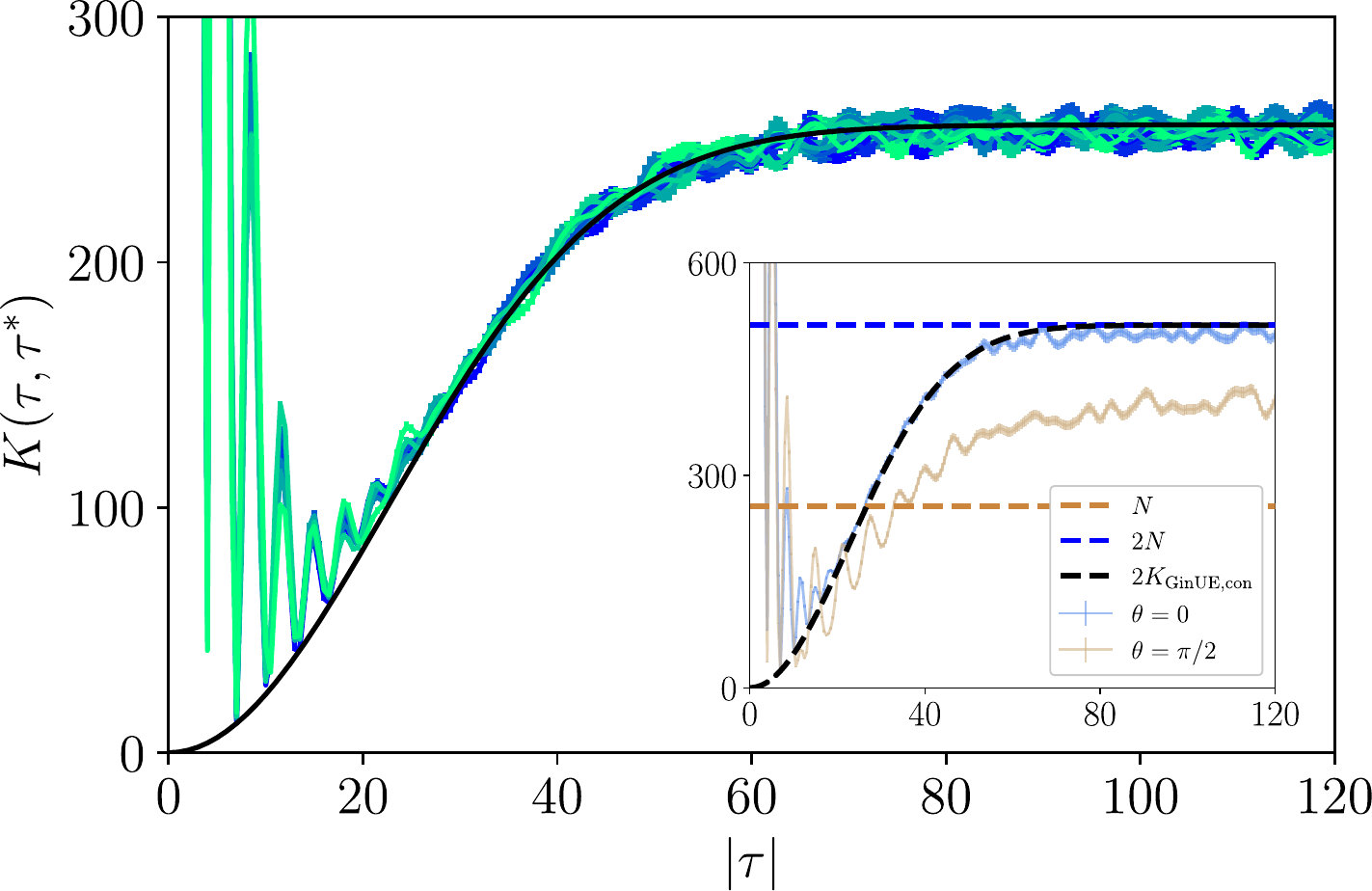}
		\end{minipage}
		\hspace*{\fill} % it's important not to leave blank lines before and after this command
		\begin{minipage}[t]{0.48\textwidth}
			\includegraphics[width=\linewidth,keepaspectratio=true]{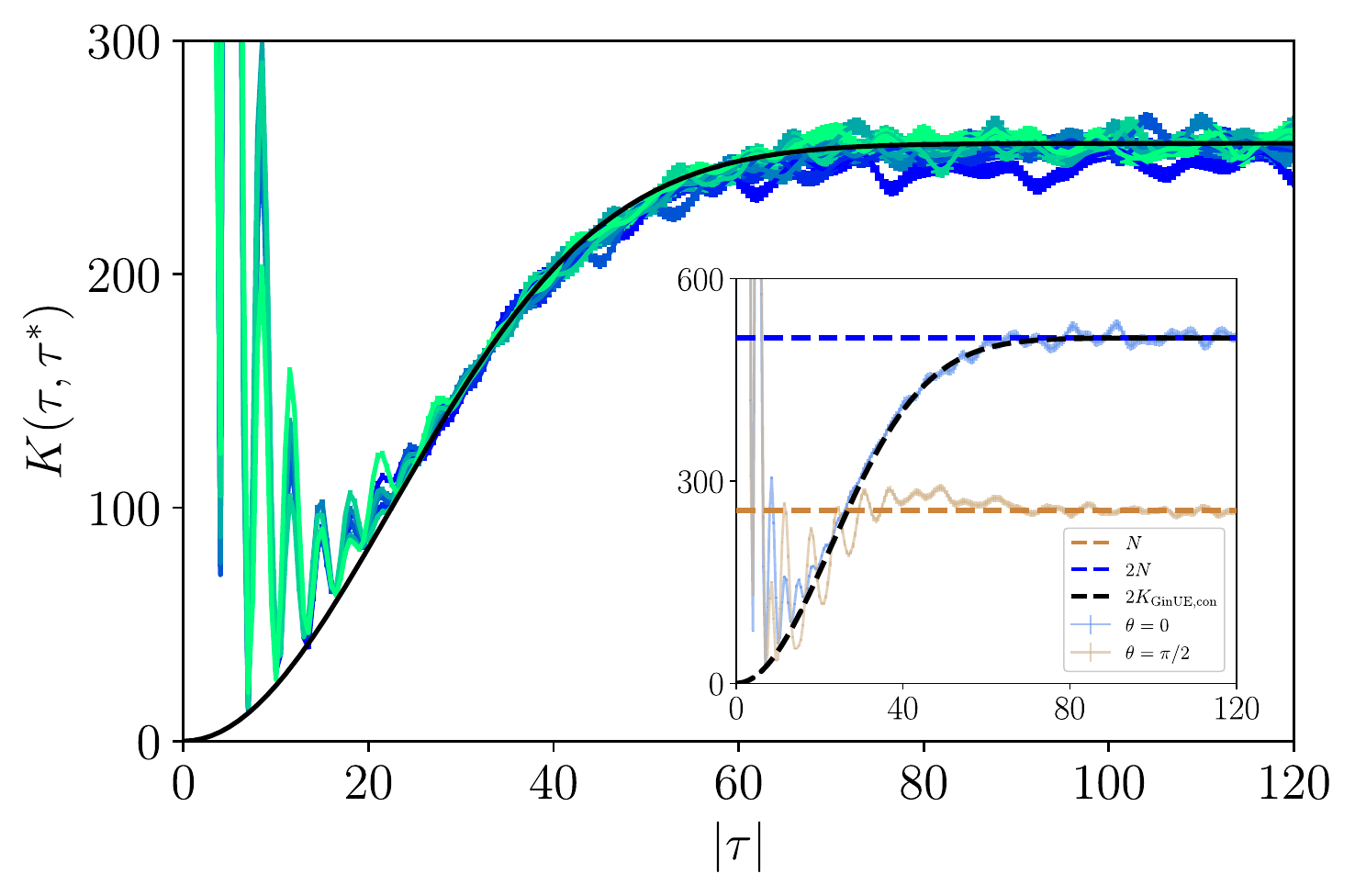}
		\end{minipage}
		\caption{Main: DSFF of the GinOE (left) and GinSE (right)  for $N= 256$ with sample sizes 5000. The blue/green colors are for fixed $\theta \in [ \pi/16 , 7\pi/16]$ in steps of $\pi/16$. The data are fitted with the connected part of GinUE (black line).
			Inset: DSFF for angle $\theta=0, \pi/2$. We find that DSFF at $\theta=0$ can be fitted with twice the connected part of DSFF for GinUE.
		}
		\label{fig:dsff_ginoe_ginse}
	\end{figure}

\begin{figure}[H]
		\centering
		\includegraphics[scale = 0.6]{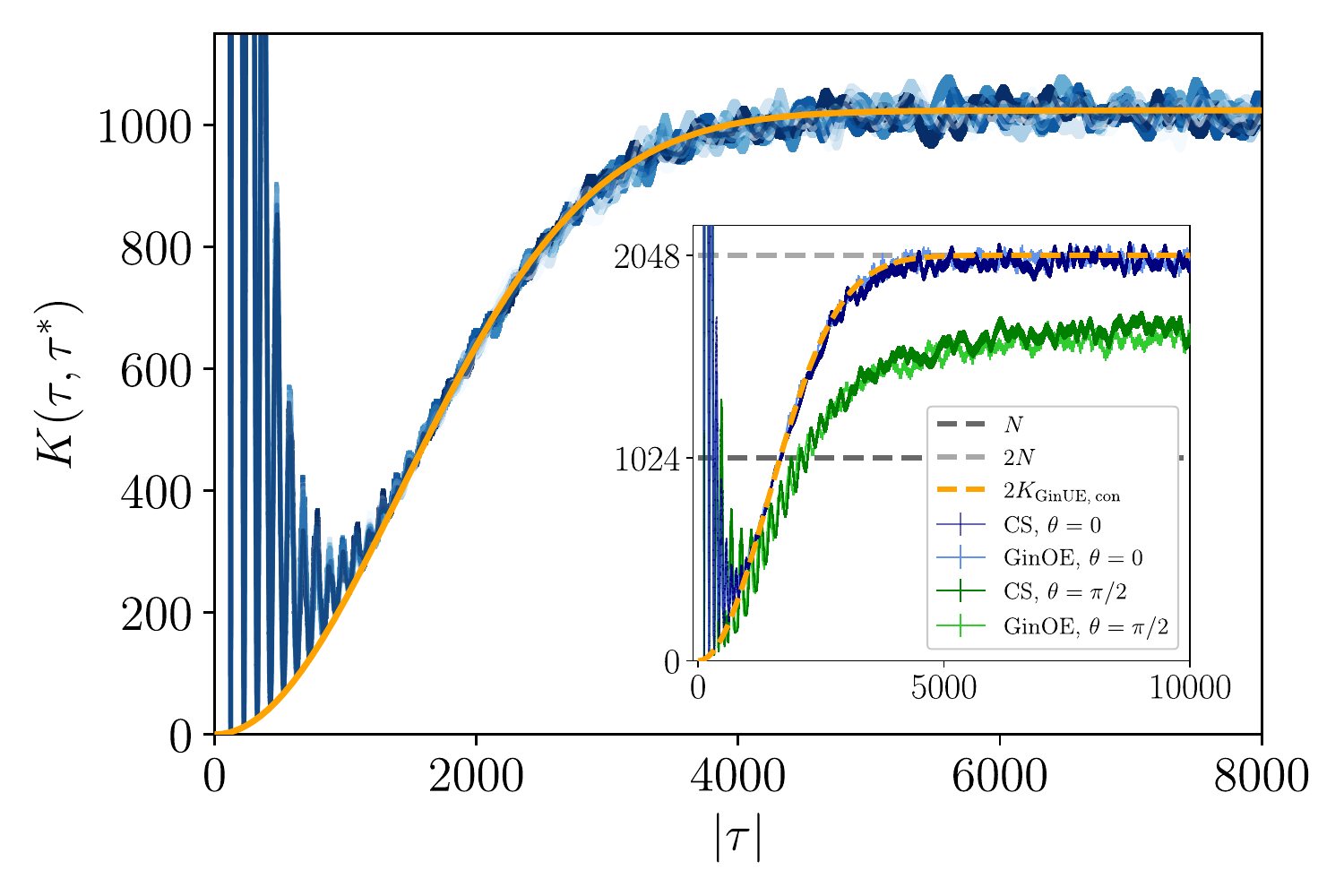}
		\caption{Main: DSFF of the ensembles of classical stochastic matrices induced by GinUE for matrix size $N= 1024$ and sample size 5000. The blue colors are for fixed $\theta \in [ \pi/16 , 7\pi/16]$ in steps of $\pi/16$. The data is fitted with the connected part of GinUE (orange line).
			Inset: DSFF for CS and GinOE at angles $\theta=0, \pi/2$. We find that DSFF at $\theta=0$ can be fitted with twice the connected part of DSFF for GinUE.
		} \label{fig:cs_GinUE}
	\end{figure}

	Here we present the DSFF for real Ginibre ensemble (GinOE), quaternion real Ginibre ensemble (GinSE) and ensembles of classical stochastic matrices induced by GinUE in Fig.~\ref{fig:dsff_ginoe_ginse} left, Fig.~\ref{fig:dsff_ginoe_ginse} right and Fig.~\ref{fig:cs_GinUE} respectively.
	To discuss DSFF at $\theta=0$ and $\pi/2$, we recall the interpretation of DSFF discussed in the main text below Eq.~\eqref{eq:dsff_def_a}:
	At fixed $\theta$ and as a function of $|\tau|$, DSFF is the SFF of the projection of complex eigenvalues onto the radial axis specified by angle $\theta$.
	At $\theta= 0$, a spectrum that is symmetric across the real axis has a rough 2-fold degeneracy in the projected set of eigenvalues (it is not an exact 2-fold degeneracy due to the real eigenvalues, which do not come in pairs).  
	This gives an explanation of the approximate fit of $2K_{\mathrm{GinUE}}$ along $\theta=0$ for GinOE, GinSE, QKT, the ensembles of classical stochastic matrices (see insets of Fig.~\ref{fig:dsff_ginoe_ginse}, \ref{fig:qkt}, \ref{fig:cs} and \ref{fig:cs_GinUE} respectively). 
	Along $\theta=\pi/2$, the real eigenvalues will be exactly degenerate after the projection onto the radial axis defined by $\theta= \pi/2$, which would lead to a different large-$|\tau|$ value for DSFF.   
	This fact is demonstrated when we compare DSFF of GinOE and GinSE at $\theta=\pi/2$: For GinOE, spectra contain real eigenvalues (see Fig.~\ref{fig:ginue_dos_1r} middle), and DSFF at $\theta=0$ approaches a value larger than matrix size $N$, while for GinSE, spectra have no real eigenvalues (see Fig.~\ref{fig:ginue_dos_1r} right) and DSFF at $\theta=0$ approaches $N$.
	We leave the study of DSFF behaviour near $\theta=0, \pi/2$ for future investigations.
	%%%%%%%%%%%%
	%%%%%%%%%%%%
	%%%%%%%%%%%%
	%%%%%%%%%%%%
	%%%%%%%%%%%%
	%%%%%%%%%%%%
	%%%%%%%%%%%%
	%%%%%%%%%%%% DSFF GinOE GinSE
	
	%%%%%%%%%%%%
	%%%%%%%%%%%%
	%%%%%%%%%%%%
	%%%%%%%%%%%%
	%%%%%%%%%%%%
	%%%%%%%%%%%%
	%%%%%%%%%%%%
	%%%%%%%%%%%% DSFF CS GinUE
	
	\section{Spectra of single realizations}\label{app:spectra}
\begin{figure}[H]
		\begin{minipage}[t]{0.31\textwidth}
			\includegraphics[width=\linewidth,keepaspectratio=true]{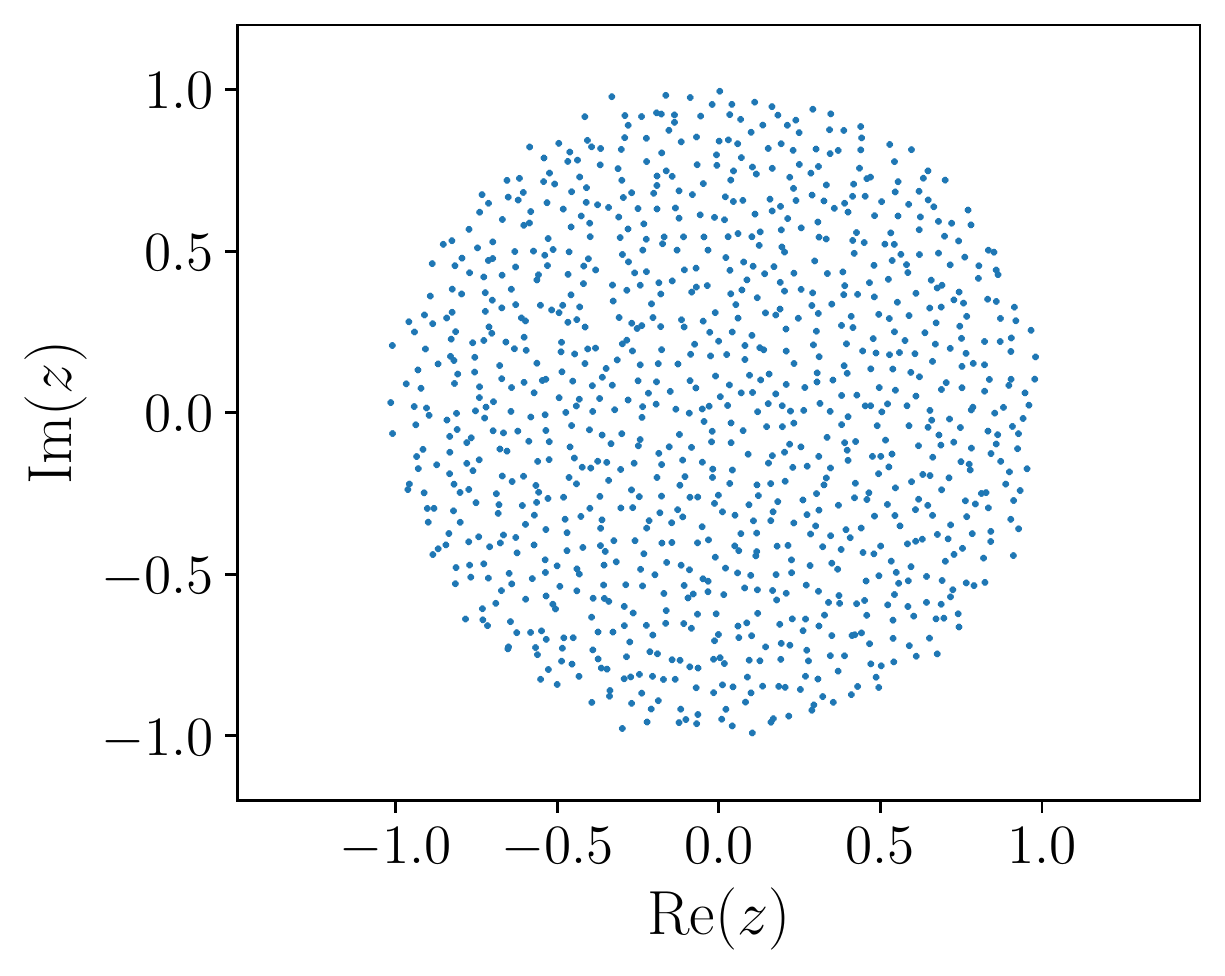}
		\end{minipage}
		\hspace*{\fill} % it's important not to leave blank lines before and after this command
		\begin{minipage}[t]{0.31\textwidth}
			\includegraphics[width=\linewidth,keepaspectratio=true]{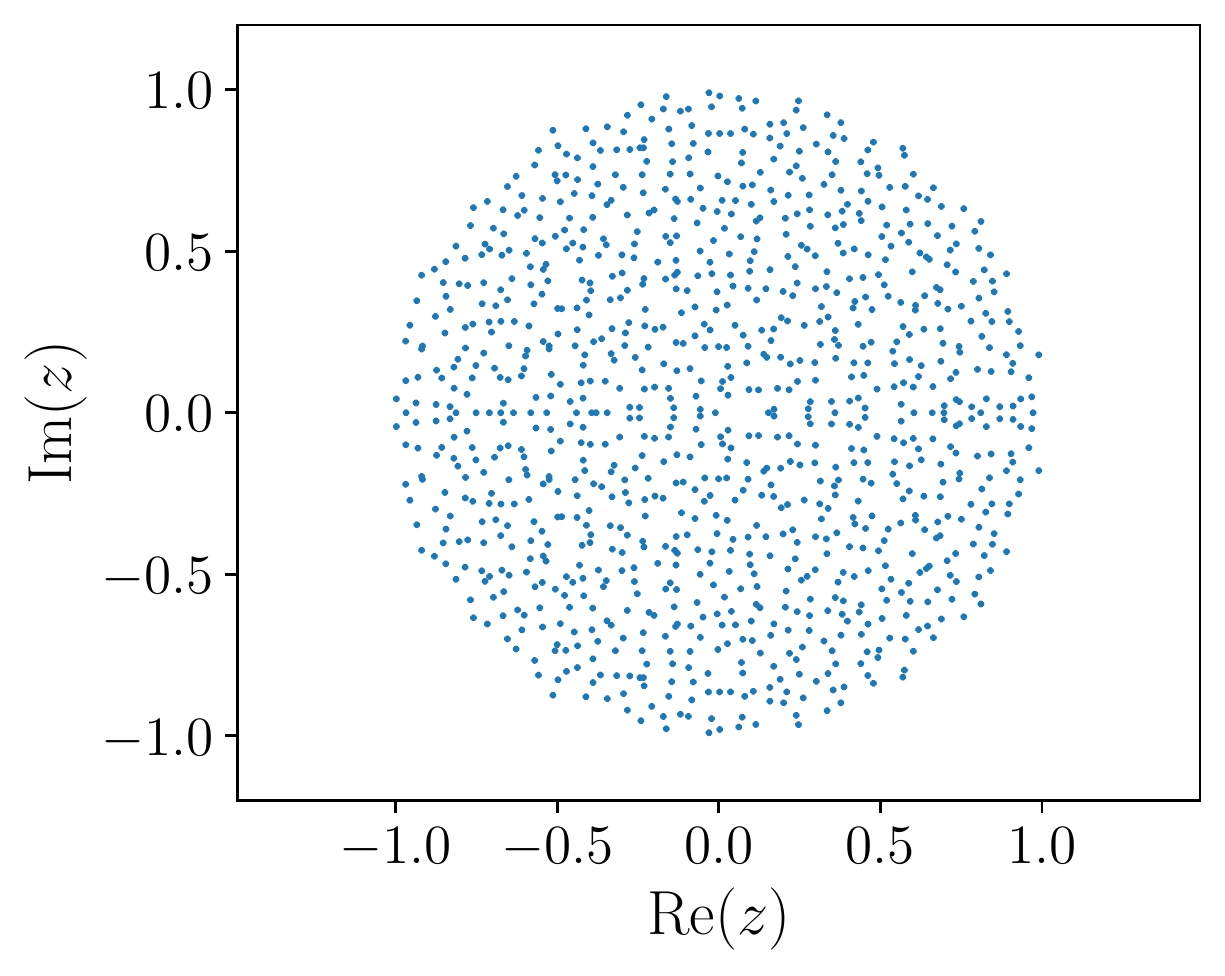}
		\end{minipage}
		\hspace*{\fill} % it's important not to leave blank lines before and after this command
		\begin{minipage}[t]{0.31\textwidth}
			\includegraphics[width=\linewidth,keepaspectratio=true]{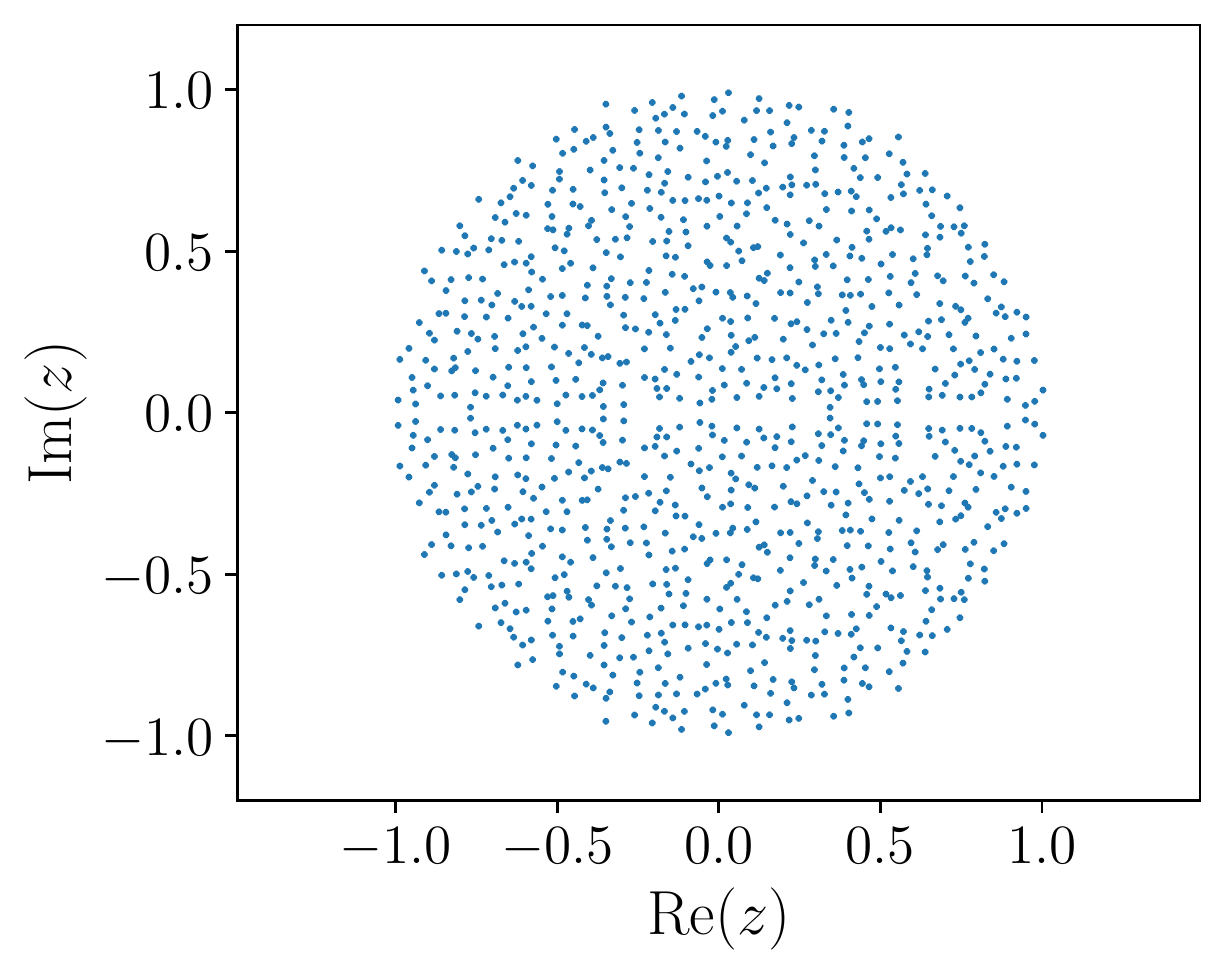}
		\end{minipage}
		\caption{
			Spectra of single realizations of the GinUE (left), GinOE (middle) and GinSE (right) with matrix size $N= 1024$. 
			Each set of eigenvalues lies within the unit circle in the complex plane, shows level repulsion, and has a uniform distribution away from the edge of the circle (and the real axis for GinOE and GinSE). 
			For GinOE (GinSE), there is a finite (zero) eigevalue density along the real axis (cf. Fig.~\ref{fig:ginue_dos_avg}).
		} \label{fig:ginue_dos_1r}
	\end{figure}

\begin{figure}[H]
		\begin{minipage}[t]{0.469\textwidth}
			\includegraphics[width=\linewidth,keepaspectratio=true]{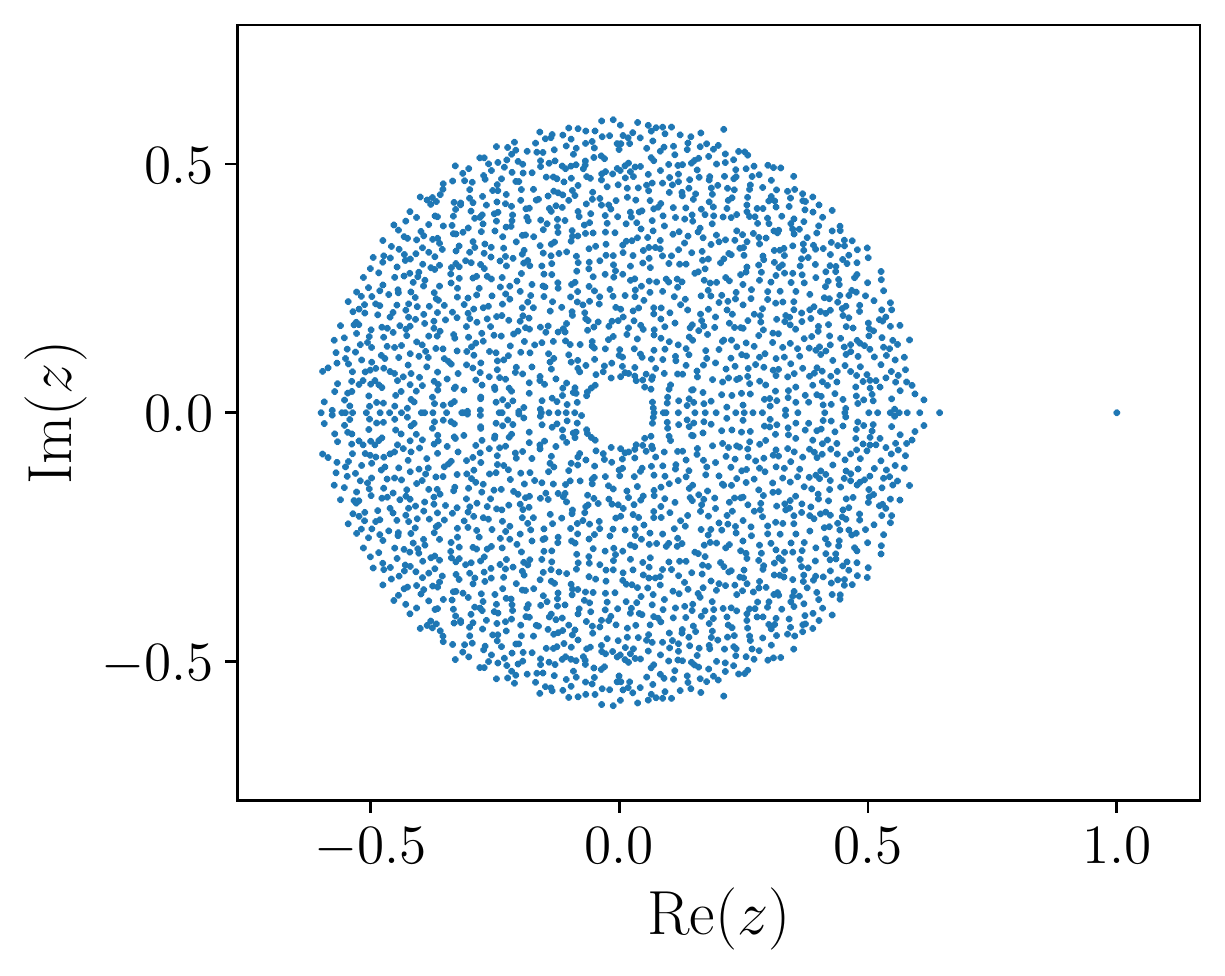}
		\end{minipage}
		\hspace*{\fill} % it's important not to leave blank lines before and after this command
		\begin{minipage}[t]{0.48\textwidth}
			\includegraphics[width=\linewidth,keepaspectratio=true]{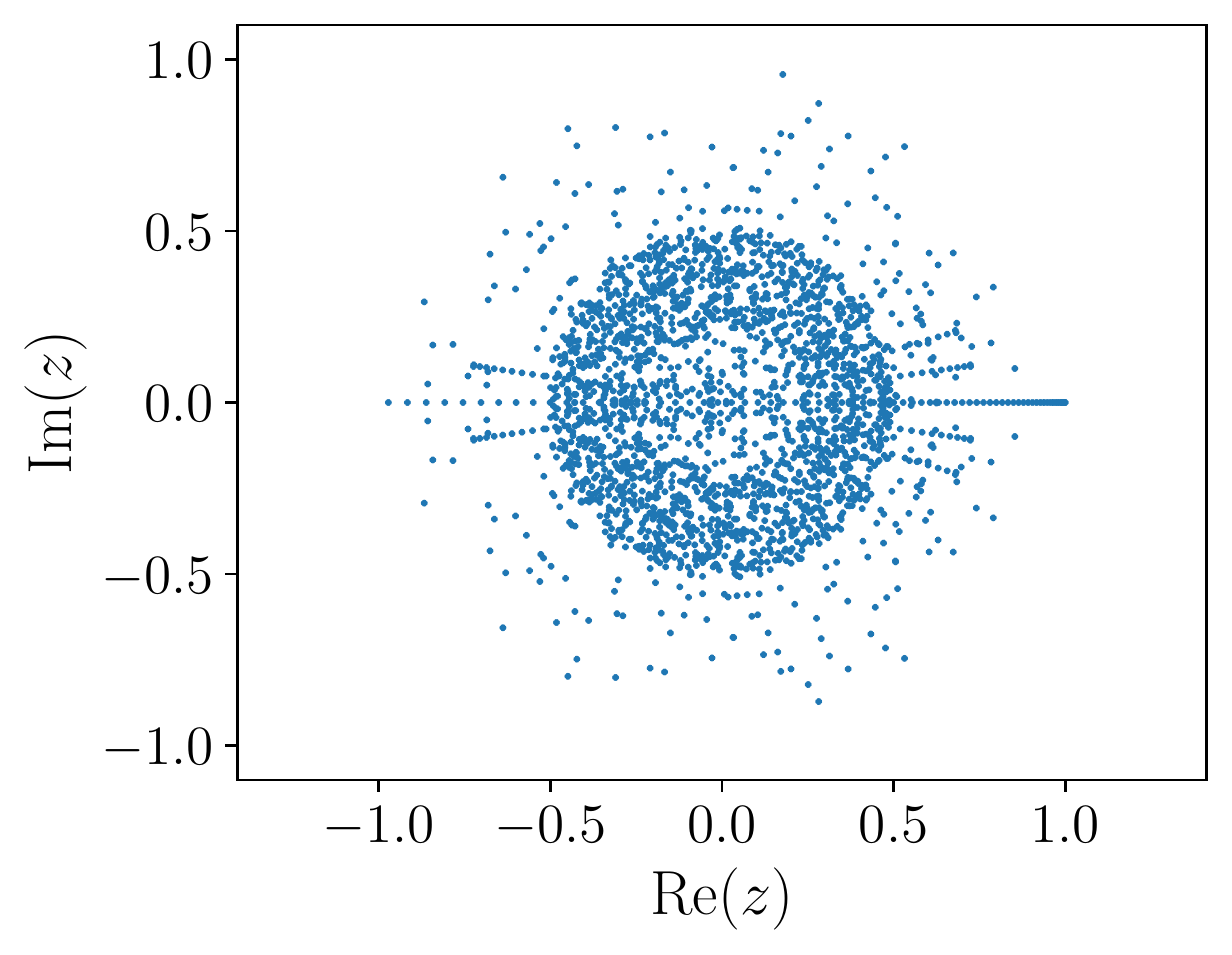}
		\end{minipage}
		\caption{Spectra of single realizations of the QKT with dissipation for $j=35$, Gaussianly-distributed $p \in \mathcal{N}(2,2/3)$ and $k_0 \in \mathcal{N}(10,3)$. 
			The left and right figures are for the QKT with the kick with $k_1 = 8$, and without the kick with $k_1 = 0$ respectively.
			Both spectra have leading eigenvalues of unity and are symmetric across the real axis.
			QKT with the kick (without the kick) is dissipative quantum chaotic (integrable), and %its complex spectrum (does not) shows level repulsion. 
			its DSFF behaviour converges to the one of GinUE  (Poisson) (see Fig.~\ref{fig:qkt} in the main text). 
		}    \label{fig:qkt_1real_dos}
	\end{figure}

\begin{figure}[H]
		\begin{minipage}[t]{0.44\textwidth}
			\includegraphics[width=\linewidth,keepaspectratio=true]{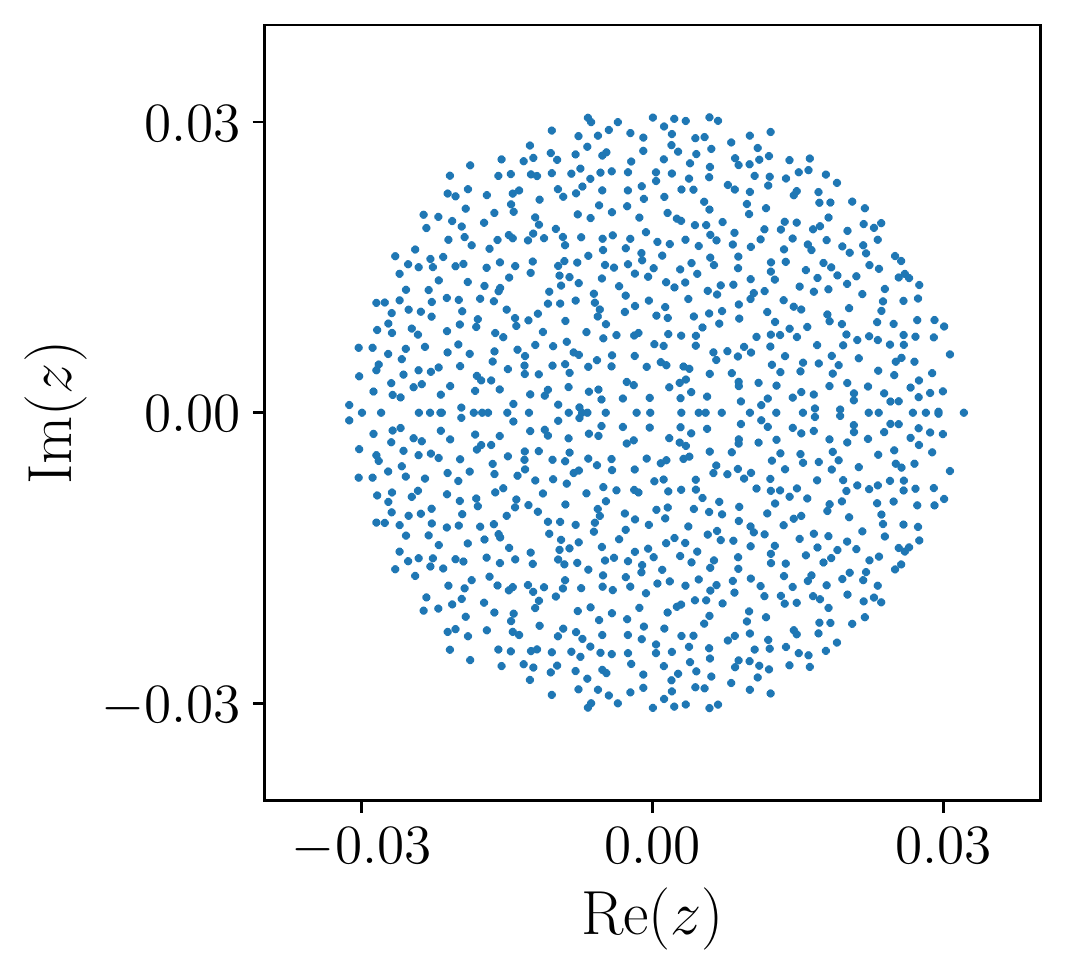}
		\end{minipage}
		\hspace*{\fill} % it's important not to leave blank lines before and after this command
		\begin{minipage}[t]{0.44\textwidth}
			\includegraphics[width=\linewidth,keepaspectratio=true]{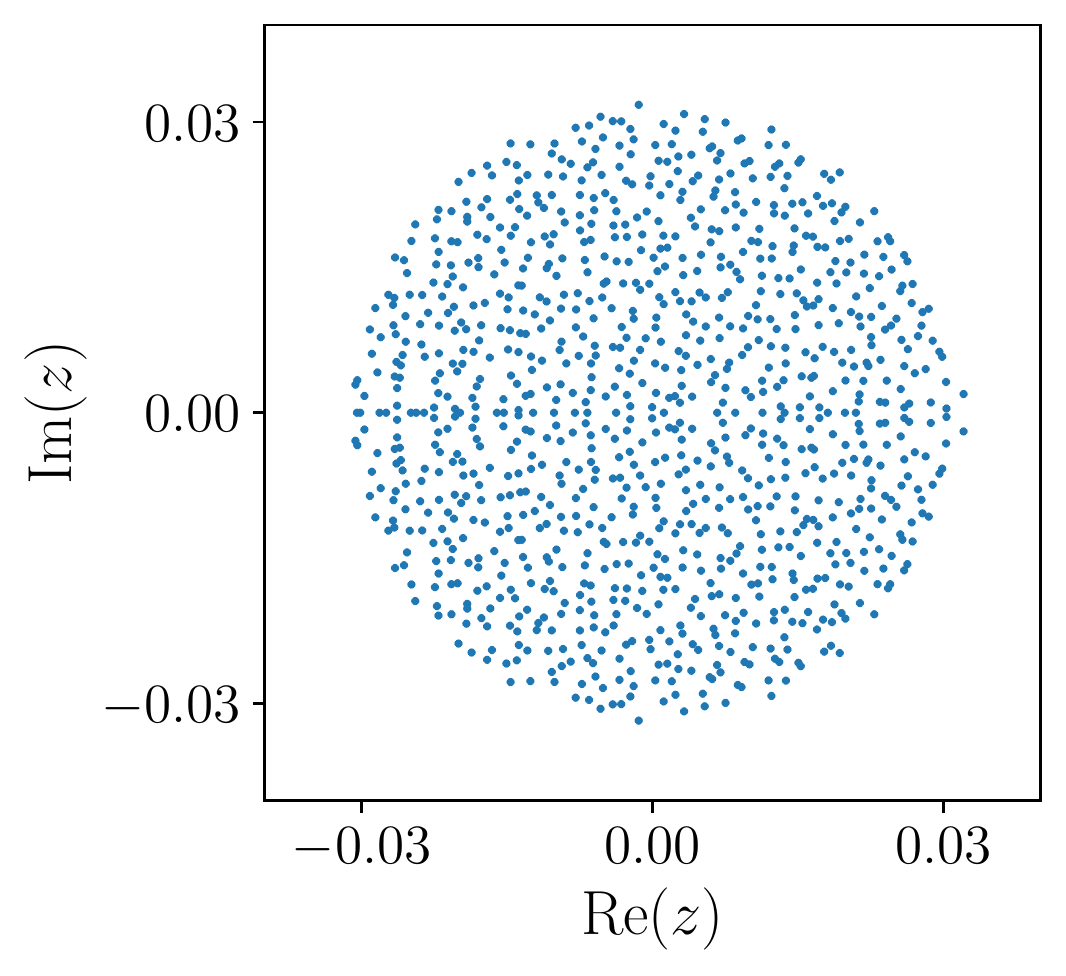}
		\end{minipage}
		\caption{Spectra of single realizations of the ensembles of classical stochastic matrices induced from the CUE (left) and GinUE (right) with matrix size $N= 1024$. 
			There is a leading eigenvalue of unity (not shown) and the spectrum is symmetric across the real axis.
			The spectrum exhibits level repulsion and its DSFF converges to the one of GinUE (see Fig.~\ref{fig:cs} and \ref{fig:cs_GinUE}).
		}
		\label{fig:cs_GinUE_dos_1r}
	\end{figure}

	In this section, we provide the plots of spectra of single realizations of the ensembles considered in the main text, namely, the DOS of GinUE, GinOE, GinSE (Fig.~\ref{fig:ginue_dos_1r}), QKT with and without the kick (Fig.~\ref{fig:qkt_1real_dos}), ensembles of classical stochastic matrices induced by CUE and GinUE (Fig.~\ref{fig:cs_GinUE_dos_1r}).  
	
	%%%%%%%%%%%%%%%%%%%%%%%%
	%%%%%%%%%%%%%%%%%%%%%%%%
	%%%%%%%%%%%%%%%%%%%%%%%%
	%%%%%%%%%%%%%%%%%%%%%%%%
	%%%%%%%%%%%%%%%%%%%%%%%% GinUE GinOE GinSE DOS 1R
	
	%%%%%%%%%%%%%%%%%%%%%%%%
	%%%%%%%%%%%%%%%%%%%%%%%%
	%%%%%%%%%%%%%%%%%%%%%%%%
	%%%%%%%%%%%%%%%%%%%%%%%%
	%%%%%%%%%%%%%%%%%%%%%%%% QKT DOS 1R
	
	%%%%%%%%%%%%%%%%%%%%%%%%%%%%%%%%%%%%%%%%%%%%%%%%%%
	
	%%%%%%%%%%%%%%%%%%%%%%%%
	%%%%%%%%%%%%%%%%%%%%%%%%
	%%%%%%%%%%%%%%%%%%%%%%%%
	%%%%%%%%%%%%%%%%%%%%%%%%
	%%%%%%%%%%%%%%%%%%%%%%%% CS DOS 1R
	
	%%%%%%%%%%%%%%%%%%%%%%%%%%%%%%%%%%%%
	%%%%%%%%%%%%%%%%%%%%%%%%%%%%%%%%%%%%
	%%%%%%%%%%%%%%%%%%%%%%%%%%%%%%%%%%%%
	%%%%%%%%%%%%%%%%%%%%%%%%%%%%%%%%%%%%
	%%%%%%%%%%%%%%%%%%%%%%%%%%%%%%%%%%%%
	%%%%%%%%%%%%%%%%%%%%%%%%%%%%%%%%%%%%
	%%%%%%%%%%%%%%%%%%%%%%%%%%%%%%%%%%%%
	%%%%%%%%%%%%%%%%%%%%%%%%%%%%%%%%%%%%
	%%%%%%%%%%%%%%%%%%%%%%%%%%%%%%%%%%%%
	%%%%%%%%%%%%%%%%%%%%%%%%%%%%%%%%%%%%
	%%%%%%%%%%%%%%%%%%%%%%%%%%%%%%%%%%%%
	%%%%%%%%%%%%%%%%%%%%%%%%%%%%%%%%%%%% DOS averaged
	
	\section{Density of states}\label{app:dos}
\begin{figure}[H]
		\begin{minipage}[t]{0.31\textwidth}
			\includegraphics[width=\linewidth,keepaspectratio=true]{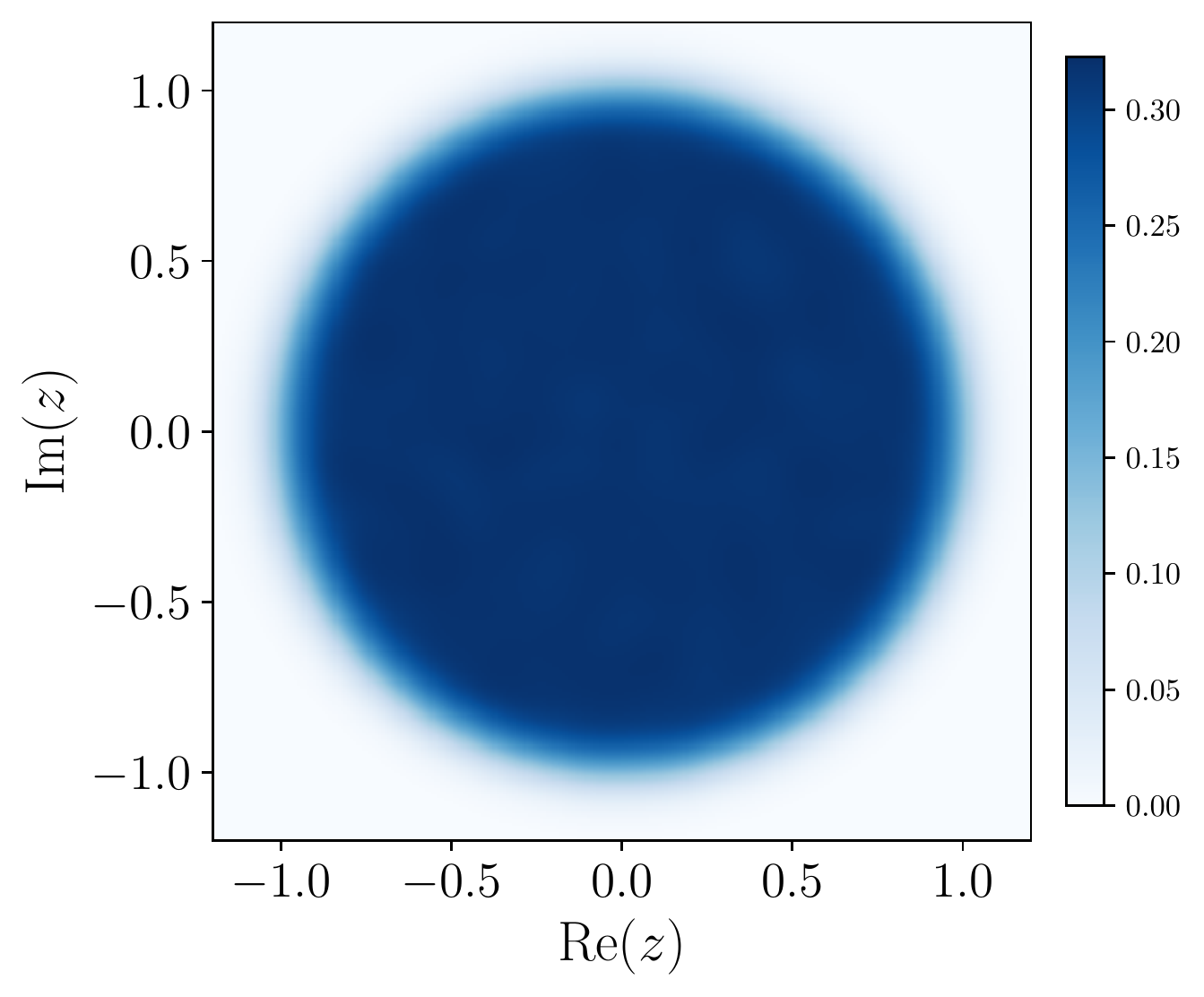}
		\end{minipage}
		\hspace*{\fill} % it's important not to leave blank lines before and after this command
		\begin{minipage}[t]{0.31\textwidth}
			\includegraphics[width=\linewidth,keepaspectratio=true]{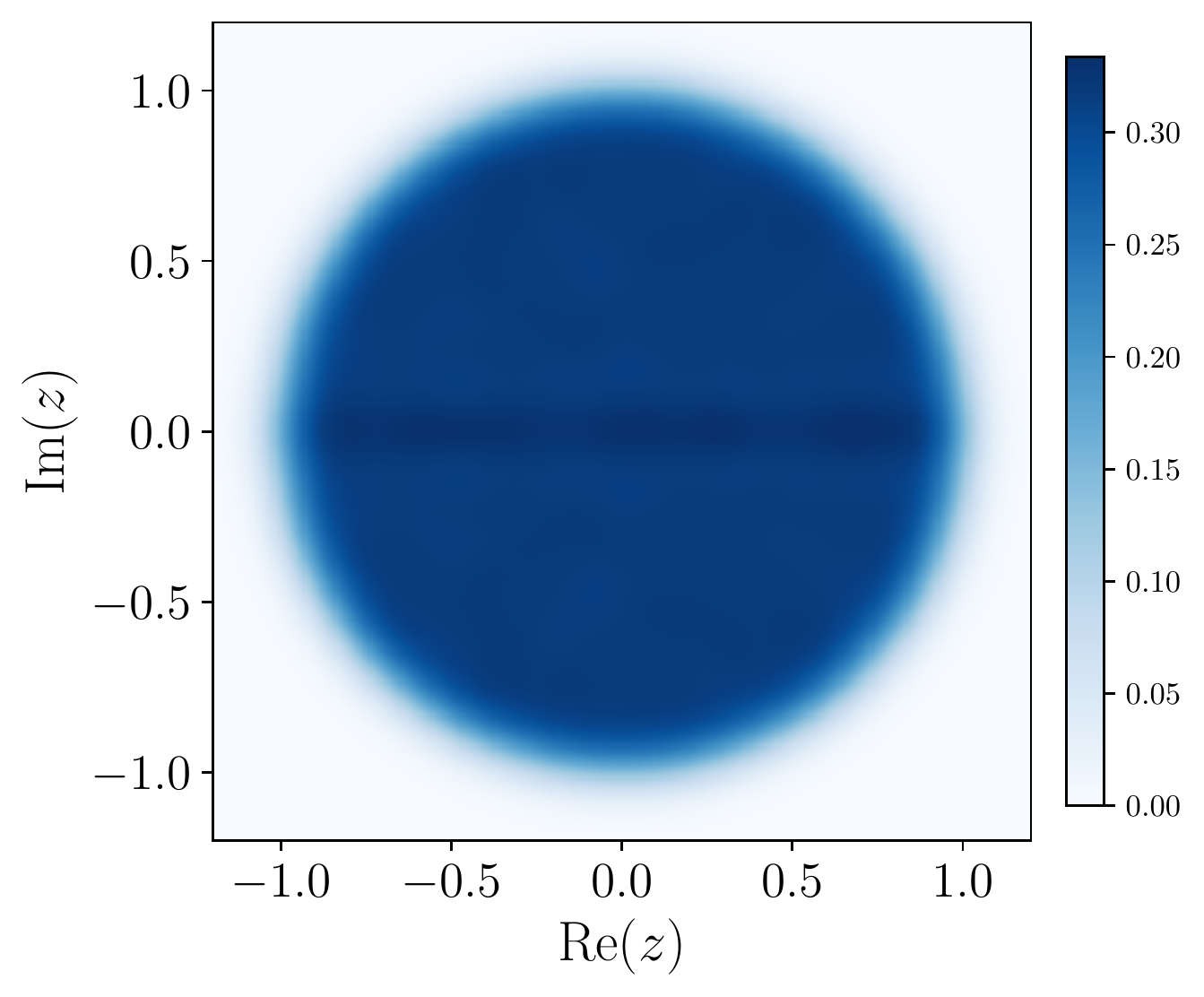}
		\end{minipage}
		\hspace*{\fill} % it's important not to leave blank lines before and after this command
		\begin{minipage}[t]{0.31\textwidth}
			\includegraphics[width=\linewidth,keepaspectratio=true]{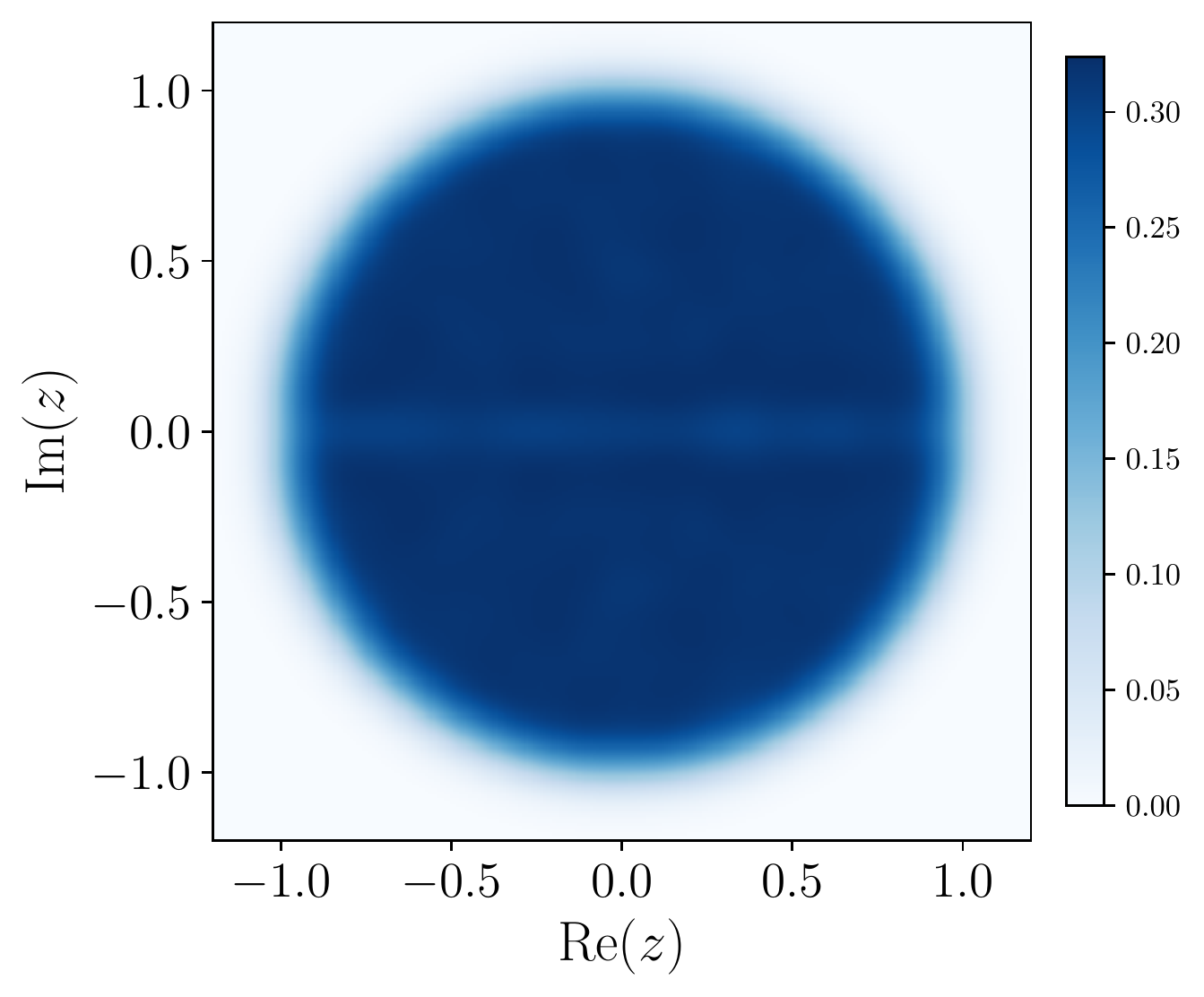}
		\end{minipage}
		\caption{DOS of GinUE (left), GinOE (middle) and GinSE (right) with matrix size $N=256$ and sample size 100.
			Each set of eigenvalues lies within the unit circle in the complex plane, shows level repulsion, and has a uniform distribution away from the edge of the circle (and the real axis for GinOE and GinSE). 
			For GinOE (GinSE), there is a finite (zero) eigevalue density along the real axis (cf. Fig.~\ref{fig:ginue_dos_1r}).
			%
			%\red{All DOS figures are obtained by averaging over a sample of 100.}
			%
		}\label{fig:ginue_dos_avg}
	\end{figure}

\begin{figure}[H]
		\begin{minipage}[t]{0.44\textwidth}
			\includegraphics[width=\linewidth,keepaspectratio=true]{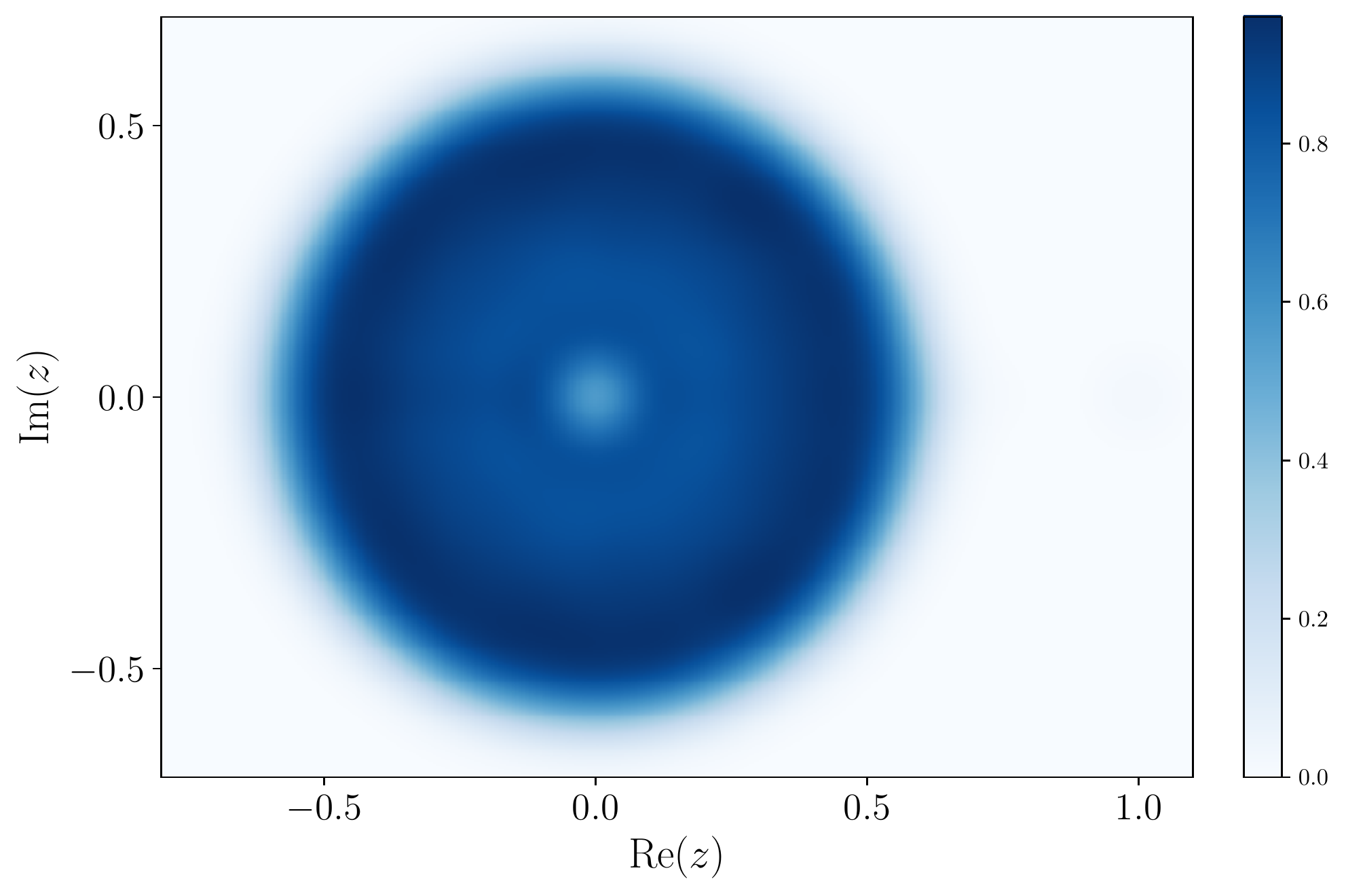}
		\end{minipage}
		\hspace*{\fill} % it's important not to leave blank lines before and after this command
		\begin{minipage}[t]{0.44\textwidth}
			\includegraphics[width=\linewidth,keepaspectratio=true]{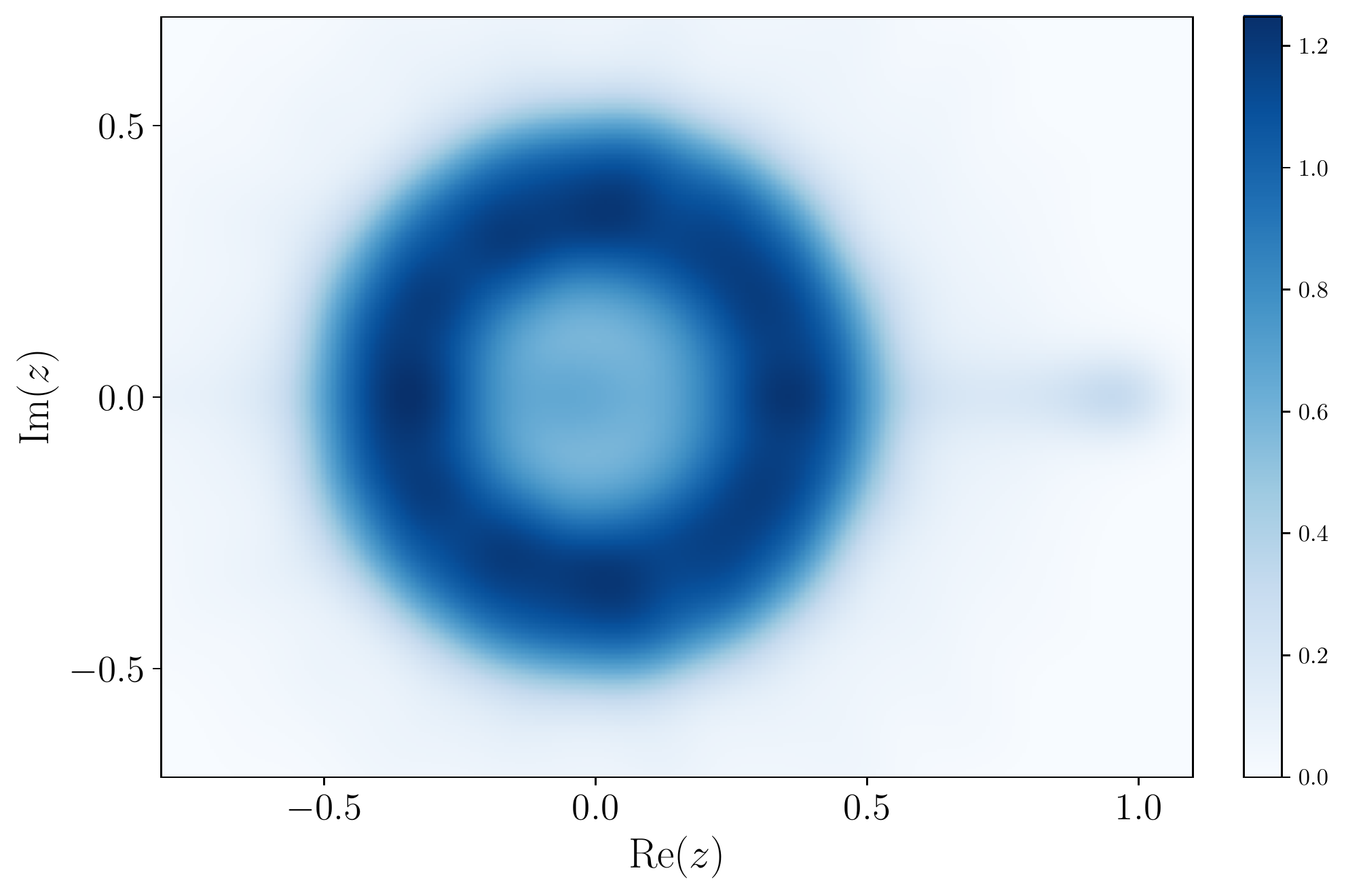}
		\end{minipage}
		\caption{DOS of the QKT with dissipation for $j=35$, %$k_1=8$ (left) $k_1 = 0$ (right), 
			and Gaussianly-distributed $p \in \mathcal{N}(2,2/3)$ and $k_0 \in \mathcal{N}(10,3)$.     %
			The left and right figures are for the QKT with the kick with $k_1 = 8$, and without the kick with $k_1 = 0$ respectively.
			Both DOS figures are obtained by averaging over a sample of 10 realizations.
		}
		\label{fig:qkt_avg_dos}
	\end{figure}

\begin{figure}[H]
		\begin{minipage}[t]{0.44\textwidth}
			\includegraphics[width=\linewidth,keepaspectratio=true]{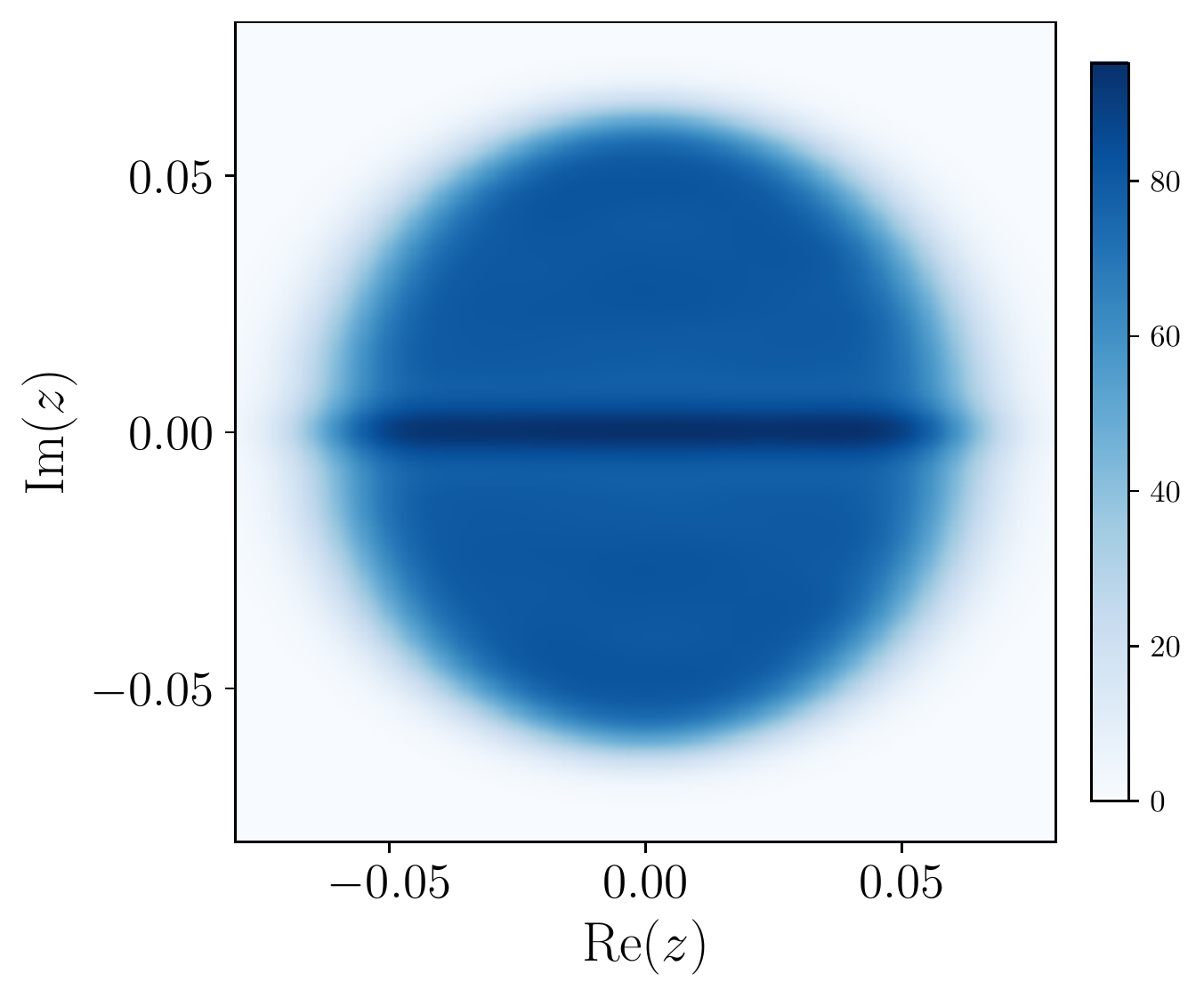}
		\end{minipage}
		\hspace*{\fill} % it's important not to leave blank lines before and after this command
		\begin{minipage}[t]{0.44\textwidth}
			\includegraphics[width=\linewidth,keepaspectratio=true]{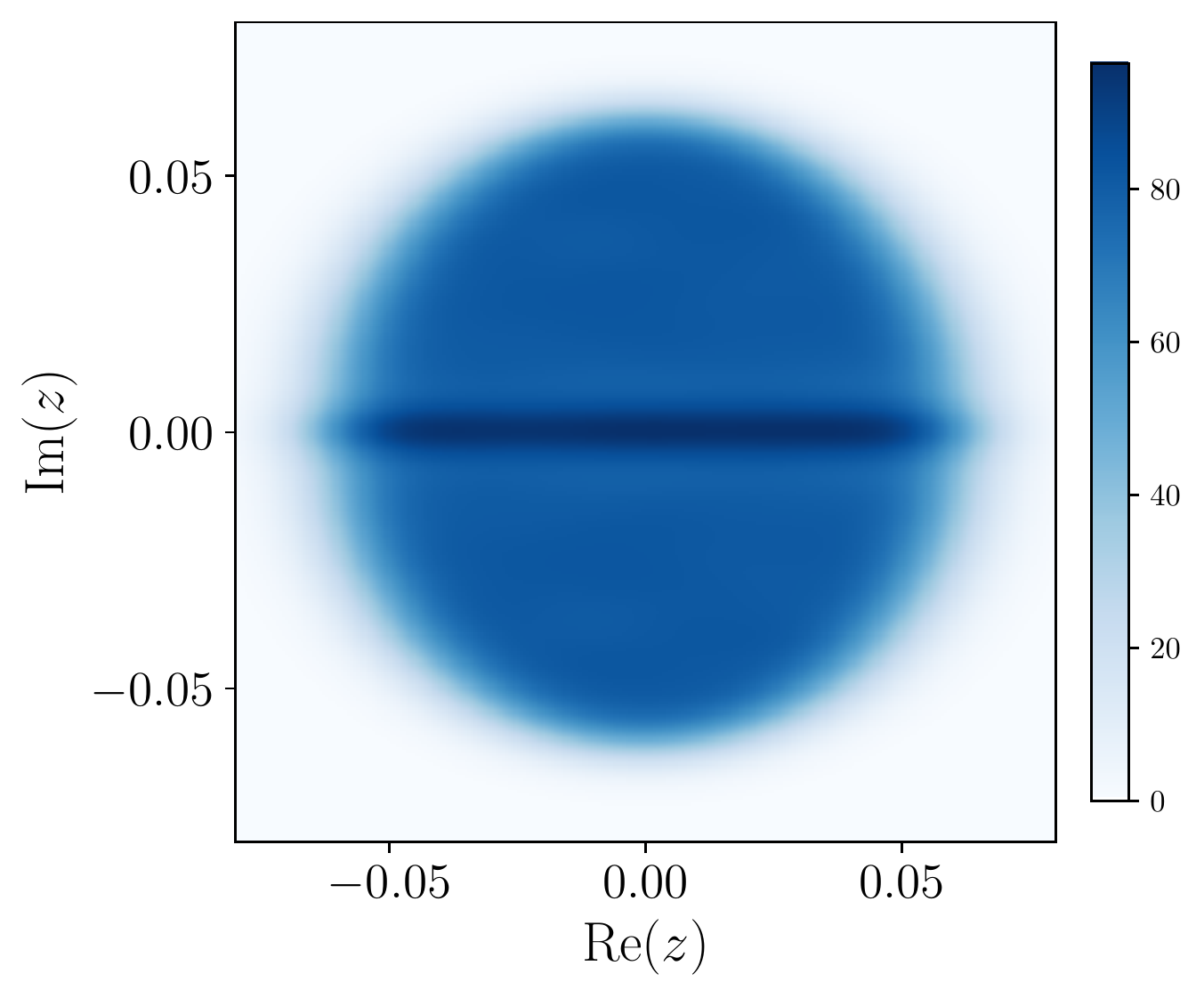}
		\end{minipage}
		\caption{DOS of ensembles of classical stochastic matrices induced from the CUE (left) and the GinUE (right) with matrix size $N=256$ and sample size 1000.
		}
		\label{fig:cs_dos_avg}
	\end{figure}

	In this section, we provide the heat maps of the density of states for the ensembles considered in the main text, namely, the DOS of GinUE, GinOE, GinSE (Fig.~\ref{fig:ginue_dos_avg}), QKT with and without the kick (Fig.~\ref{fig:qkt_avg_dos}), ensembles of classical stochastic matrices induced by CUE and GinUE (Fig.~\ref{fig:cs_dos_avg}).  
	
	%%%%%%%%%%%%%%%%%%%%%%%%
	%%%%%%%%%%%%%%%%%%%%%%%%
	%%%%%%%%%%%%%%%%%%%%%%%%
	%%%%%%%%%%%%%%%%%%%%%%%%
	%%%%%%%%%%%%%%%%%%%%%%%% Ginibre DOS avg
	
	%%%%%%%%%%%%%%%%%%%%%%%%
	%%%%%%%%%%%%%%%%%%%%%%%%
	%%%%%%%%%%%%%%%%%%%%%%%%
	%%%%%%%%%%%%%%%%%%%%%%%%
	%%%%%%%%%%%%%%%%%%%%%%%%
	%%%%%%%%%%%%%%%%%%%%%%%%
	%%%%%%%%%%%%%%%%%%%%%%%%
	%%%%%%%%%%%%%%%%%%%%%%%%
	%%%%%%%%%%%%%%%%%%%%%%%% QKT DOS avg
	
	%%%%%%%%%%%%%%%%%%%%%%%%
	%%%%%%%%%%%%%%%%%%%%%%%%
	%%%%%%%%%%%%%%%%%%%%%%%%
	%%%%%%%%%%%%%%%%%%%%%%%%
	%%%%%%%%%%%%%%%%%%%%%%%% CS DOS avg
	
	%%%%%%%%%%%%%%%%%%%%%%%%
	%%%%%%%%%%%%%%%%%%%%%%%%
	%%%%%%%%%%%%%%%%%%%%%%%%
	%%%%%%%%%%%%%%%%%%%%%%%%
	%%%%%%%%%%%%%%%%%%%%%%%%
	%%%%%%%%%%%%%%%%%%%%%%%%
	%%%%%%%%%%%%%%%%%%%%%%%%
	%%%%%%%%%%%%%%%%%%%%%%%%
	%%%%%%%%%%%%%%%%%%%%%%%% 

	\section{Scaling of Heisenberg time $\tthei$}
\begin{figure}[H]
		\begin{minipage}[t]{0.31\textwidth}
			\includegraphics[width=\linewidth,keepaspectratio=true]{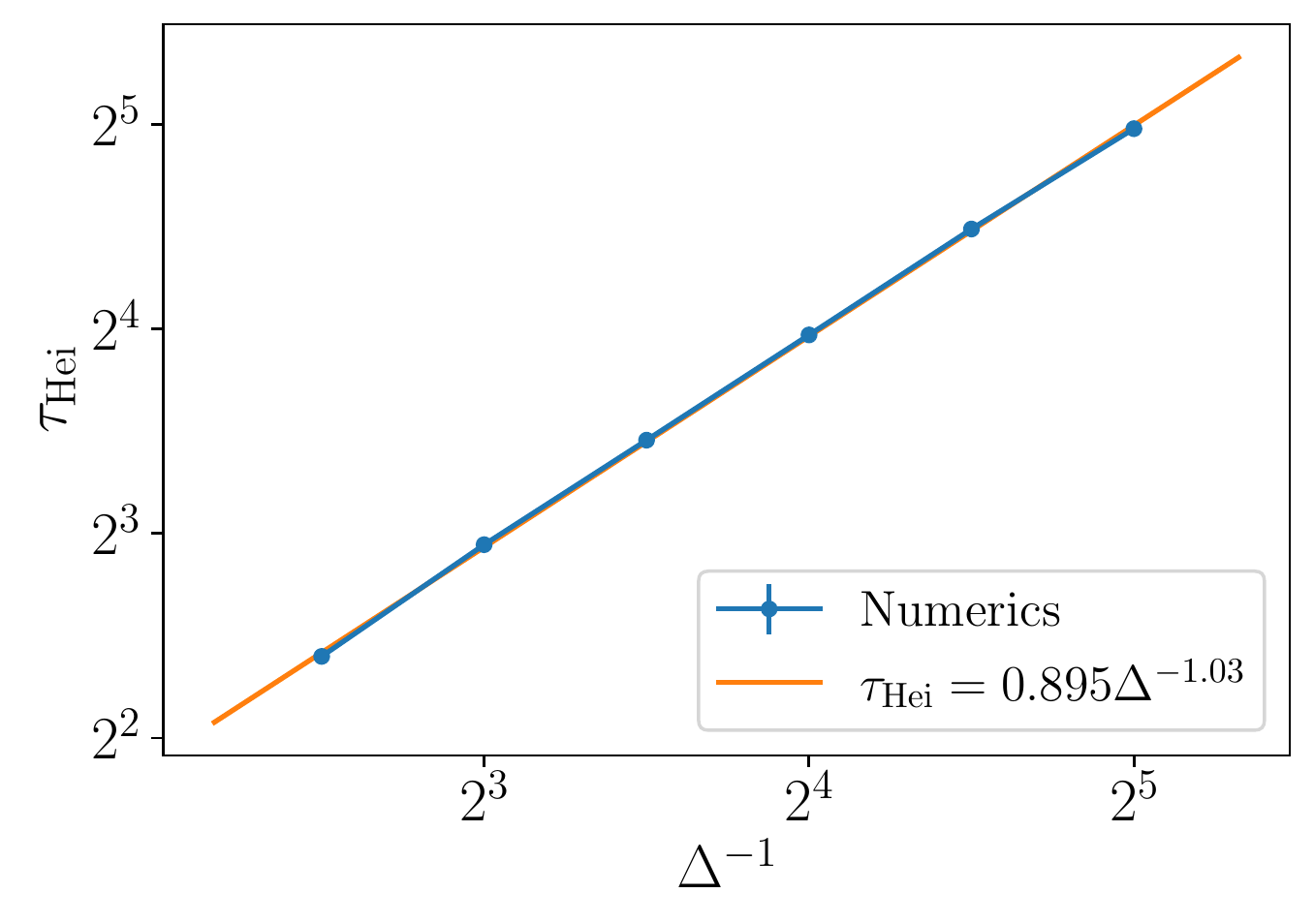}
			\label{subfig:ginue_th}
		\end{minipage}
		\hspace*{\fill} % it's important not to leave blank lines before and after this command
		\begin{minipage}[t]{0.31\textwidth}
			\includegraphics[width=\linewidth,keepaspectratio=true]{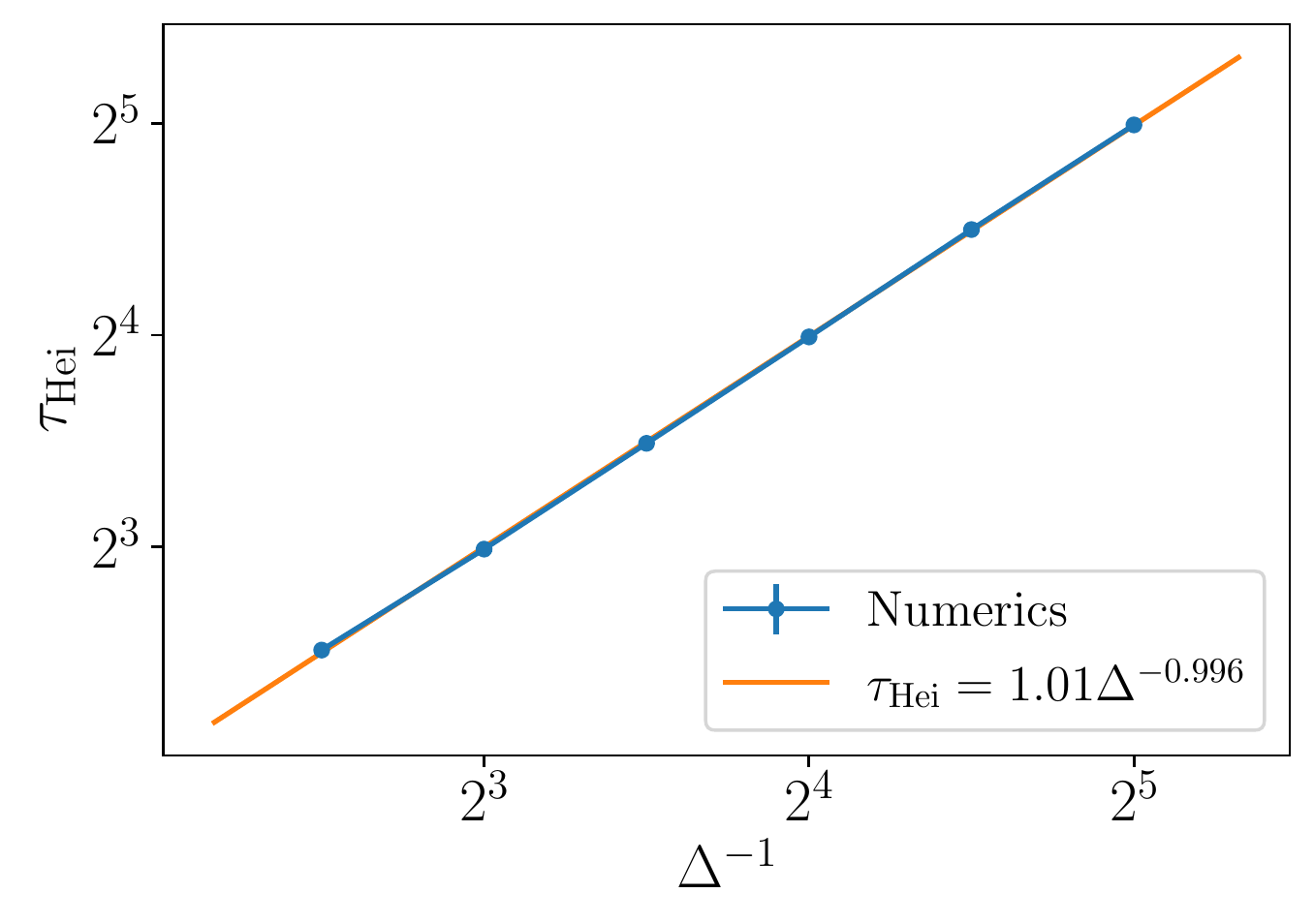}
		\end{minipage}
		\hspace*{\fill} % it's important not to leave blank lines before and after this command
		\begin{minipage}[t]{0.31\textwidth}
			\includegraphics[width=\linewidth,keepaspectratio=true]{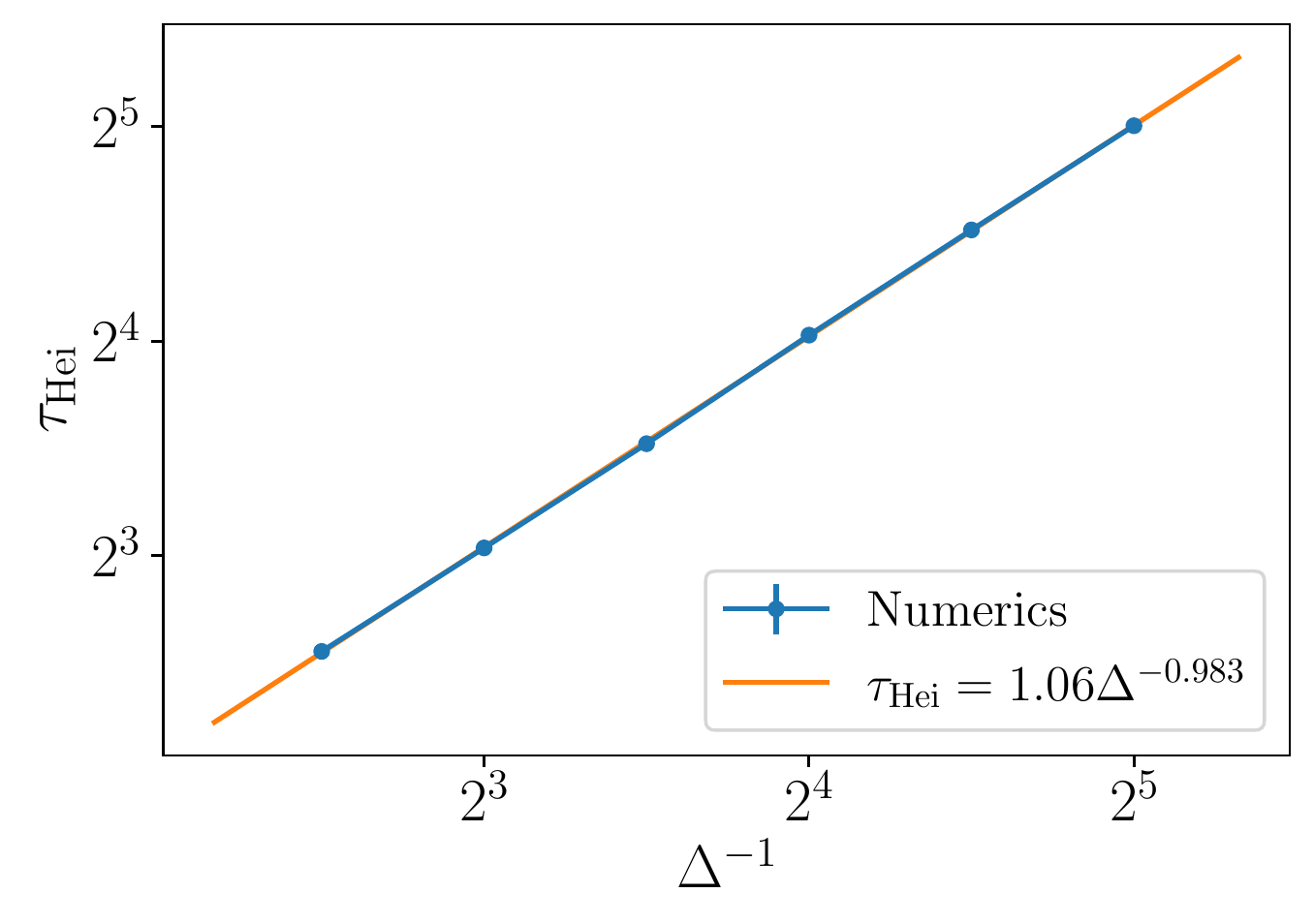}
		\end{minipage}
		\caption{Heisenberg time $\tthei$ vs. inverse level spacing $\Delta^{-1}$ for GinUE (left), GinOE (middle) and GinSE (right).  
			We see that $\tthei$ scales as $\tthei \propto \Delta^{-1} \propto \sqrt{N}$, as described by the analytical results from GinUE.
			Note that the error bars are smaller than the data points.
		}\label{fig:thei_ginue}
	\end{figure}

\begin{figure}[H]
		\centering
		\includegraphics[scale = 0.6]{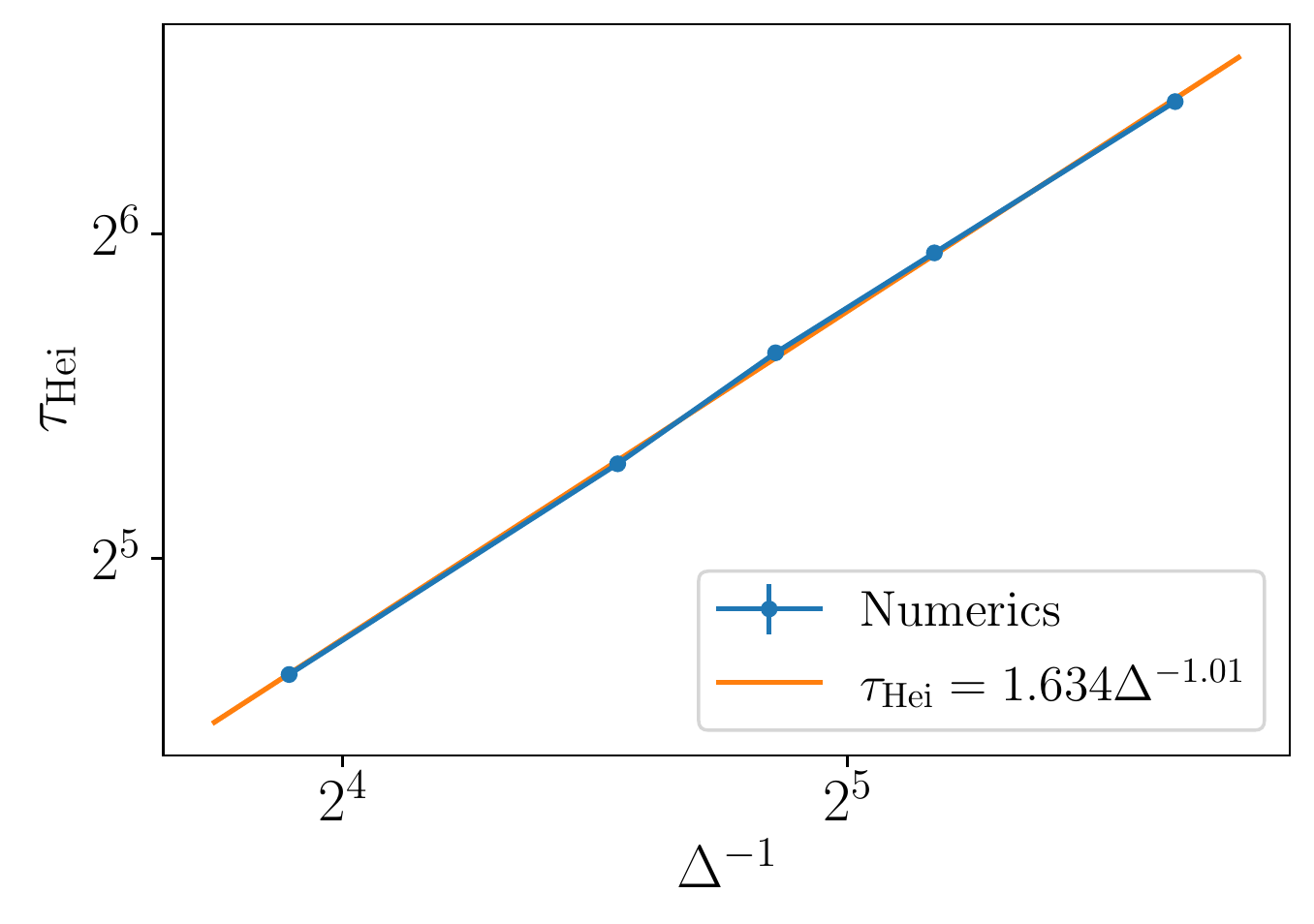}
		\caption{Heisenberg time $\tthei$ vs. inverse level spacing $\Delta^{-1}$ for the QKT with dissipation for $j=35$, $k_1 = 8$, $p \in \mathcal{N}(2,2/3)$ and $k_0 \in \mathcal{N}(10,3)$.
			The Heisenberg time scales as $\tthei \propto \Delta^{-1} \propto\sqrt{N}$, as described by the GinUE results.
			Note that the error bars are smaller than the data points.
		}
		\label{fig:qkt_thei}
	\end{figure}

\begin{figure}[H]
		\begin{minipage}[t]{0.44\textwidth}
			\includegraphics[width=\linewidth,keepaspectratio=true]{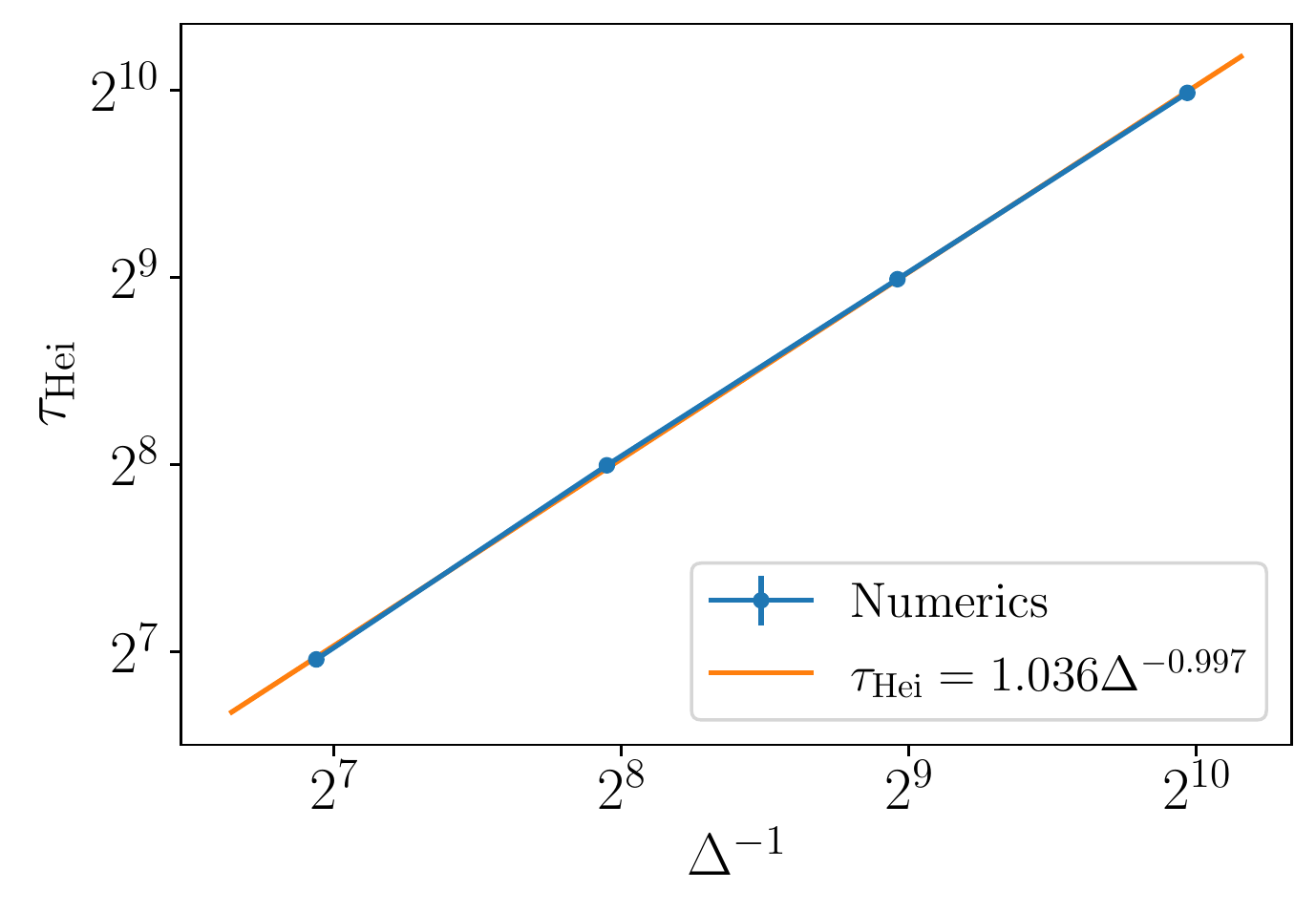}
		\end{minipage}
		\hspace*{\fill} % it's important not to leave blank lines before and after this command
		\begin{minipage}[t]{0.44\textwidth}
			\includegraphics[width=\linewidth,keepaspectratio=true]{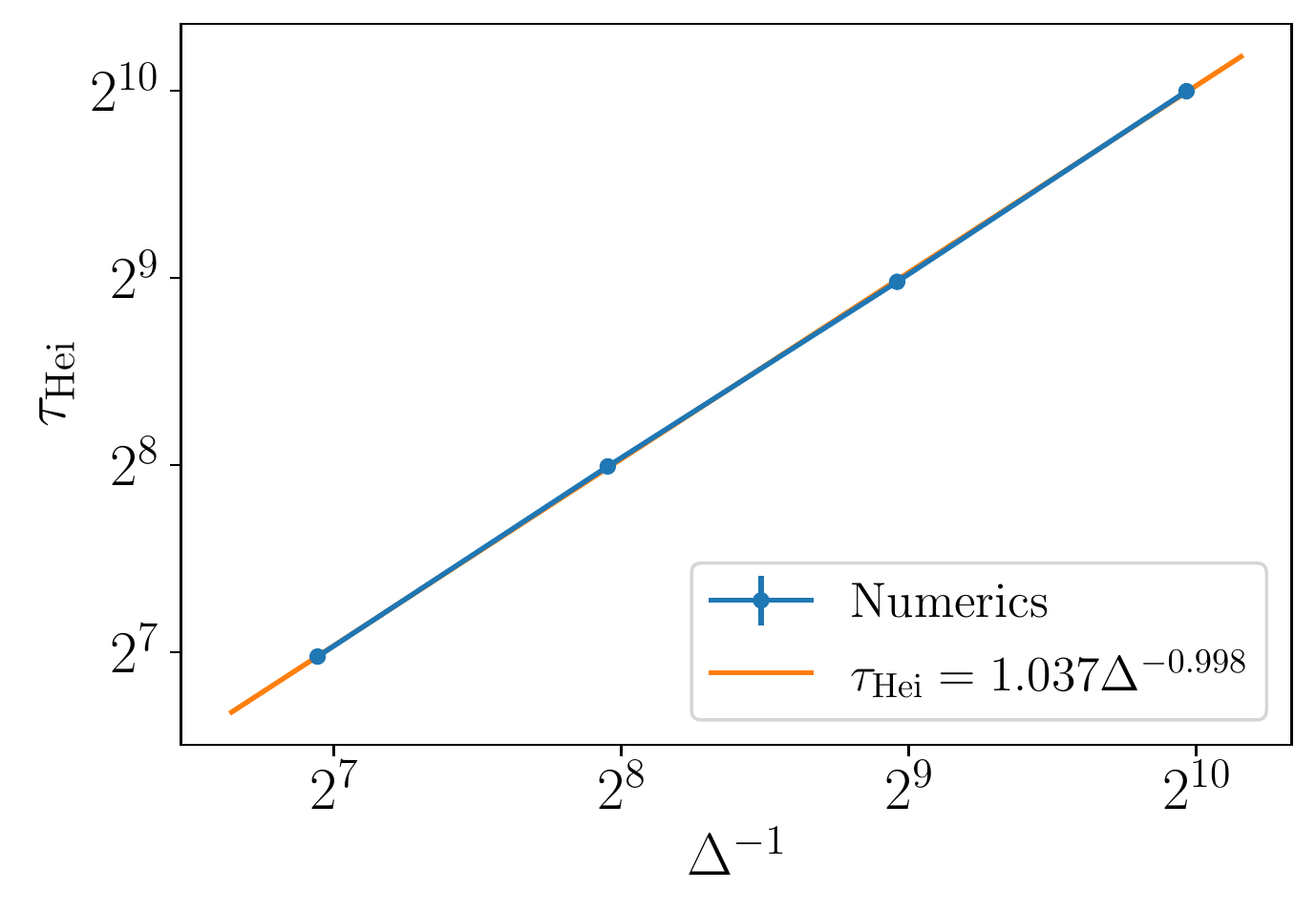}
		\end{minipage}
		\caption{Heisenberg time $\tthei$ vs. inverse level spacing $\Delta^{-1}$ for ensembles of classical stochastic matrices induced from the CUE (left) or GinUE (right). 
			Note that $\Delta^{-1} \propto \sqrt{N} /r(N)$ where $N$ and $r(N)$ are the matrix size and radius of the bulk of the DOS.
			For both cases, the Heisenberg time scales as $\tthei \propto \Delta^{-1}\propto \sqrt{N} /r(N)$, as described by the GinUE results.
			Note that the error bars are smaller than the data points.
		}
		\label{fig:cs_thei}
	\end{figure}

	In this section, we provide the scalings of the Heisenberg time, $\tthei$, for GinUE, GinOE, GinSE (Fig.~\ref{fig:thei_ginue}), QKT with the kick (Fig.~\ref{fig:qkt_thei}), and ensembles of classical stochastic matrices (Fig.~\ref{fig:cs_thei}).
	We show that $\tthei$ scale as the inverse level spacing $\Delta^{-1} \propto \sqrt{N}$, as predicted by the GinUE results. 
	Our fitting protocol is as follows. 
	We consider the fucntional form of the analytical solution of the connected DSFF for GinUE (the first and third terms in Eq.~\eqref{eq:dsff_ginue_largeN}) given by 
	$f(|\tau|, N) = N - N\exp(- A(N) |\tau|^2)$, where $N$ and $A(N)$  are the matrix size (restricted to the relevant symmetry sector) and a fitting parameter respectively. 
	We numerically identify the time scale $\tau^*$ %$\tau_{\mathrm{edge}}$ 
	after which the contribution of disconnected DSFF is small compare to the connected DSFF. 
	We fit $f(|\tau|, N)$ with the data for $|\tau | > \tau^*$ for multiple $N$ and extract $A(N)$.
	We plot $\tthei \propto A(N)^{1/2}$ vs. $\Delta^{-1} $ in the log-log scale, and find a linear fit.
	%%%%%%%%%%%%%%%%%%%%%%%%
	%%%%%%%%%%%%%%%%%%%%%%%%
	%%%%%%%%%%%%%%%%%%%%%%%%
	%%%%%%%%%%%%%%%%%%%%%%%%
	%%%%%%%%%%%%%%%%%%%%%%%% GinUE GinOE GinSE t_Hei
	
	%%%%%%%%%%%%%%%%%%%%%%%%
	%%%%%%%%%%%%%%%%%%%%%%%%
	%%%%%%%%%%%%%%%%%%%%%%%%
	%%%%%%%%%%%%%%%%%%%%%%%% CS thei
	
	%%%%%%%%%%%%%%%%%%%%%%%%
	%%%%%%%%%%%%%%%%%%%%%%%%
	%%%%%%%%%%%%%%%%%%%%%%%%
	%%%%%%%%%%%%%%%%%%%%%%%%
	%%%%%%%%%%%%%%%%%%%%%%%%
	
	%%%%%%%%%%%%%%%%%%%%%%%%
	%%%%%%%%%%%%%%%%%%%%%%%%
	%%%%%%%%%%%%%%%%%%%%%%%%
	%%%%%%%%%%%%%%%%%%%%%%%%
	%%%%%%%%%%%%%%%%%%%%%%%%
	%%%%%%%%%%%%%%%%%%%%%%%%
	%%%%%%%%%%%%%%%%%%%%%%%%
	%%%%%%%%%%%%%%%%%%%%%%%%
	%%%%%%%%%%%%%%%%%%%%%%%%
	%%%%%%%%%%%%%%%%%%%%%%%%
	%%%%%%%%%%%%%%%%%%%%%%%%
	%%%%%%%%%%%%%%%%%%%%%%%%
	%%%%%%%%%%%%%%%%%%%%%%%%
	%%%%%%%%%%%%%%%%%%%%%%%%
	%%%%%%%%%%%%%%%%%%%%%%%%
	
	\section{DSFF for modified spectra with ``projected degeneracies'' removed} \label{app:halfplane}
\begin{figure}[H]
		\begin{minipage}[t]{0.31\textwidth}
			\includegraphics[width=\linewidth,keepaspectratio=true]{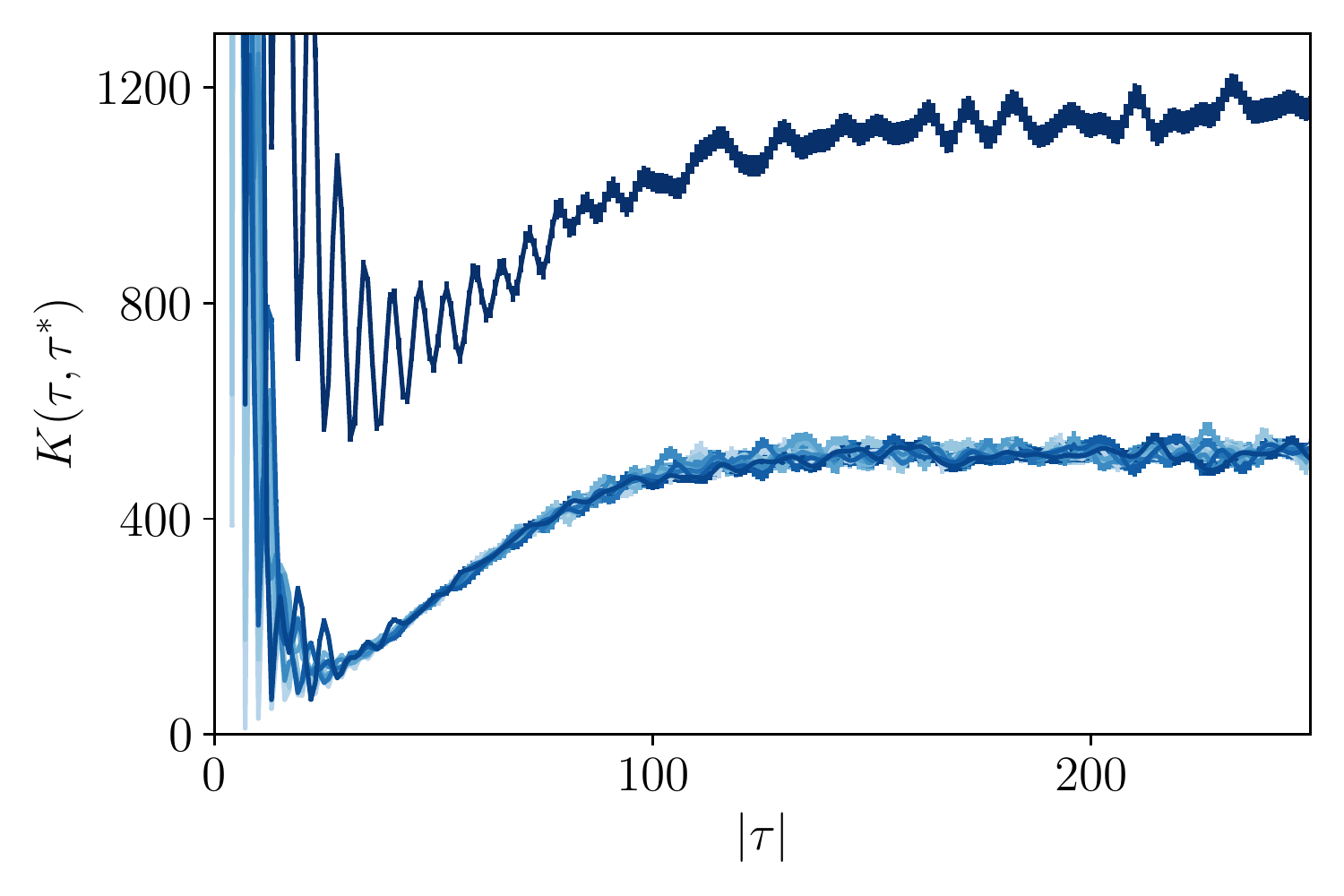}
			\label{subfig:ginue_th}
		\end{minipage}
		\hspace*{\fill} % it's important not to leave blank lines before and after this command
		\begin{minipage}[t]{0.31\textwidth}
			\includegraphics[width=\linewidth,keepaspectratio=true]{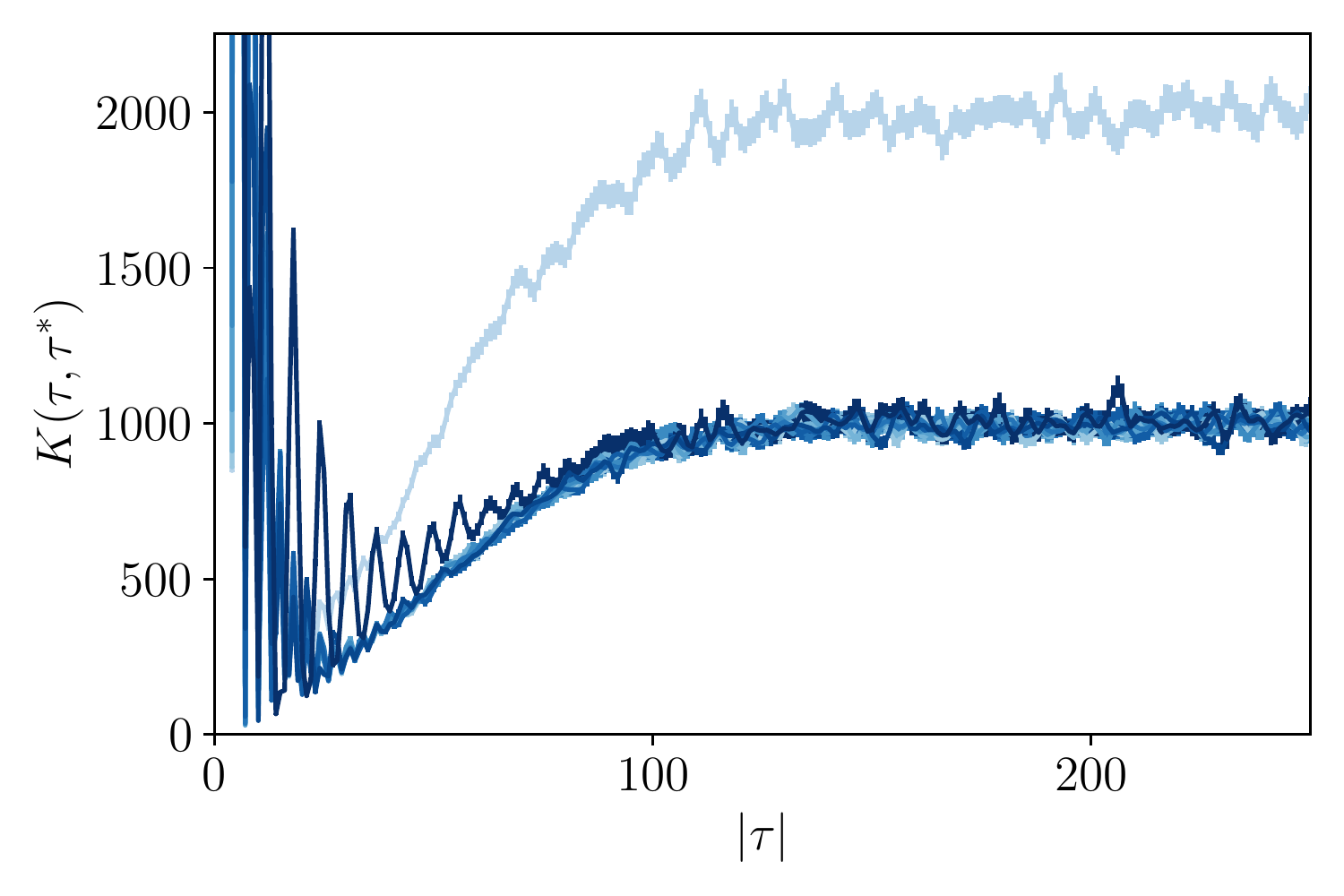}
		\end{minipage}
		\hspace*{\fill} % it's important not to leave blank lines before and after this command
		\begin{minipage}[t]{0.31\textwidth}
			\includegraphics[width=\linewidth,keepaspectratio=true]{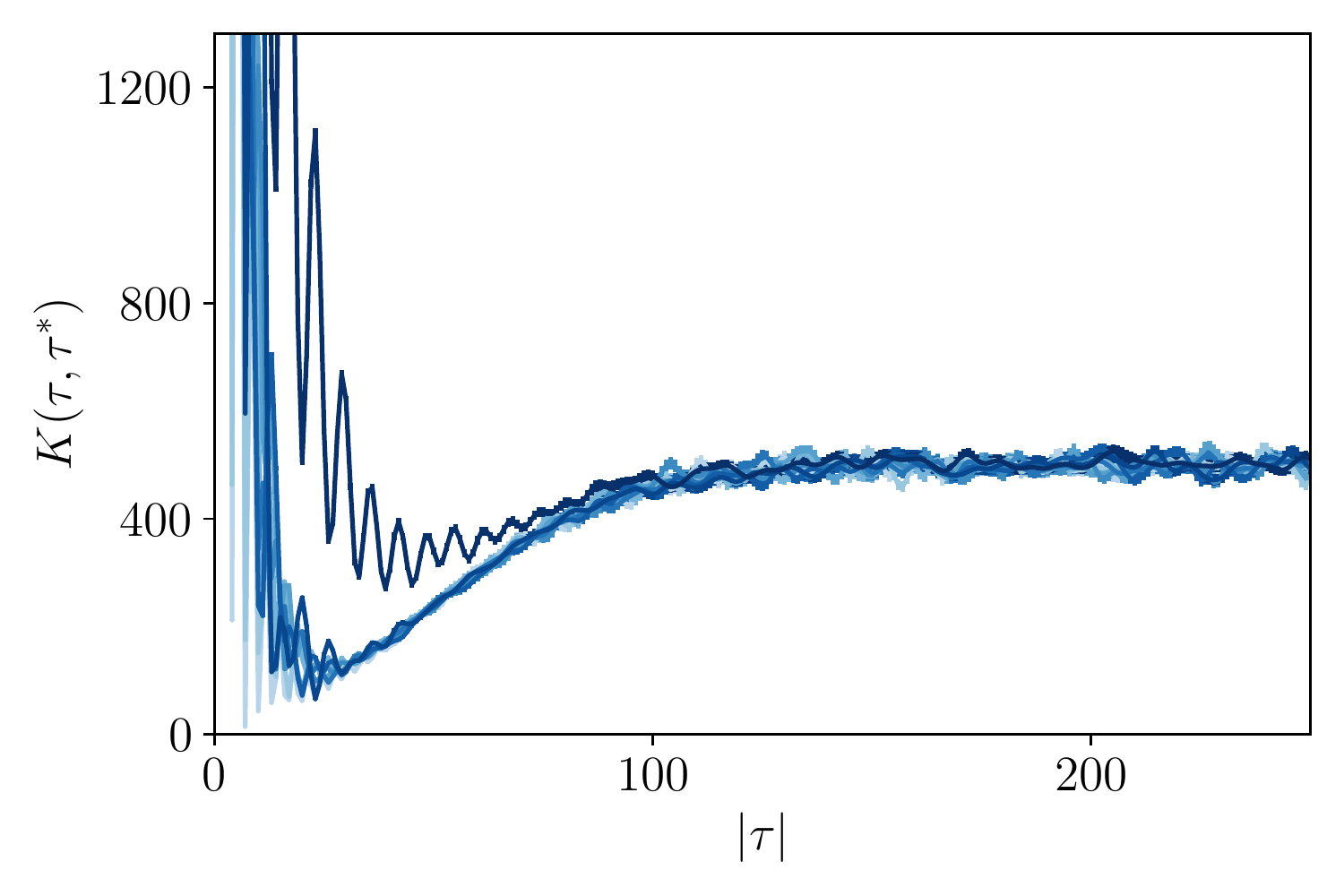}
		\end{minipage}
		\caption{DSFF $K(\tau, \tau^*)$ against $|\tau|$ at different values of $\theta$ from $0$ to $\pi/2$ in steps of $\pi/16$ (from light to dark blue) for the following cases:
			(i) GinOE with bottom half of the complex plane removed but with real eigenvalues kept (left);
			(ii) GinOE with real eigenvalues removed but with the bottom half plane kept (middle);
			and (iii) GinOE with both the real eigenvalues and the bottom half plane removed (right).  
			We see that without the eigenvalues in the bottom half plane ((i) and (iii)), DSFF at $\theta=0$ (light blue) behaves like the one at $\theta = \pi/4$. 
			Without the real eigenvalues ((ii) and (iii)), DSFF at $\theta=\pi/2$ (dark blue) after $|\tau| > \tthei$ saturates to the same plateau as DSFF at other angles, but the approach to the plateau is different (we suspect $\tte$ is large due to the disconnected DSFF of the modified spectra).
			Note that 
			the matrix size $N=1024$ with sample size 2000.
		}\label{fig:ginoe_modified_spectra}
	\end{figure}

\begin{figure}[H]
		\centering
		\includegraphics[scale = 0.6]{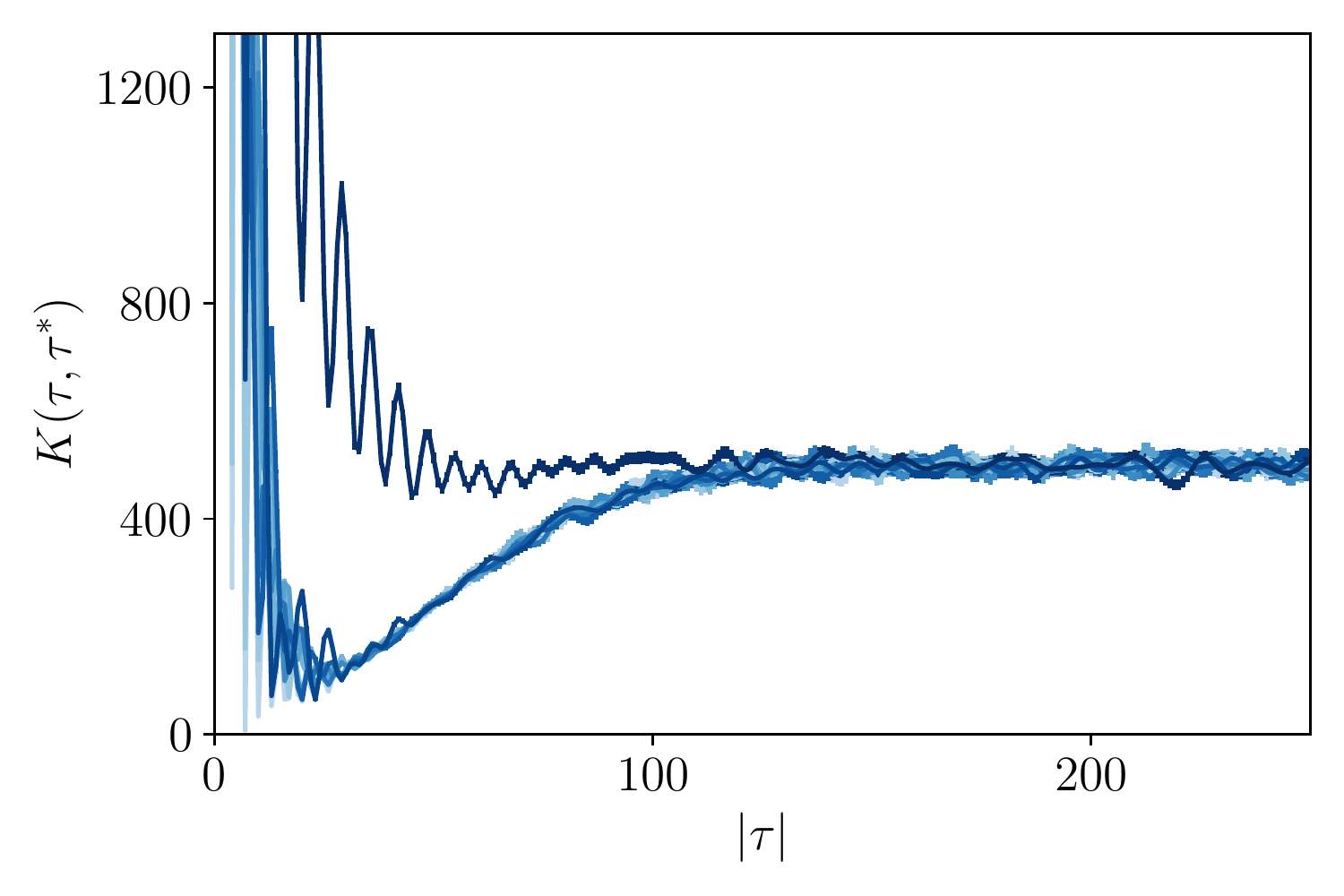}
		\caption{DSFF $K(\tau, \tau^*)$ against $|\tau|$ at different values of $\theta$ from $0$ to $\pi/2$ in steps of $\pi/16$ (from light to dark blue) for the  GinSE with the eigenvalues in the bottom half complex plane removed.
			We see that without the real eigenvalues, DSFF at $\theta=\pi/2$ after $|\tau| > \tthei$ saturates to the same plateau as DSFF at other angles, but the approach to the plateau is different (we suspect $\tte$ is large due to the disconnected DSFF of the modified spectra).
			Note that 
			the matrix size $N=1024$ with sample size 2000. 
		}
		\label{fig:ginse_modified_spectra}
	\end{figure}

	The spectra of GinOE, QKT and CS contain complex conjugate pairs of eigenvalues and real eigenvalues, while the one of GinSE contain complex conjugate pairs only. 
	These features lead to degeneracies in the set of eigenvalues projected along the axis defined by $\theta$, the ``projected degeneracies'' (c.f. interpretation of DSFF in the main text), which in turn make the DSFF behaviour special near $\theta =0, \pi/2$. 
	Here we investigate the behaviour of DSFF of the modified spectra of GinOE and GinSE in Fig.~\ref{fig:ginoe_modified_spectra} and Fig.~\ref{fig:ginse_modified_spectra} respectively.
	%
	%For GinOE, we 
	%after we remove the bottom half plane of the spectrum 
	
	%%%%%%%%%%%%%%%%%%%%%%%%
	%%%%%%%%%%%%%%%%%%%%%%%%
	%%%%%%%%%%%%%%%%%%%%%%%%
	%%%%%%%%%%%%%%%%%%%%%%%%
	%%%%%%%%%%%%%%%%%%%%%%%%
	%%%%%%%%%%%%%%%%%%%%%%%%
	%%%%%%%%%%%%%%%%%%%%%%%%
	%%%%%%%%%%%%%%%%%%%%%%%%
	%%%%%%%%%%%%%%%%%%%%%%%%
	%%%%%%%%%%%%%%%%%%%%%%%%
	%%%%%%%%%%%%%%%%%%%%%%%%
	%%%%%%%%%%%%%%%%%%%%%%%%
	%%%%%%%%%%%%%%%%%%%%%%%%
	%%%%%%%%%%%%%%%%%%%%%%%%
	%%%%%%%%%%%%%%%%%%%%%%%%
	
	\section{DSFF for $\theta$ near $0$ and $\pi/2$}
\begin{figure}[H]
		\begin{minipage}[t]{0.44\textwidth}
			\includegraphics[width=\linewidth,keepaspectratio=true]{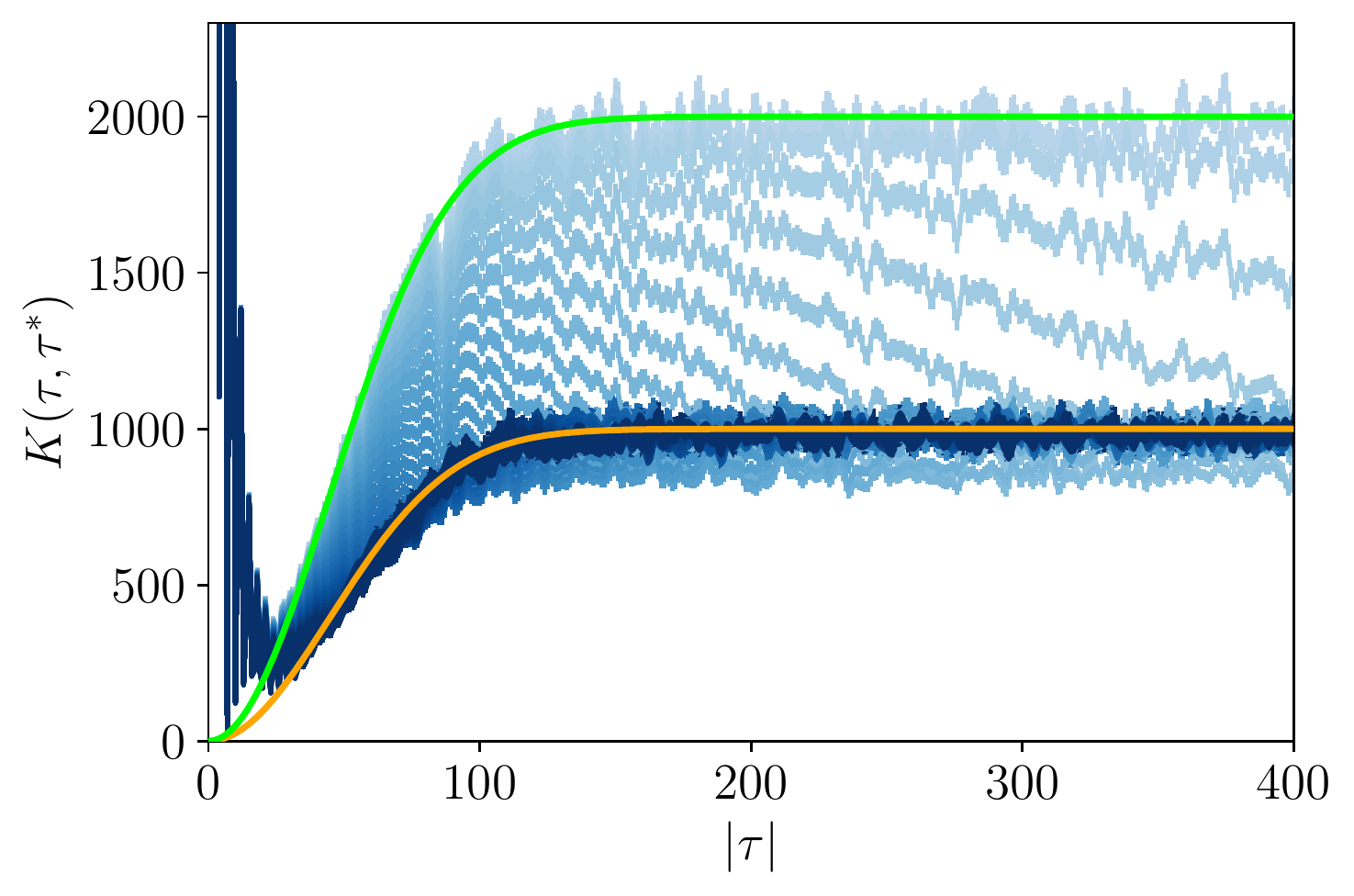}
		\end{minipage}
		\hspace*{\fill} % it's important not to leave blank lines before and after this command
		\begin{minipage}[t]{0.44\textwidth}
			\includegraphics[width=\linewidth,keepaspectratio=true]{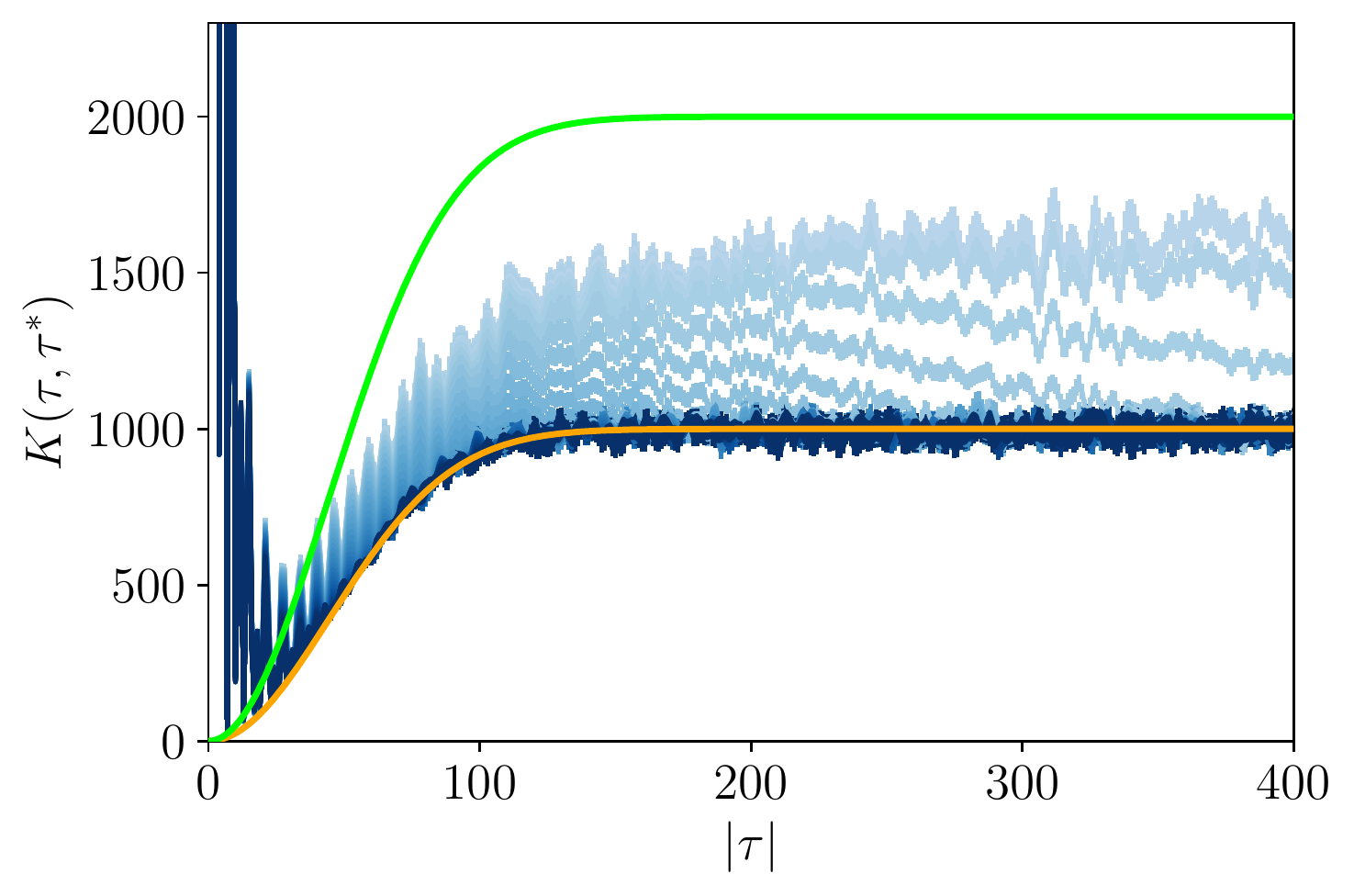}
		\end{minipage}
		\caption{
			Left:
			DSFF $K(\tau, \tau^*)$ of GinOE  against $|\tau|$ from $\theta=0^\circ $ (light blue) to $\theta = 2.64^\circ$ (dark blue) in steps of  $0.24^\circ$.
			Right: 
			DSFF $K(\tau, \tau^*)$ of GinOE  against $|\tau|$ from $\theta=86.48^\circ $ (dark blue) to $\theta = 90^\circ$ (light blue) in steps of  $0.32^\circ$.
			The orange (green) lines is (twice) the connected DSFF behaviour for GinUE (1st and 3rd terms in Eq.~\eqref{eq:dsff_ginue}).
			Note that the matrix size $N=1000$ with sample size 2000.
		}
		\label{fig:dsff_ginoe_near_0}
	\end{figure}

\begin{figure}[H]
		\begin{minipage}[t]{0.44\textwidth}
			\includegraphics[width=\linewidth,keepaspectratio=true]{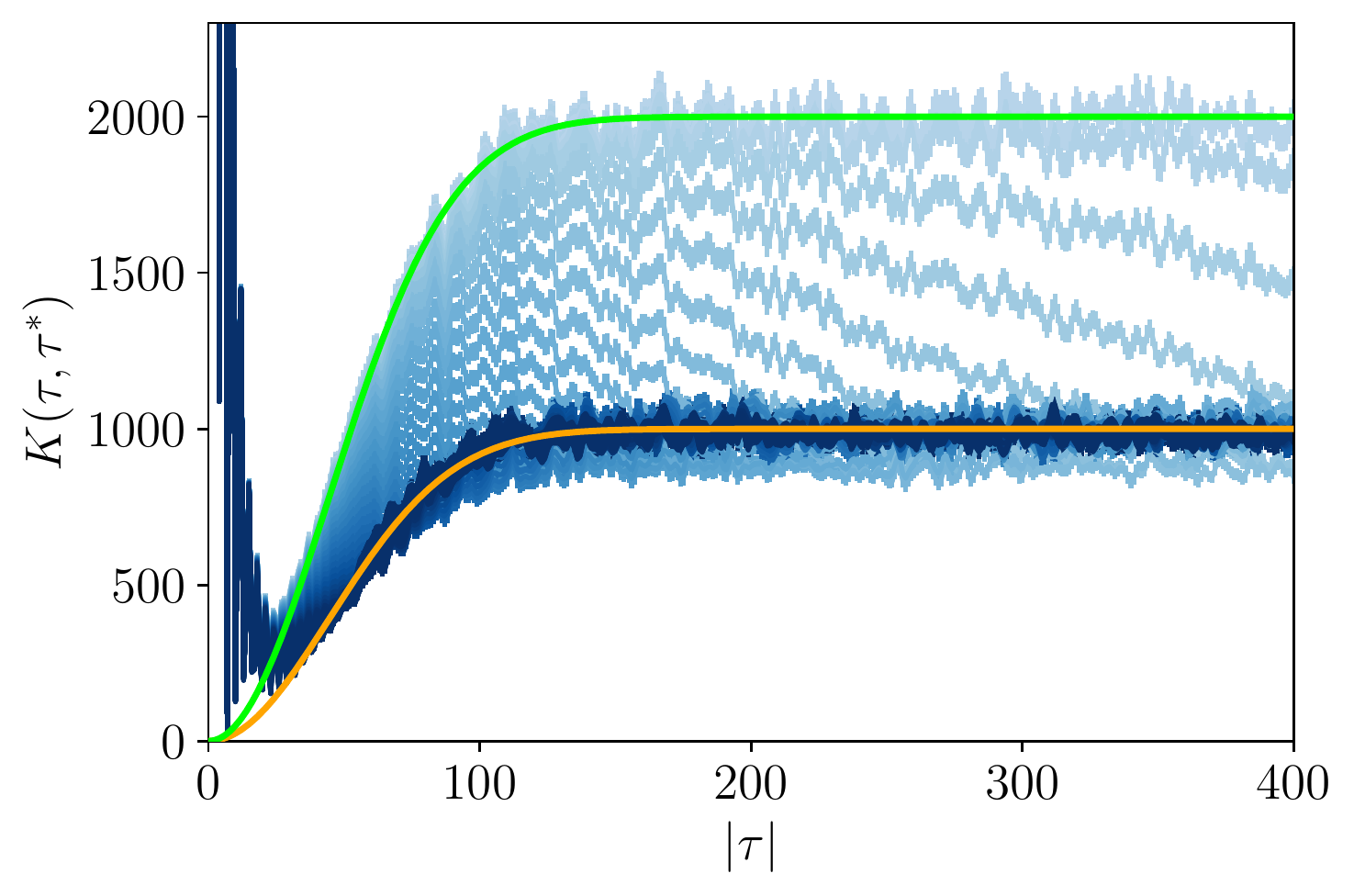}
		\end{minipage}
		\hspace*{\fill} % it's important not to leave blank lines before and after this command
		\begin{minipage}[t]{0.44\textwidth}
			\includegraphics[width=\linewidth,keepaspectratio=true]{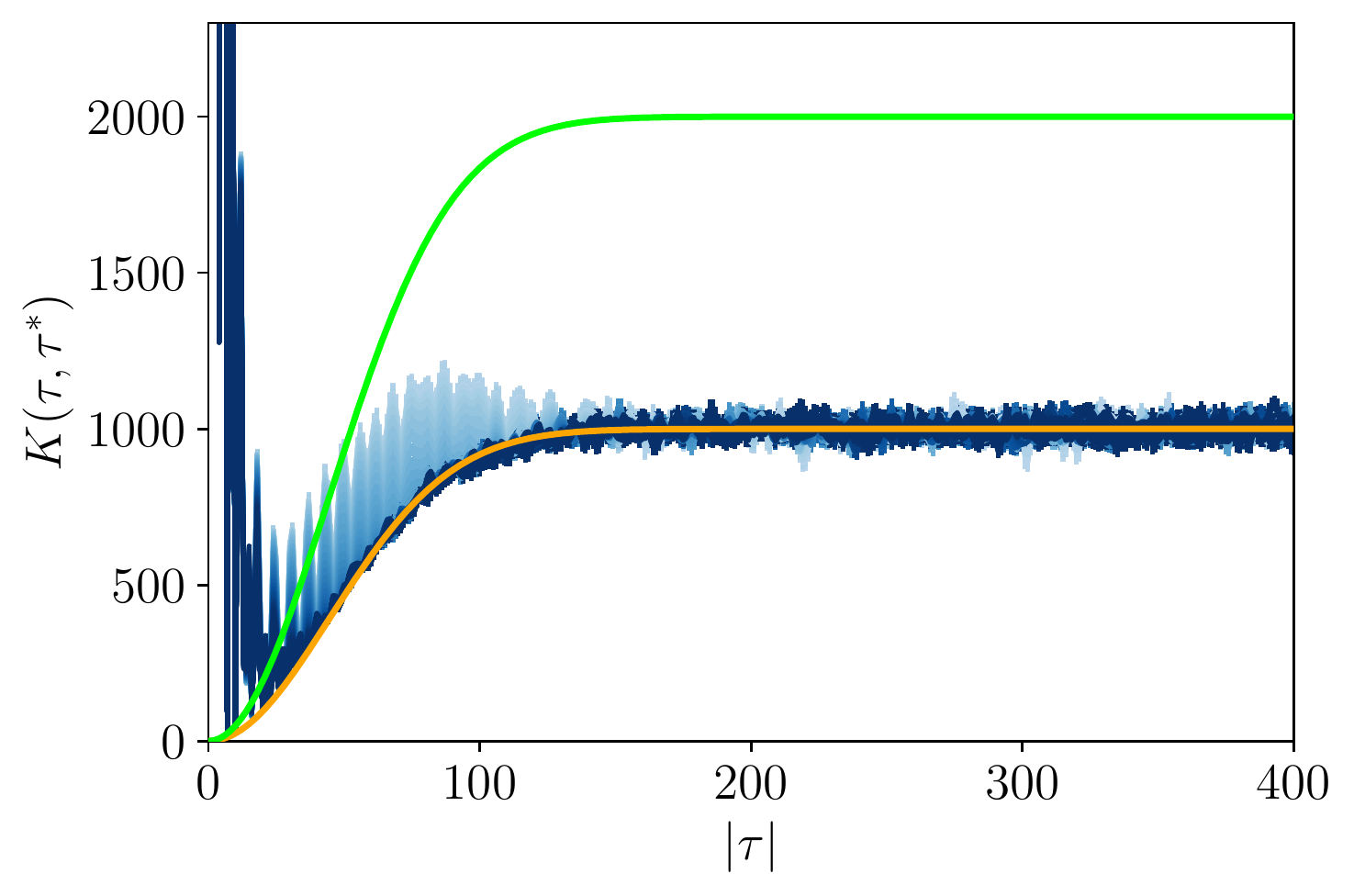}
		\end{minipage}
		\caption{
			Left:
			DSFF $K(\tau, \tau^*)$ of GinSE against $|\tau|$  from $\theta=0^\circ $ (light blue) to $\theta = 2.64^\circ$ (dark blue) in steps of  $0.24^\circ$.
			Right: 
			DSFF $K(\tau, \tau^*)$ of GinSE against $|\tau|$ from $\theta=86.48^\circ $ (dark blue) to $\theta = 90^\circ$ (light blue) in steps of  $0.32^\circ$.
			The orange (green) lines is (twice) the connected DSFF behaviour for GinUE (1st and 3rd terms in Eq.~\eqref{eq:dsff_ginue}).
			Note that the matrix size $N=1000$ with sample size 2000.
		}
		\label{fig:dsff_ginse_near_0}
	\end{figure}

\begin{figure}[H]
		\begin{minipage}[t]{0.44\textwidth}
			\includegraphics[width=\linewidth,keepaspectratio=true]{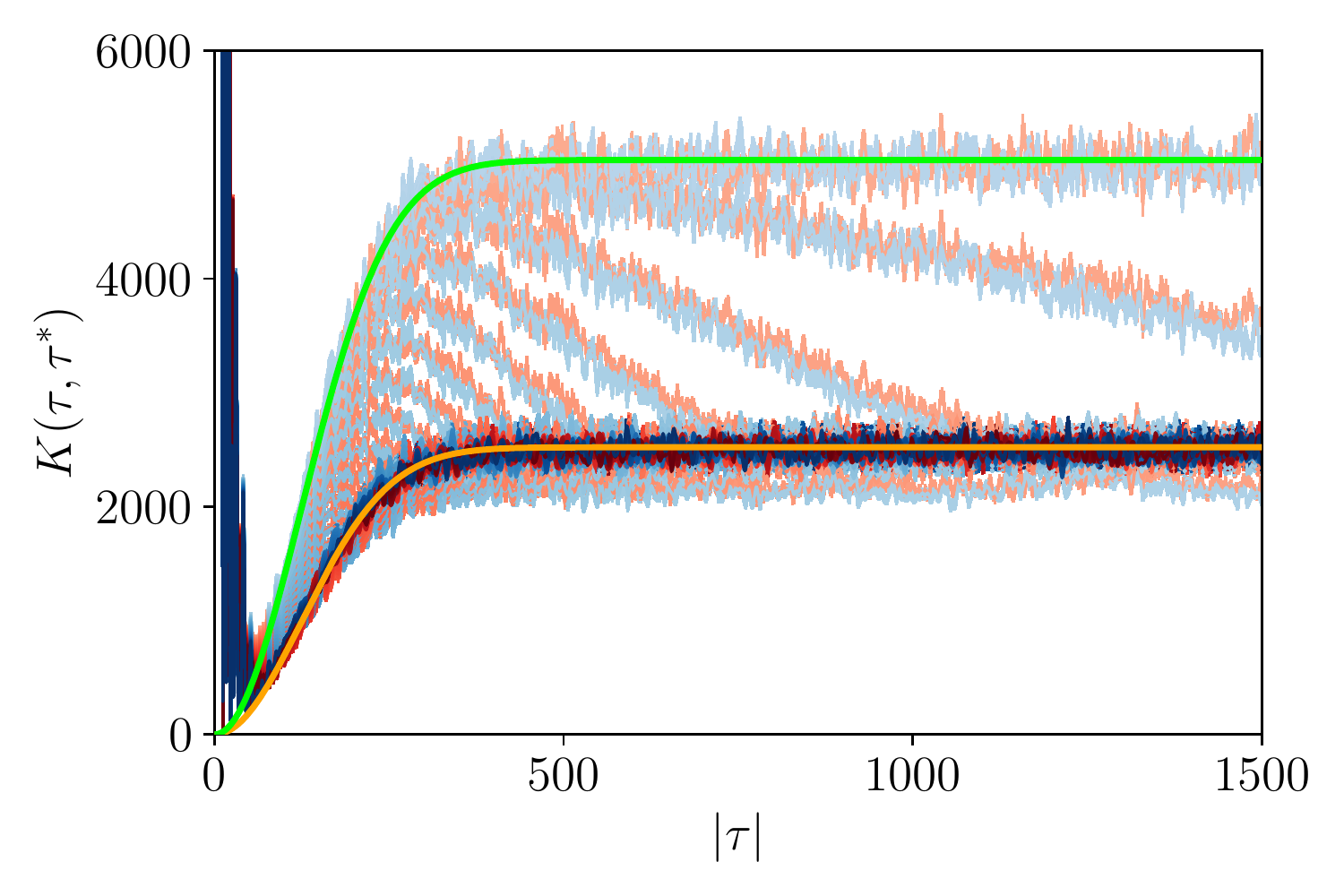}
		\end{minipage}
		\hspace*{\fill} % it's important not to leave blank lines before and after this command
		\begin{minipage}[t]{0.44\textwidth}
			\includegraphics[width=\linewidth,keepaspectratio=true]{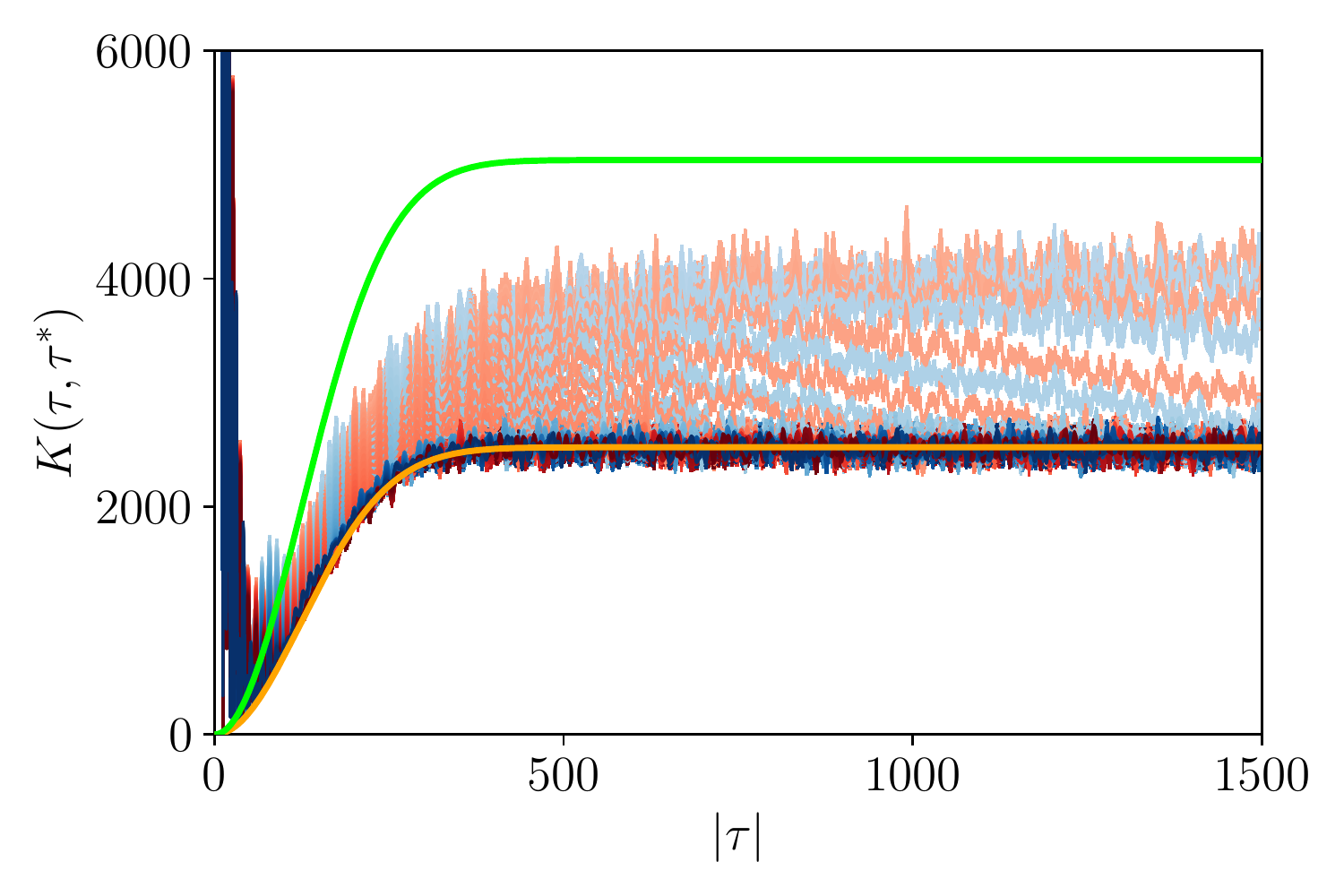}
		\end{minipage}
		\caption{       Left:
			DSFF $K(\tau, \tau^*)$ of QKT with the kick with $j=35$ (blue) and DSFF of GinOE (red) against $|\tau|$  from $\theta=0^\circ $ (light) to $\theta = 4^\circ$ (dark) in steps of  $0.08^\circ$.
			Note that, as in the main text, the QKT with the kick is specified by Gaussianly-distributed $p \in \mathcal{N}(2,2/3)$ and $k_0 \in \mathcal{N}(10,3)$.
			Right: 
			DSFF $K(\tau, \tau^*)$ of QKT (blue) with the same parameters and DSFF of GinOE (red) against $|\tau|$ from $\theta=87^\circ $ (light) to $\theta = 90^\circ$ (dark) in steps of  $0.06^\circ$.
			The orange (green) lines is (twice) the connected DSFF behaviour for GinUE (1st and 3rd terms in Eq.~\eqref{eq:dsff_ginue}).
			Note that the matrix size is $N=2521$ with sample size 2000. 
		}
		\label{fig:dsff_qkt_near_0}
	\end{figure}

\begin{figure}[H]
		\begin{minipage}[t]{0.44\textwidth}
			\includegraphics[width=\linewidth,keepaspectratio=true]{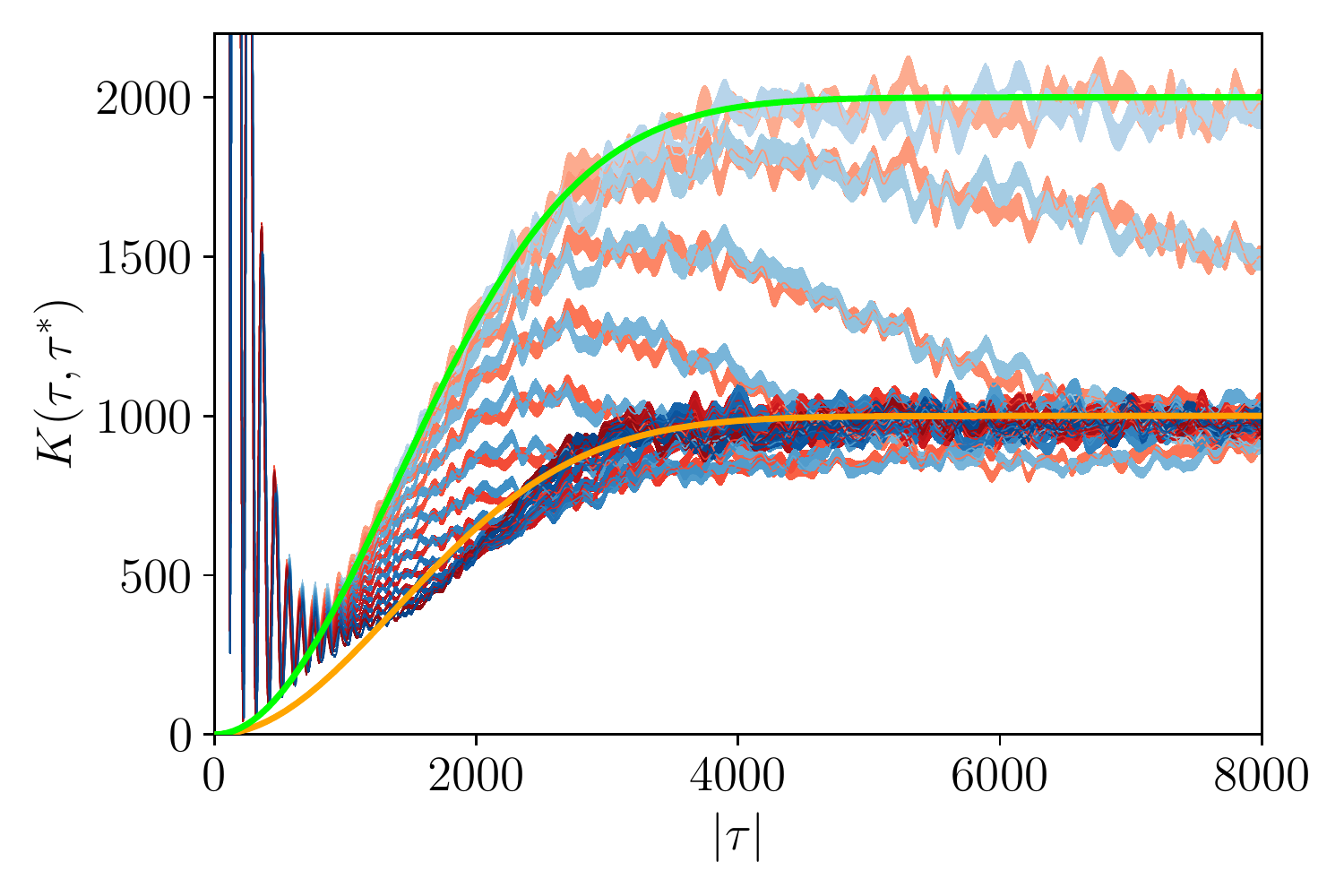}
		\end{minipage}
		\hspace*{\fill} % it's important not to leave blank lines before and after this command
		\begin{minipage}[t]{0.44\textwidth}
			\includegraphics[width=\linewidth,keepaspectratio=true]{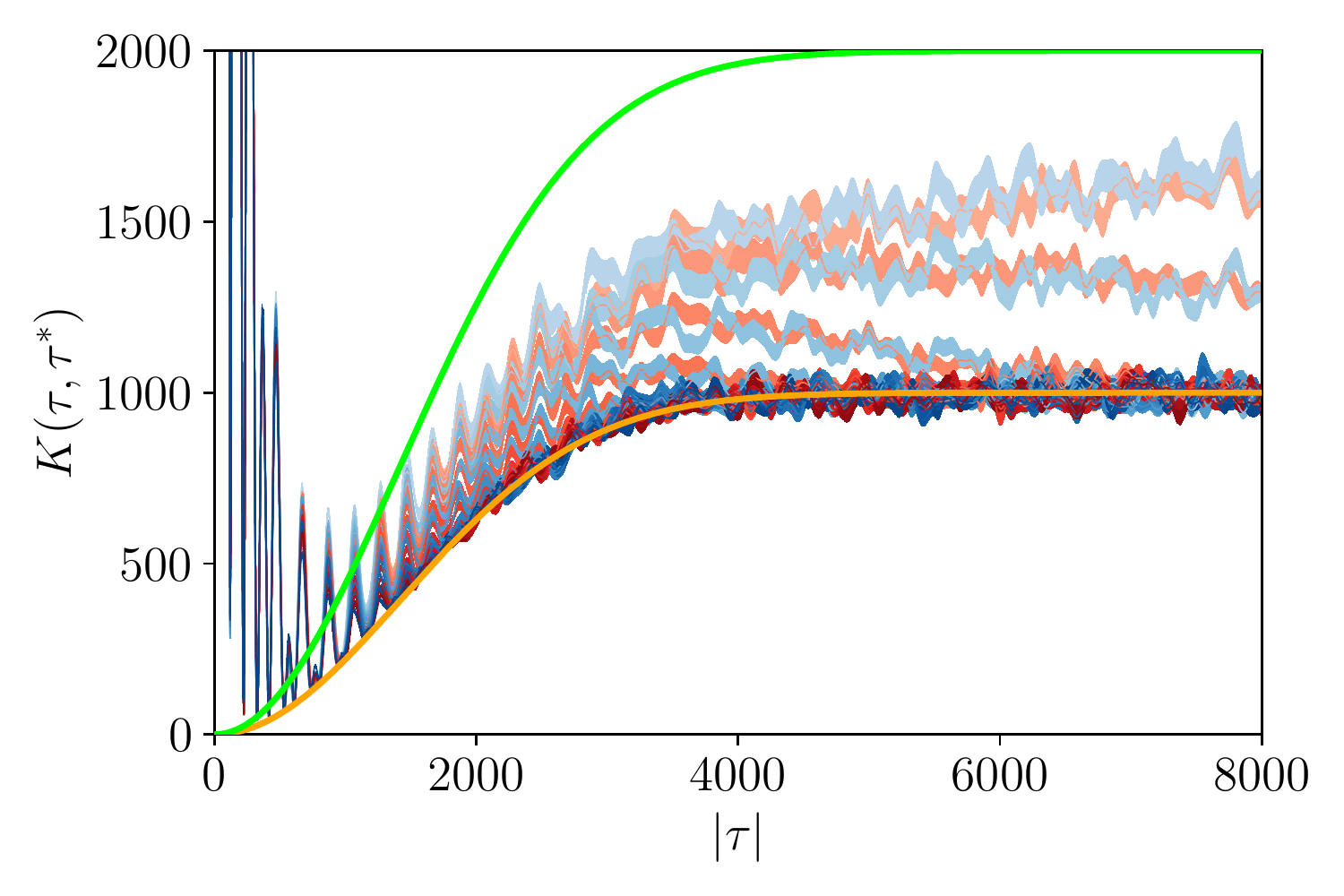}
		\end{minipage}
		\caption{
			Left:
			DSFF $K(\tau, \tau^*)$ of CS induced by CUE (blue) and DSFF of GinOE (red) against $|\tau|$  from $\theta=0^\circ $ (light) to $\theta = 2.64^\circ$ (dark) in steps of  $0.24^\circ$.
			Right: 
			DSFF $K(\tau, \tau^*)$ of CS induced by CUE (blue) and DSFF of GinOE (red) against $|\tau|$ from $\theta=86.48^\circ $ (light) to $\theta = 90^\circ$ (dark) in steps of  $0.32^\circ$.
			The orange (green) lines is (twice) the connected DSFF behaviour for GinUE (1st and 3rd terms in Eq.~\eqref{eq:dsff_ginue}).
			Note that the matrix size is $N=1000$ with sample size 2000.
		}
		\label{fig:dsff_cs_cue_near_0}
	\end{figure}

	As discussed in the previous section, the spectra of GinOE, QKT and CS contain complex conjugate pairs of eigenvalues and real eigenvalues, while the one of GinSE contain complex conjugate pairs only. 
	These features lead to degeneracies in the set of eigenvalues projected along the axis defined by $\theta$, the ``projected degeneracies'' (c.f. interpretation of DSFF in the main text), which in turn make the DSFF behaviour special near $\theta =0, \pi/2$. 
	Here we investigate the behaviour of DSFF near $\theta =0$  and $\theta= \pi/2$, firstly in the universality classes of GinOE (Fig.~\ref{fig:dsff_ginoe_near_0}) and GinSE (Fig.~\ref{fig:dsff_ginse_near_0}), and secondly in the examples of QKT (Fig.~\ref{fig:dsff_qkt_near_0}) and CS (Fig.~\ref{fig:dsff_cs_cue_near_0} and ~\ref{fig:dsff_cs_ginue_near_0}). 
	Note also that, to compare between the DSFF of non-RMT ensemble and DSFF of RMT, we have rescaled the width of the spectrum / the DOS in the complex plane of the non-RMT ensemble (see Appendix \ref{app:spectra} and \ref{app:dos}) to unity, such that the transformed mean level spacing does not scale with matrix size $N$.

	%
	%in Fig.~\ref{fig:ginoe_modified_spectra} and Fig.~\ref{fig:ginse_modified_spectra} respectively.
	%%%%%%%%%%%%%%%%%%%%%%%%%%
	%%%%%%%%%%%%%%%%%%%%%%%%%%
	%%%%%%%%%%%%%%%%%%%%%%%%%%
	%
	%%%%%%%%%%%%%%
	%%%%%%%%%%%%%%
	\begin{figure}[H]
		\begin{minipage}[t]{0.44\textwidth}
			\includegraphics[width=\linewidth,keepaspectratio=true]{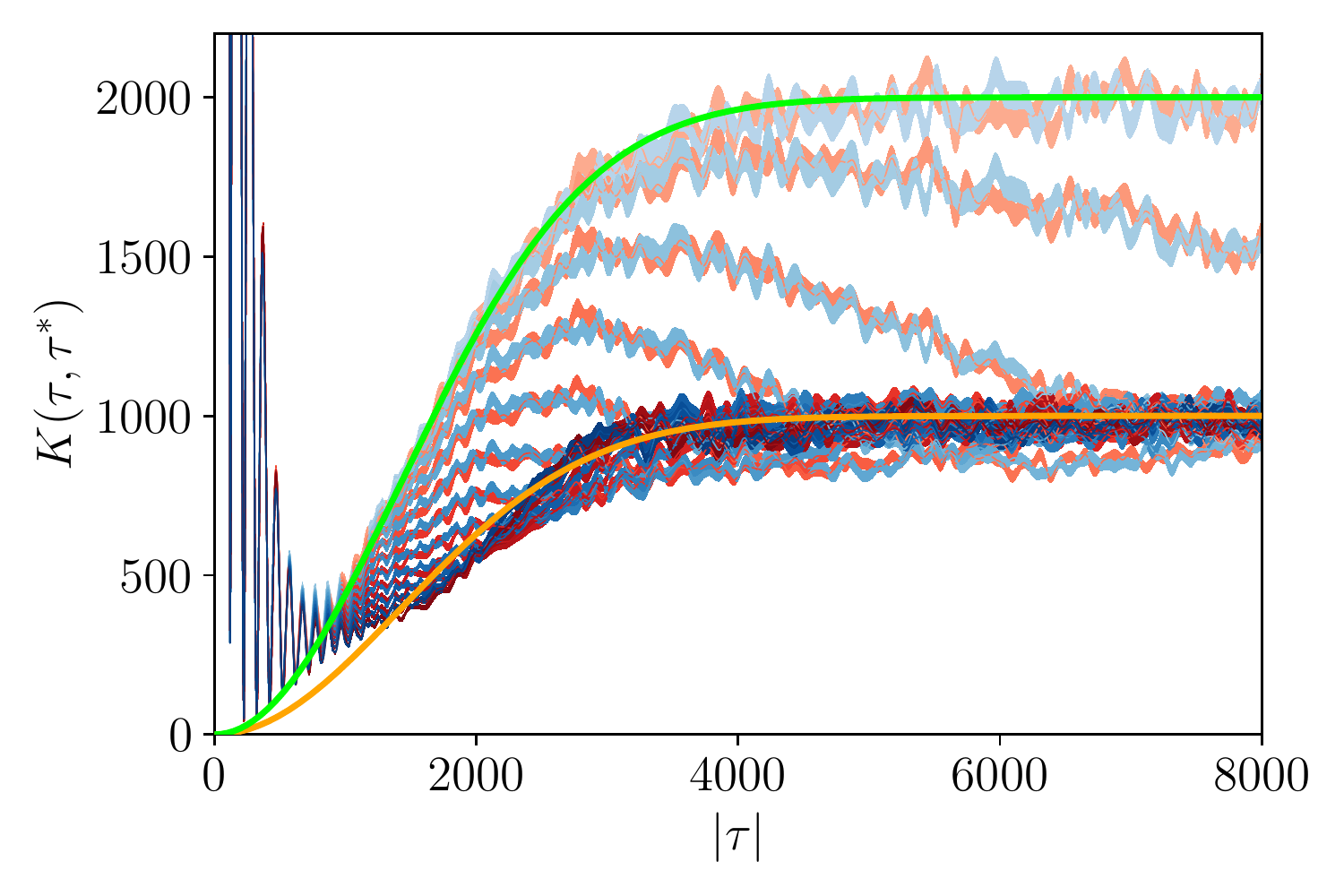}
		\end{minipage}
		\hspace*{\fill} % it's important not to leave blank lines before and after this command
		\begin{minipage}[t]{0.44\textwidth}
			\includegraphics[width=\linewidth,keepaspectratio=true]{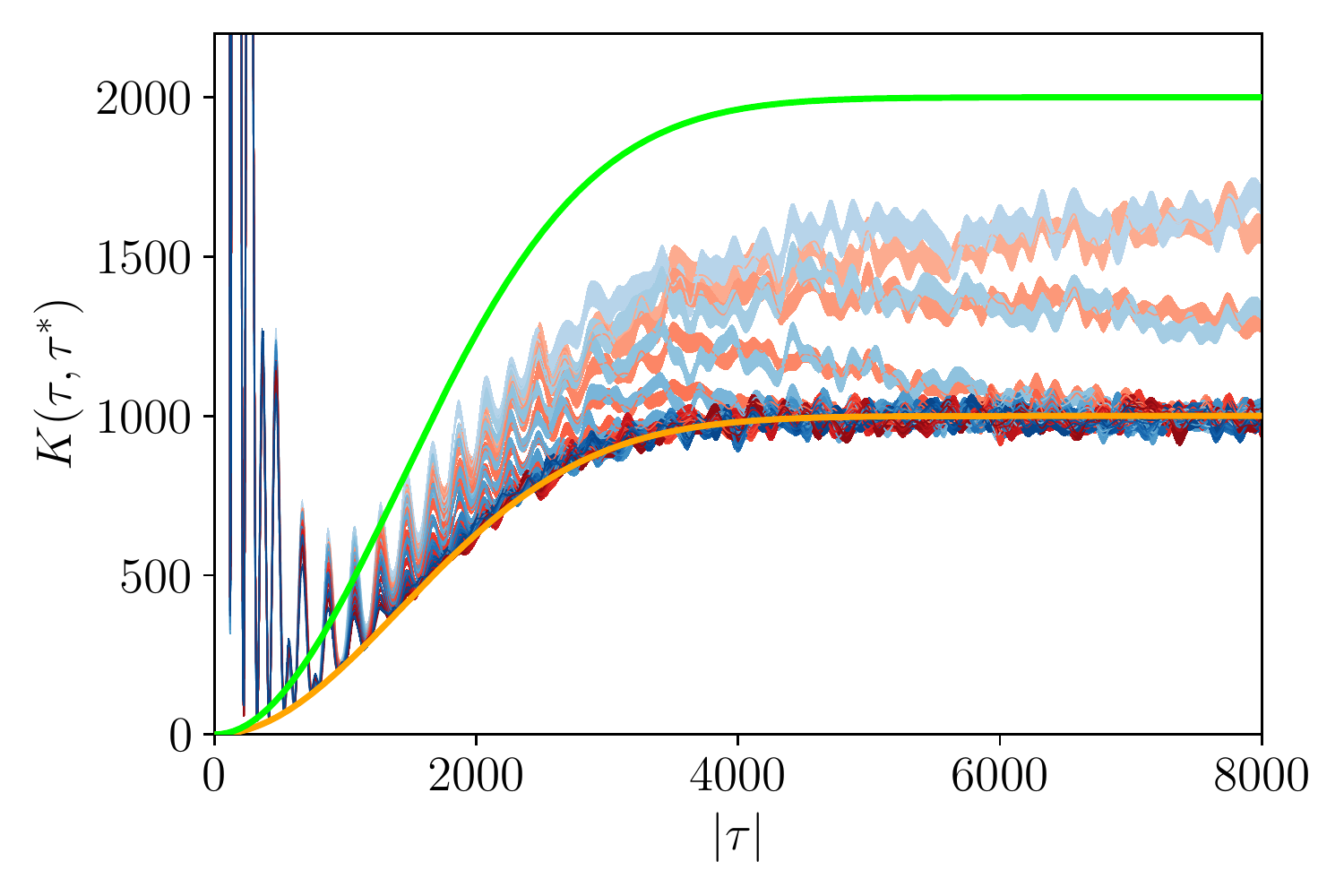}
		\end{minipage}
		\caption{
			Left:
			DSFF $K(\tau, \tau^*)$ of CS induced by GinUE (blue) and DSFF of GinOE (red) against $|\tau|$  from $\theta=0^\circ $ (light) to $\theta = 2.64^\circ$ (dark) in steps of  $0.24^\circ$.
			Right: 
			DSFF $K(\tau, \tau^*)$ of CS induced by CUE (blue) and DSFF of GinOE (red) against $|\tau|$ from $\theta=86.48^\circ $ (light) to $\theta = 90^\circ$ (dark) in steps of  $0.32^\circ$.
			The orange (green) lines is (twice) the connected DSFF behaviour for GinUE (1st and 3rd terms in Eq.~\eqref{eq:dsff_ginue}).
			Note that the matrix size is $N=1000$ with sample size 2000.
		}
		\label{fig:dsff_cs_ginue_near_0}
	\end{figure}

	\section{Critical angle $\thetas$}
	\subsection{Definition of $\thetas$}
	To investigate the deviation of DSFF near $\theta =0$ or $\theta=\pi/2$, 
	we want to define a critical angle $\theta^*_\phi$ with $\phi=0,\pi/2$, such that for $K(|\tau|, \thetasone < \theta < \pi/2 - \thetastwo) \approx K(|\tau|, \theta = \pi/4)$, i.e.
	the behaviour of DSFF along  $ \theta \in [\thetasone, \pi/2 - \thetastwo]$ falls under the GinUE universality class.
	To this end, we first define an error function $\chi_{\phi}(\theta) \in [0,1]$, 
	\begin{equation}
	\chi_{\phi} (\theta) = 
	\frac{
		\sum_{|\tau| \leq \tauhei } 
		\Big| K(|\tau|,\theta) - K(|\tau|,\theta= \pi/4)
		\Big|^2
	}{
		\sum_{|\tau| \leq \tauhei }
		\Big| K(|\tau|,\theta=\phi) - K(|\tau|,\theta= \pi/4)
		\Big|^2
	}
	\;,
	\end{equation}
	with $\phi=0$ or  $\phi= \pi/2$. We have defined
	$\chi_\phi $ such that  $\chi_\phi(\theta=\phi) = 1$ and $\chi_0(\theta=\pi/4) = 0$.
	To understand the typical behaviours of $\chi_\phi(\theta)$, we plot $\chi_0(\theta)$ for GinOE for $\theta\in [ 0, \pi/4]$ below.
	
	\begin{figure}[H]
		\begin{minipage}[t]{0.44\textwidth}
			\includegraphics[width=\linewidth,keepaspectratio=true]{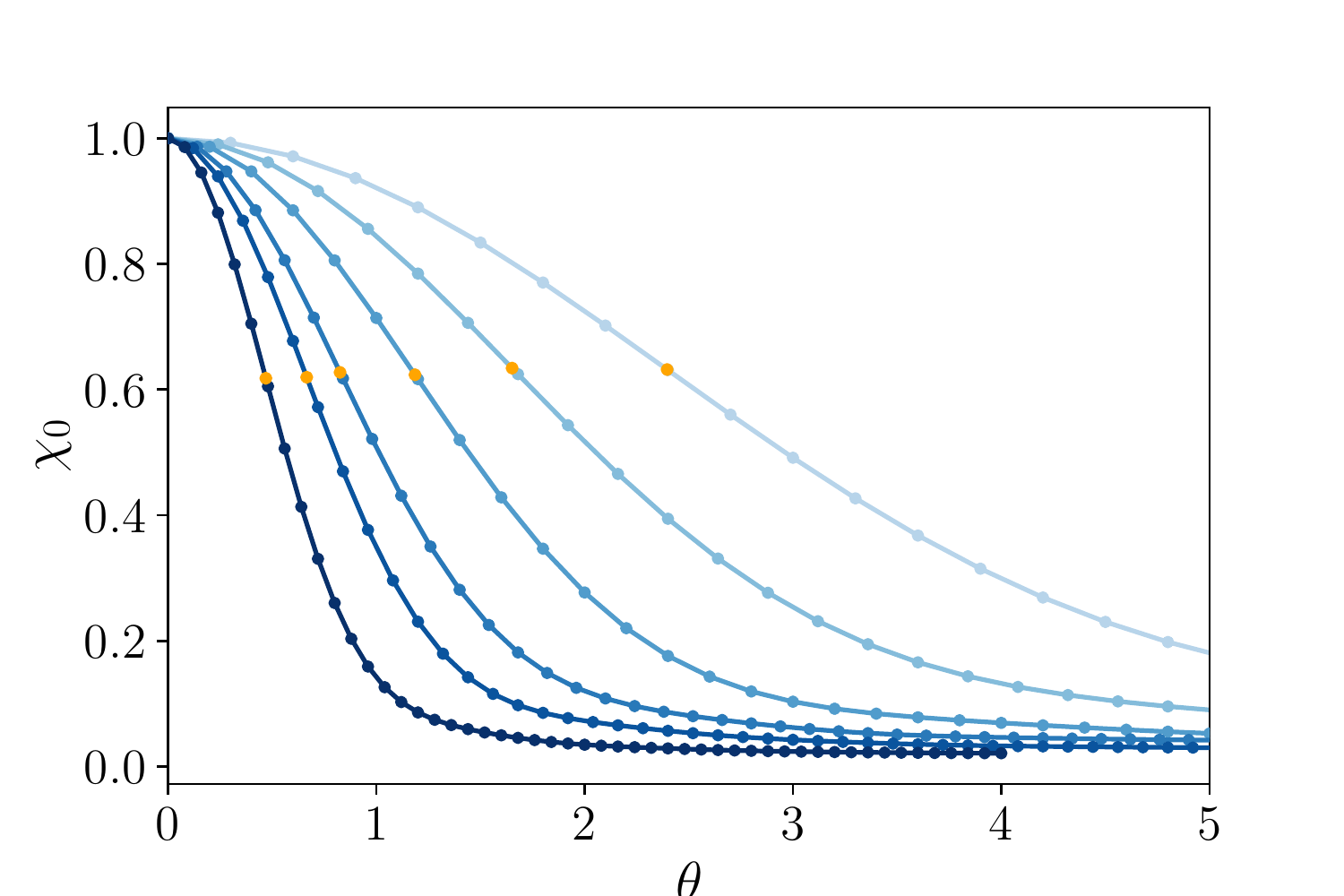}
		\end{minipage}
		\hspace*{\fill} % it's important not to leave blank lines before and after this command
		\begin{minipage}[t]{0.44\textwidth}
			\includegraphics[width=\linewidth,keepaspectratio=true]{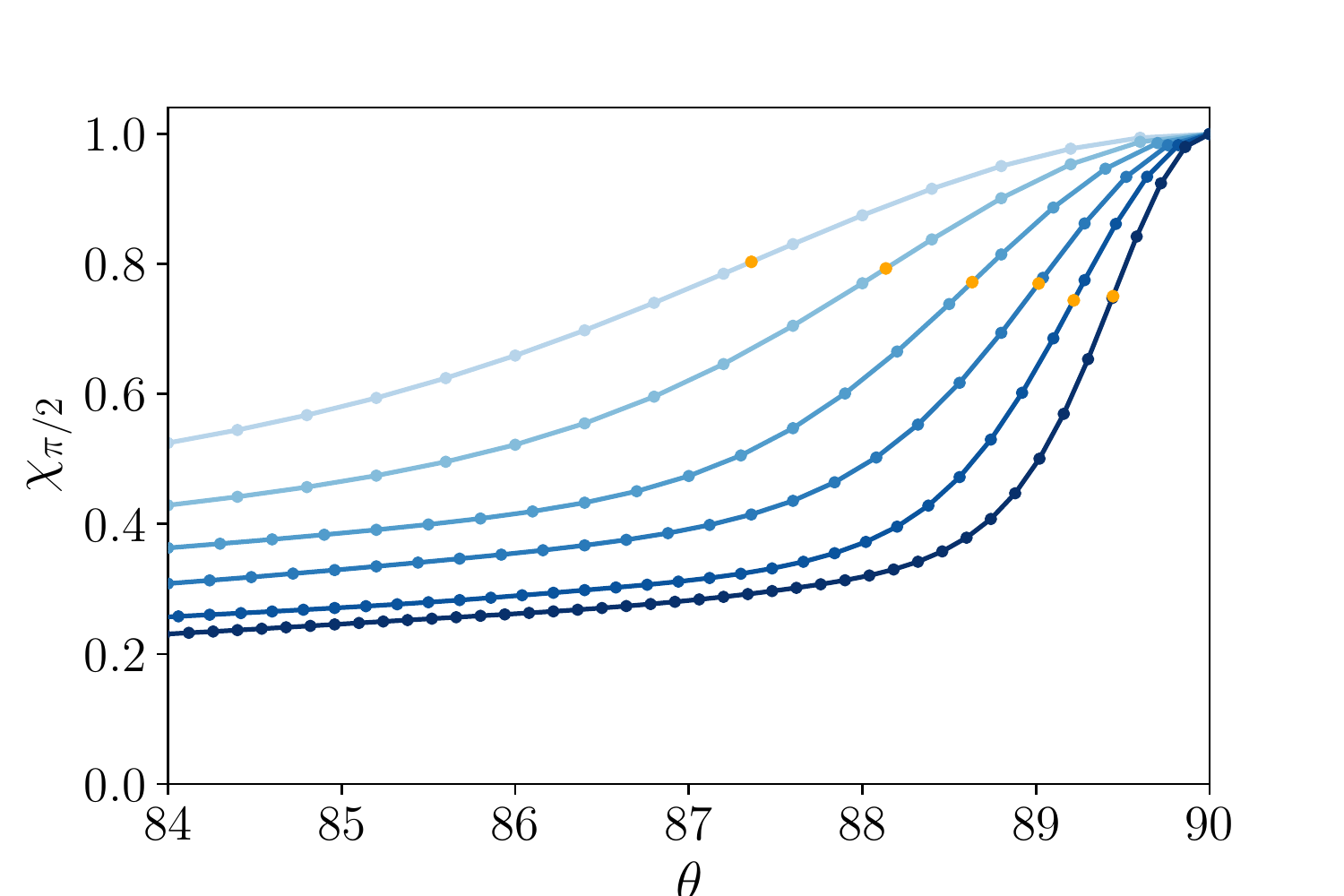}
		\end{minipage}
		\caption{
			Left: $\chi_0(\theta)$ against $\theta$ for increasing matrix sizes from light blue to dark blue. 
			Right: $\chi_{\pi/2}(\theta)$ against $\theta$ for increasing matrix sizes from light blue to dark blue. 
			$\thetasphi(N)$ is marked with orange dots. 
			The matrix sizes are $N \in \{ 32,64,128,256,400,800 \}$, and the sample size is 2000.
		}
		\label{fig:dsff_cs_ginue_near_0}
	\end{figure}
	As a function of $\theta$, $\chi_0(\theta)$ first has a faster dip, then a slower decay. This shape is found in the DSFF of RMT, QKT (with the kick) and CS for multiple systems sizes.
	To define the critical angle $\thetas_\phi$, we track the position of steepest point of the faster dip, by writing
	\begin{equation}
	\begin{split}
	\theta^*_0(N) & = 
	\left\{ \theta : 
	%\left.
	\partial^2_\theta \chi_\phi(\theta) 
	%\right|_{\theta= \theta^*}
	= 0
	\right\} 
	\\
	\theta^*_{\pi/2}(N)  & = 
	\left\{ \theta : 
	%\left.
	\partial^2_\theta \chi_\phi( \pi/2 - \theta) 
	%\right|_{\theta= \theta^*}
	= 0
	\right\}     
	\end{split}
	\;,
	\end{equation}
	where $N$ is the size of the matrices in the ensemble of interest.
	Alternative definitions of $\theta^*_\phi$ that tracks the position of the faster dip in $\chi_\phi$ will give rise to similar scaling of $\theta^*_\phi$ in $N$ below.

	\subsection{Scaling of $\thetas$}\label{app:scaling_thetas}
	We compute $\thetasphi$  for all ensembles discussed in the paper, and show that $\thetasphi \approx \alpha N^{-1/2}$   for both $\phi=0$ and $ \phi= \pi/2$. The one exception is the scaling of $\thetastwo$ for GinSE, which is atypical because GinSE has no real eigenvalues, and thus no projected degeneracies at $\theta = \pi/2$ as discussed in the main text and in Appendix \ref{app:halfplane}. Nonetheless, there is deviation between DSFF at $\theta= \pi/2$ and the one at $\theta = \pi/4$ (see Appendix \ref{app:dsff_ginoe_ginse}). 
	Note also that for CS induced by CUE and GinUE, $\partial^2_\theta \chi_{\pi/2}(\theta)$ shows a large amount fluctuation relative to the other ensembles (which we expect to go away if the sample sizes are increased). We instead use the definition, $
	\theta^*_{\pi/2}(N)   = 
	\max \left\{ \theta : 
	\chi_\phi( \pi/2 - \theta) 
	< \delta
	\right\}     
	$. 

	\begin{figure}[H]
		\begin{minipage}[t]{0.44\textwidth}
			\includegraphics[width=\linewidth,keepaspectratio=true]{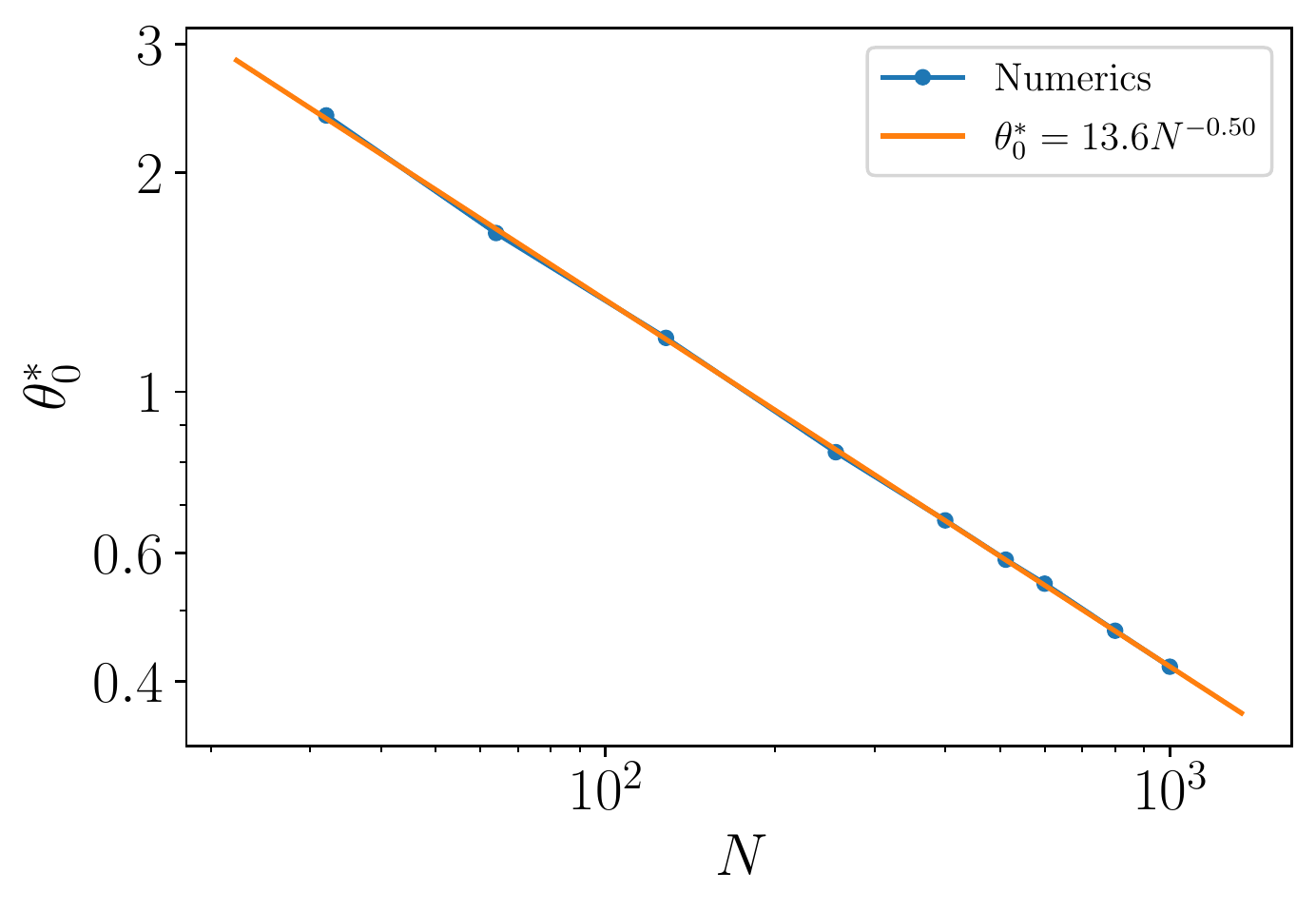}
		\end{minipage}
		\hspace*{\fill} % it's important not to leave blank lines before and after this command
		\begin{minipage}[t]{0.44\textwidth}
			\includegraphics[width=\linewidth,keepaspectratio=true]{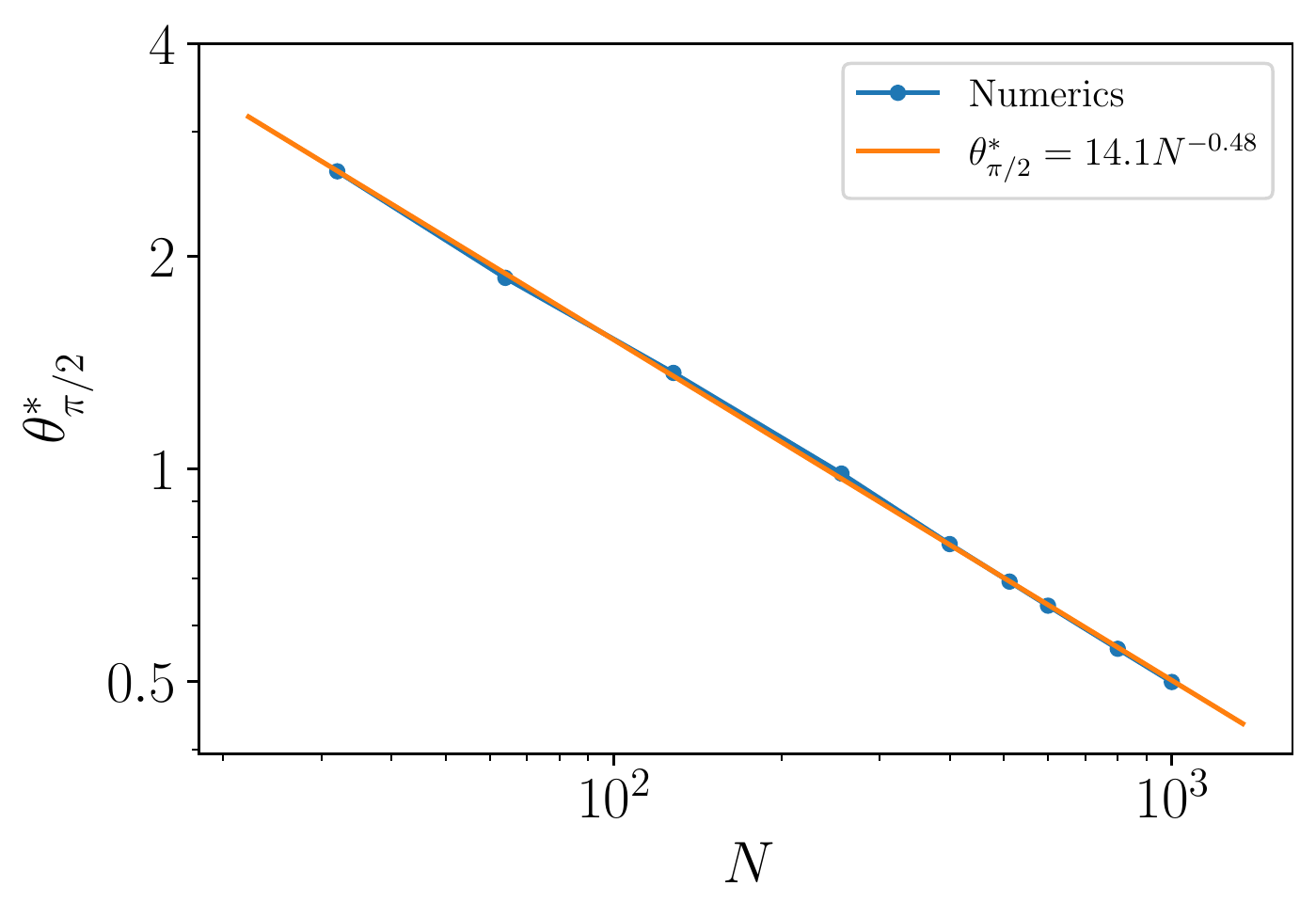}
		\end{minipage}
		\caption{
			$\thetasone(N)$ (left) and $\thetastwo(N)$ (right) of the GinOE for $N$ from $32$ to $1000$. 
			The sample size is 2000.
		}
		\label{fig:dsff_cs_ginue_near_0}
	\end{figure}
	
	\begin{figure}[H]
		\begin{minipage}[t]{0.44\textwidth}
			\includegraphics[width=\linewidth,keepaspectratio=true]{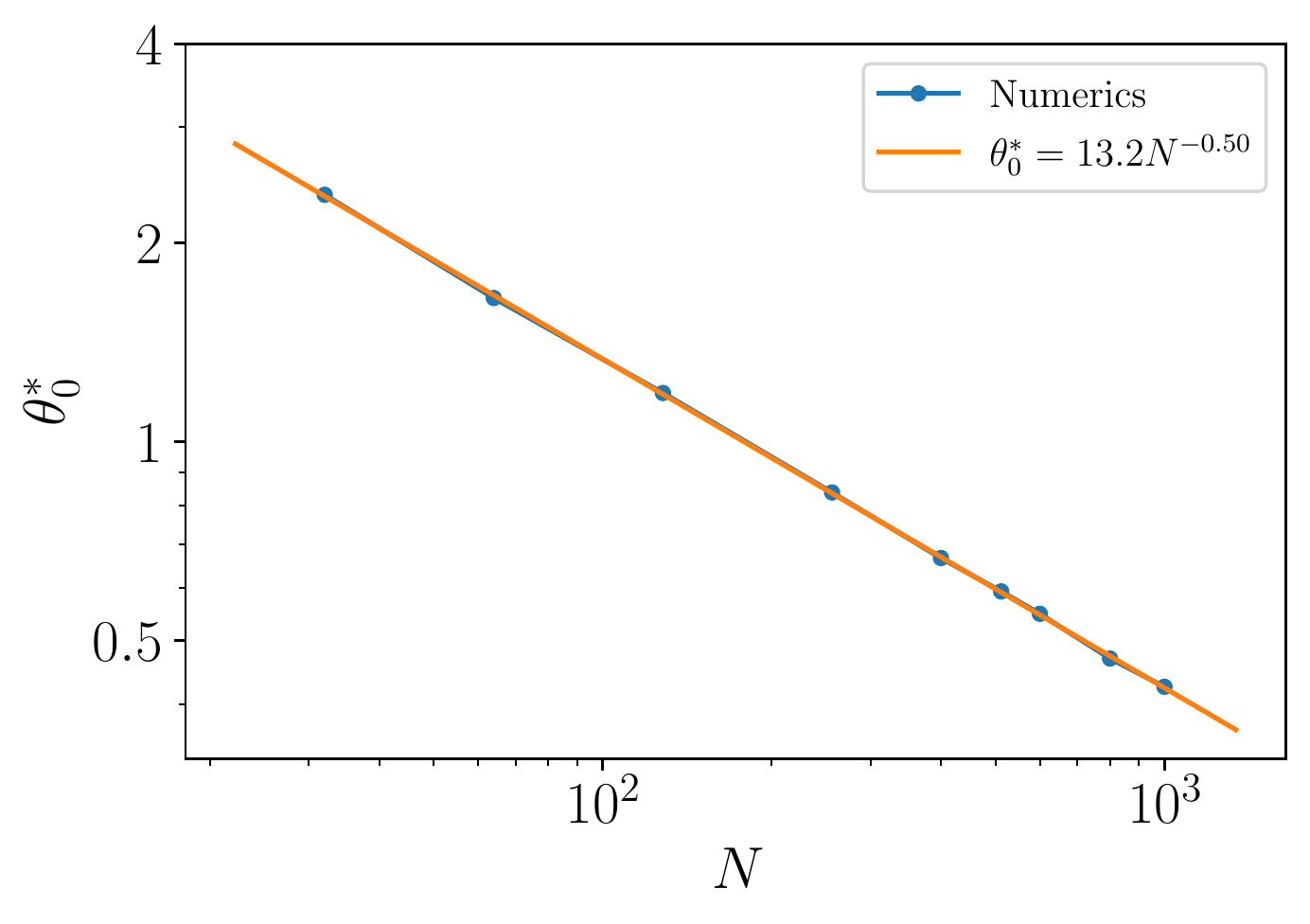}
		\end{minipage}
		\hspace*{\fill} % it's important not to leave blank lines before and after this command
		\begin{minipage}[t]{0.44\textwidth}
			\includegraphics[width=\linewidth,keepaspectratio=true]{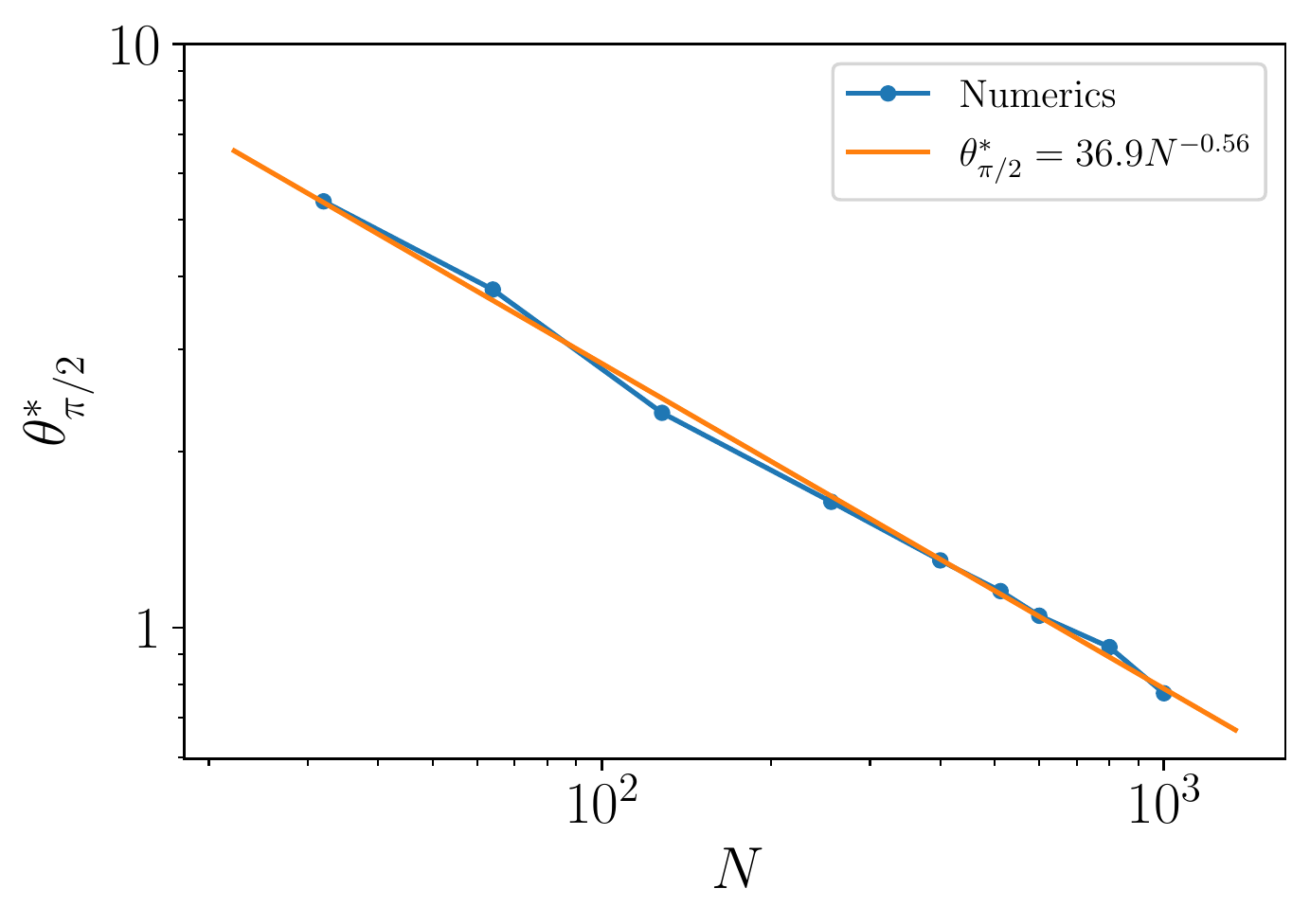}
		\end{minipage}
		\caption{
			$\thetasone(N)$ (left) and $\thetastwo(N)$ (right) of the GinSE for $N$ from $32$ to $1000$. 
			The sample size is 2000.
		}
		\label{fig:dsff_cs_ginue_near_0}
	\end{figure}
	
	\begin{figure}[H]
		\begin{minipage}[t]{0.44\textwidth}
			\includegraphics[width=\linewidth,keepaspectratio=true]{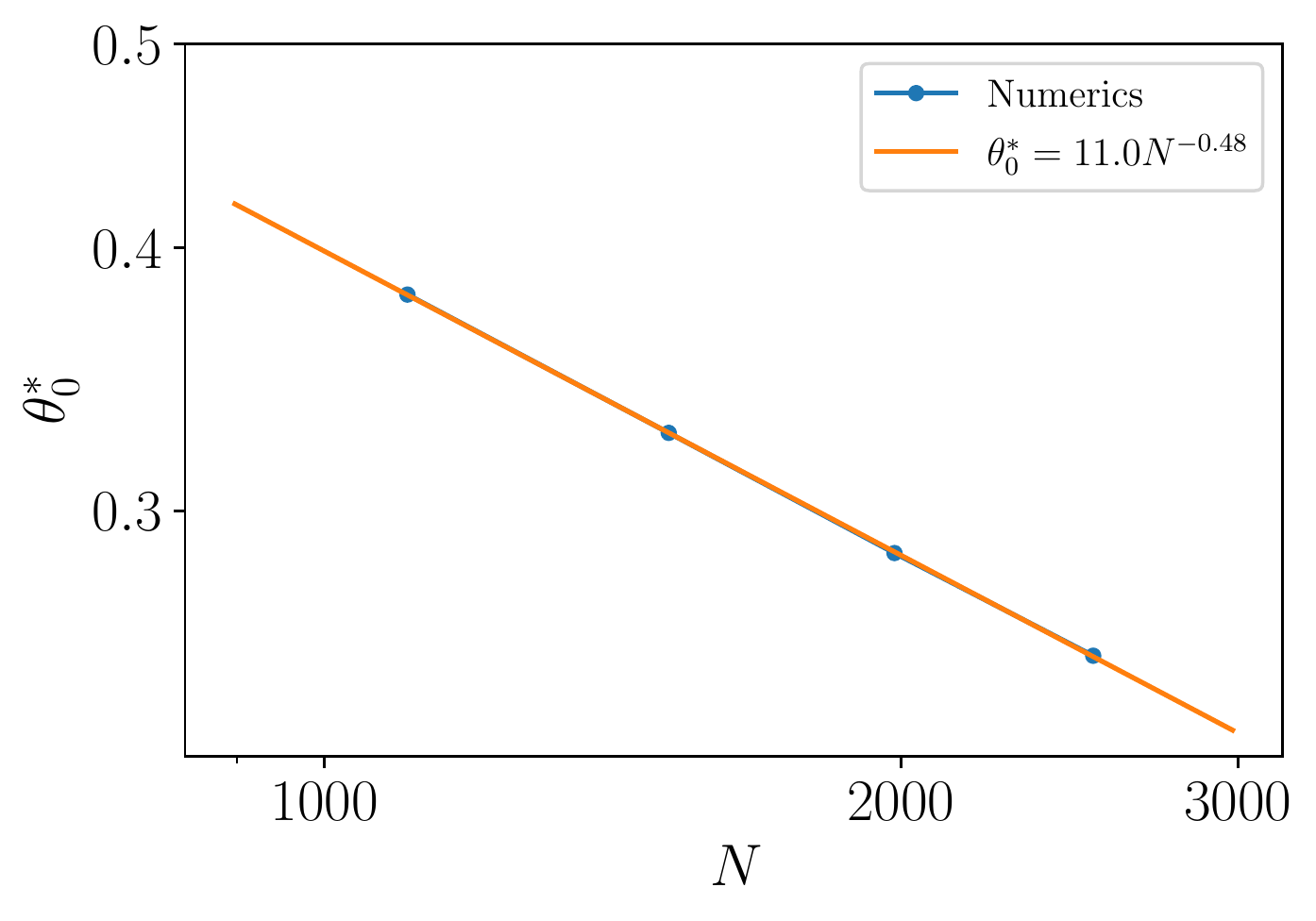}
		\end{minipage}
		\hspace*{\fill} % it's important not to leave blank lines before and after this command
		\begin{minipage}[t]{0.44\textwidth}
			\includegraphics[width=\linewidth,keepaspectratio=true]{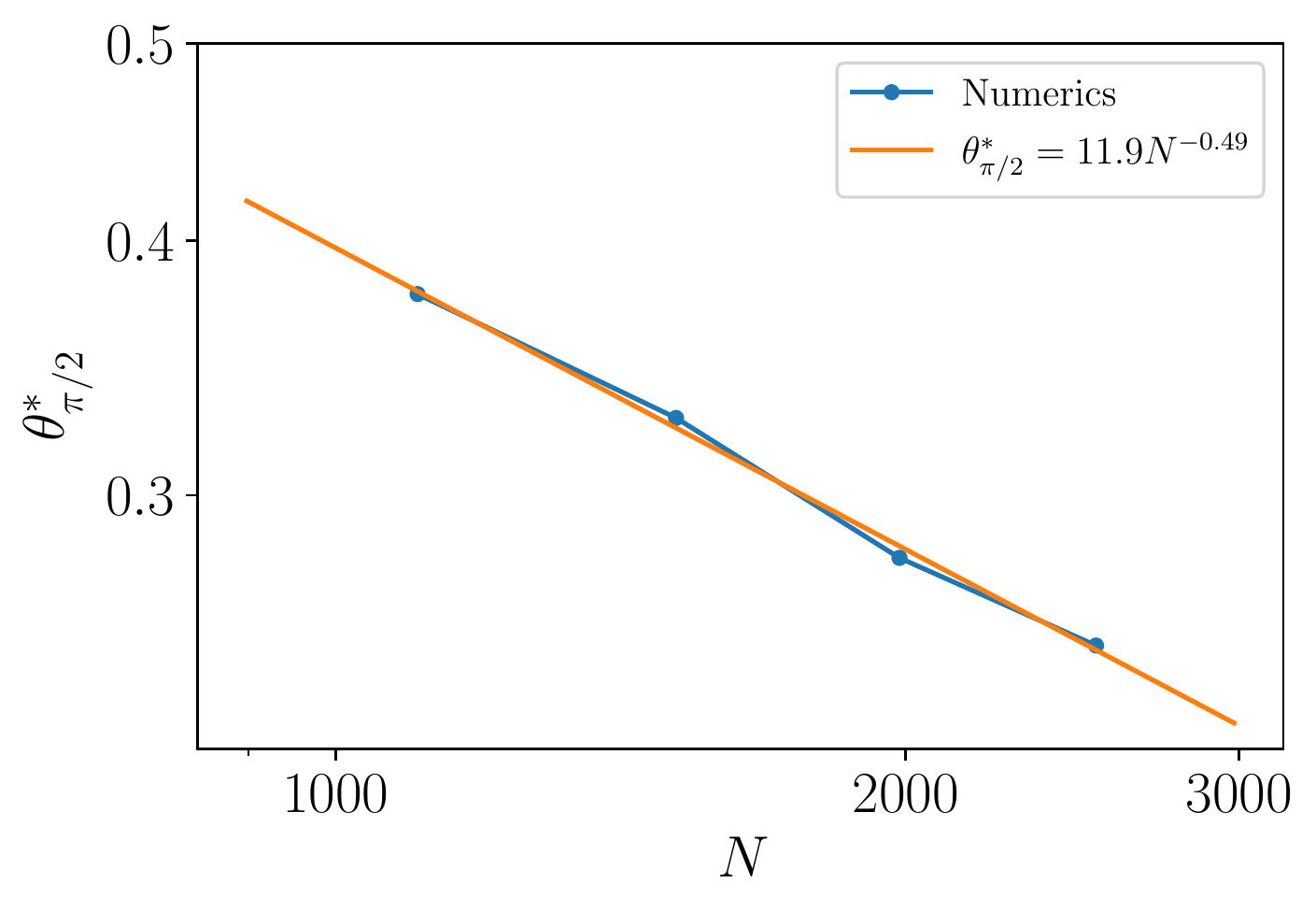}
		\end{minipage}
		\caption{
			$\thetasone(N)$ (left) and $\thetastwo(N)$ (right) of the QKT with the kick for $j=23,27,31$ and $35$, Gaussianly-distributed $p \in \mathcal{N}(2,2/3)$ and $k_0 \in \mathcal{N}(10,3)$. 
			The sample size is 2000.
		}
		\label{fig:dsff_cs_ginue_near_0}
	\end{figure}
	
	\begin{figure}[H]
		\begin{minipage}[t]{0.44\textwidth}
			\includegraphics[width=\linewidth,keepaspectratio=true]{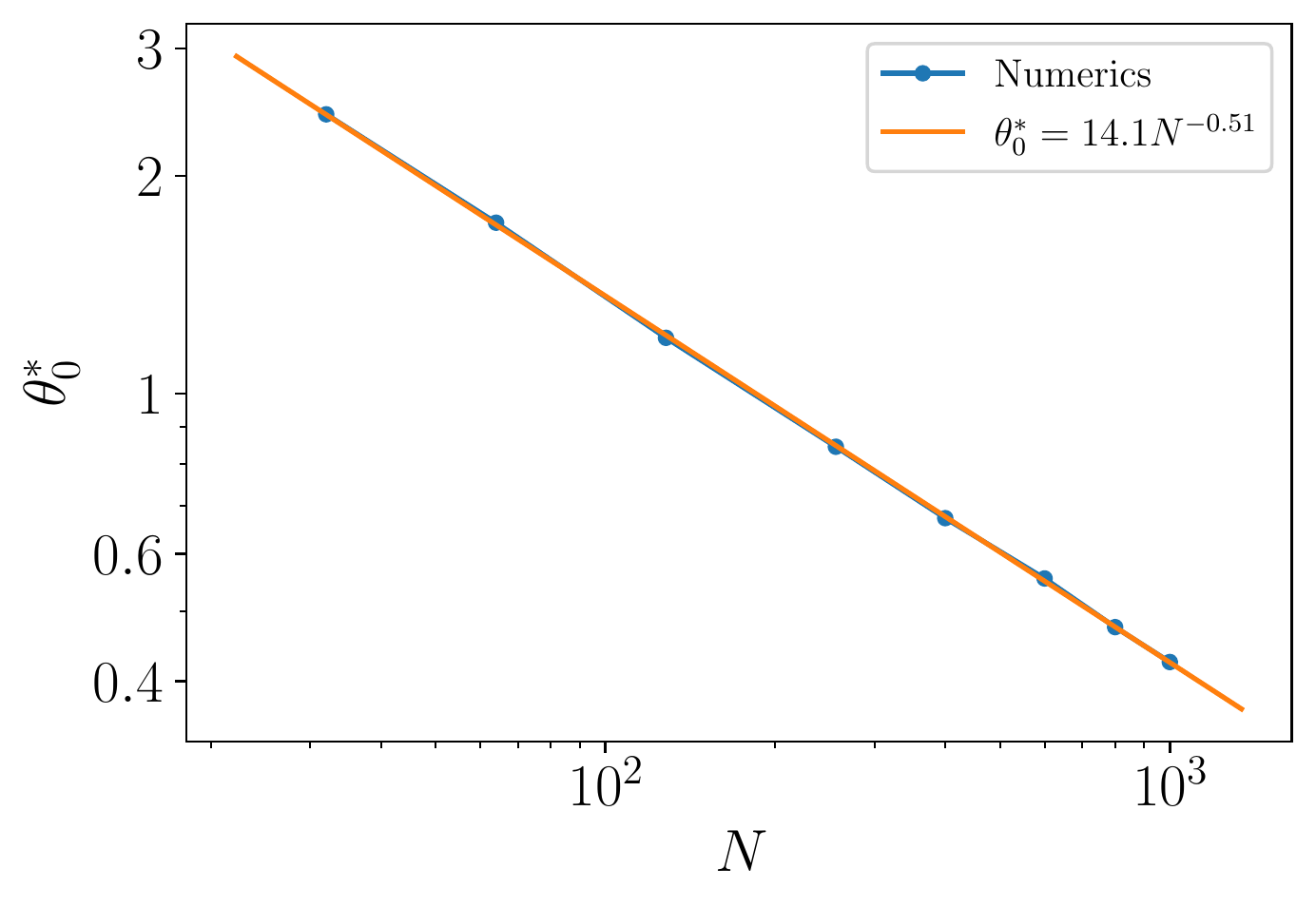}
		\end{minipage}
		\hspace*{\fill} % it's important not to leave blank lines before and after this command
		\begin{minipage}[t]{0.44\textwidth}
			\includegraphics[width=\linewidth,keepaspectratio=true]{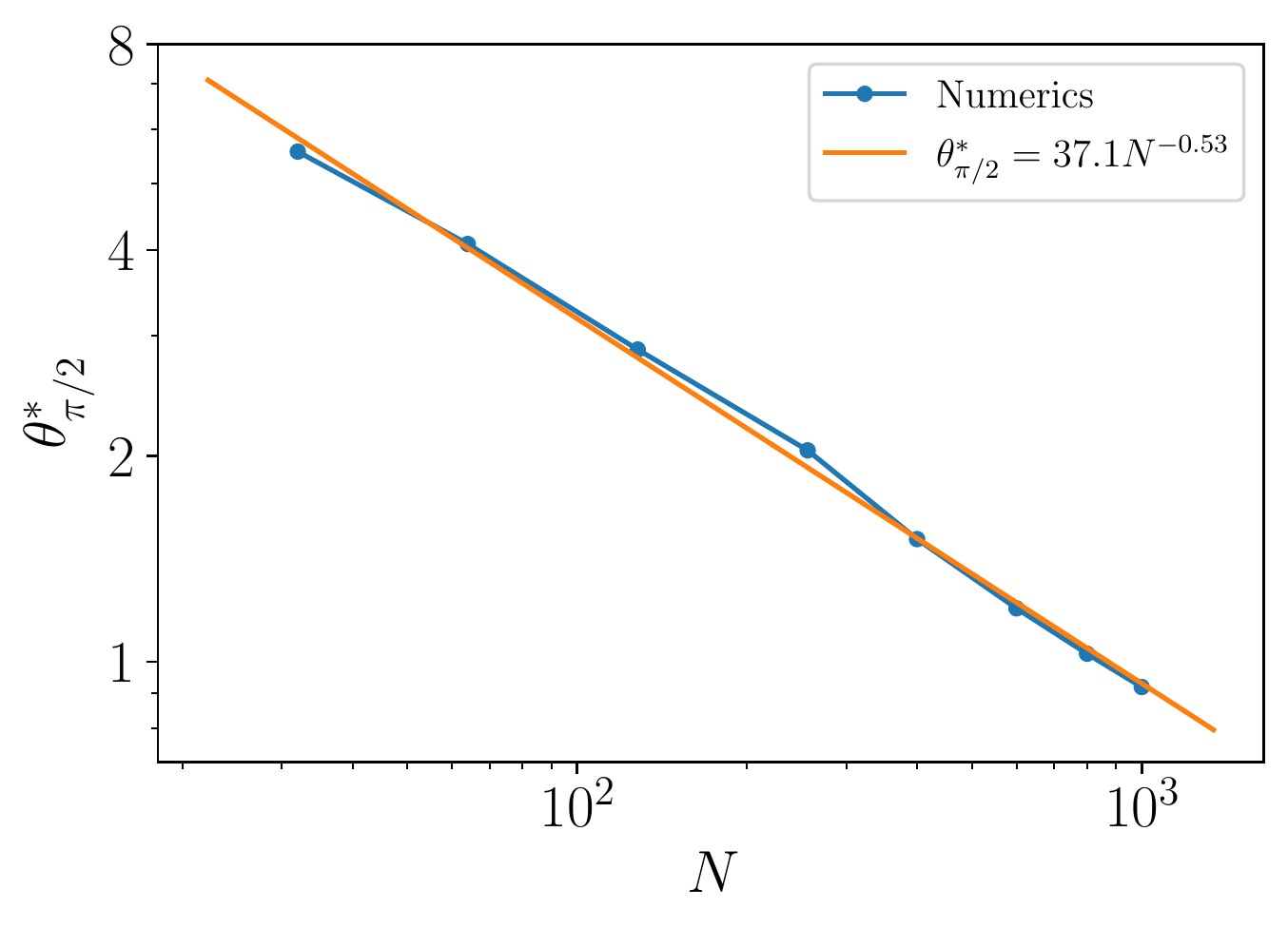}
		\end{minipage}
		\caption{
			$\thetasone(N)$ (left) and $\thetastwo(N)$ (right) of CS induced by CUE for $N$ from $32$ to $1000$. 
			The sample size is 2000.
		}
		\label{fig:dsff_cs_ginue_near_0}
	\end{figure}

	\begin{figure}[H]
		\begin{minipage}[t]{0.44\textwidth}
			\includegraphics[width=\linewidth,keepaspectratio=true]{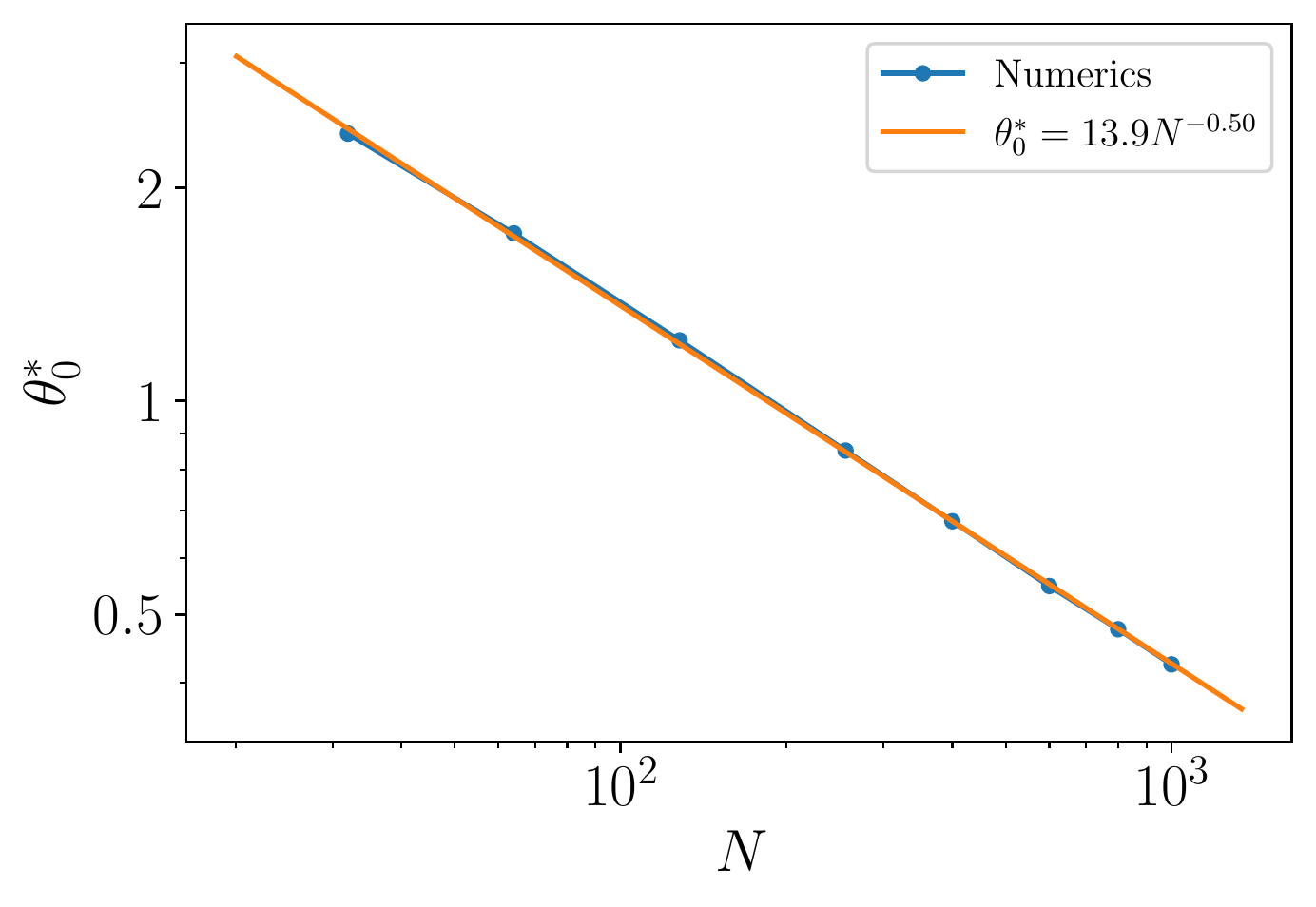}
		\end{minipage}
		\hspace*{\fill} % it's important not to leave blank lines before and after this command
		\begin{minipage}[t]{0.44\textwidth}
			\includegraphics[width=\linewidth,keepaspectratio=true]{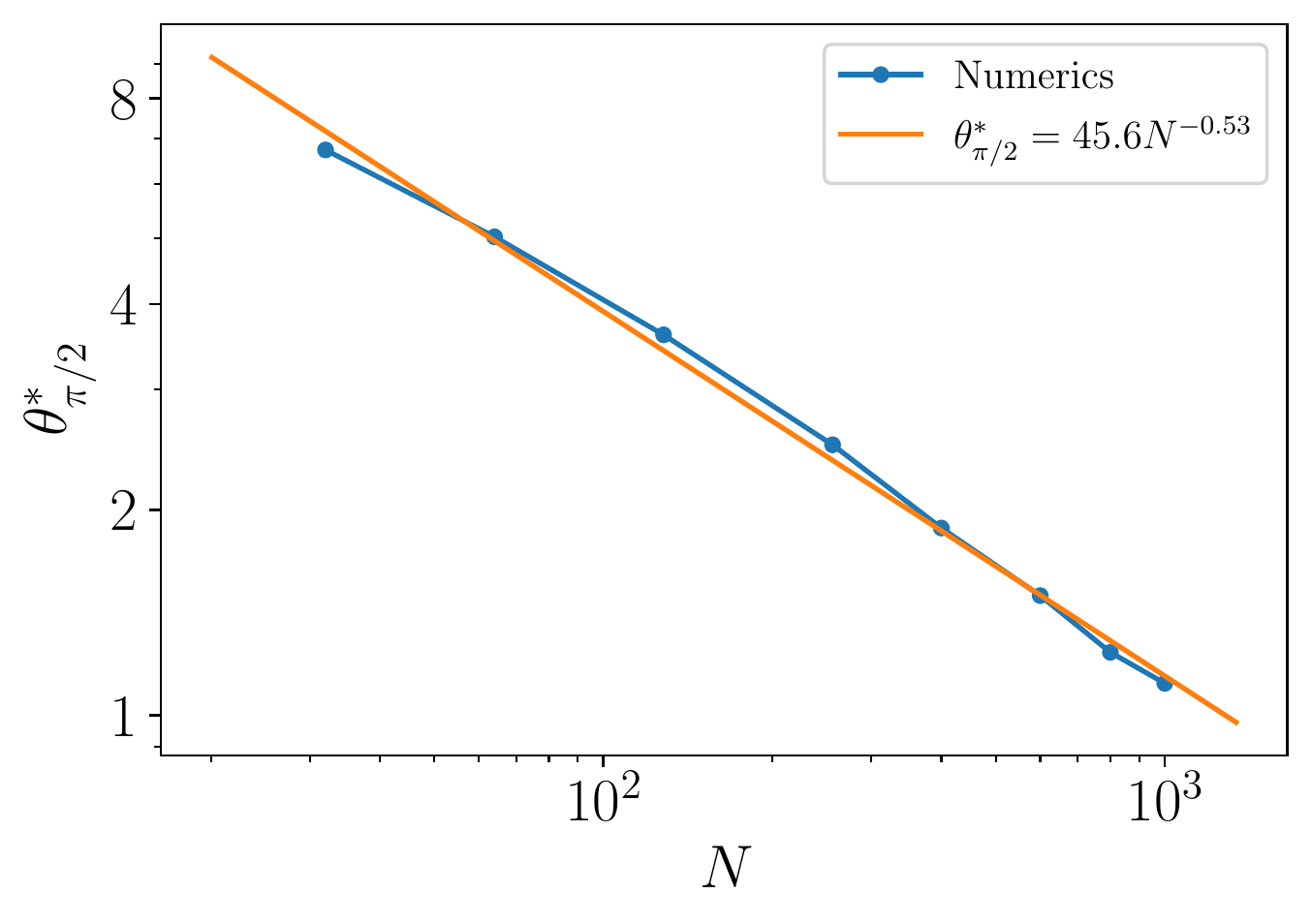}
		\end{minipage}
		\caption{
			$\thetasone(N)$ (left) and $\thetastwo(N)$ (right) of CS induced by GinUE for $N$ from $32$ to $1000$. 
			The sample size is 2000.
		}
		\label{fig:dsff_cs_ginue_near_0}
	\end{figure}

	\subsection{Arguments for the scaling of $\thetas$}
	\begin{figure}[H]
		\centering
		\includegraphics[scale = 0.6]{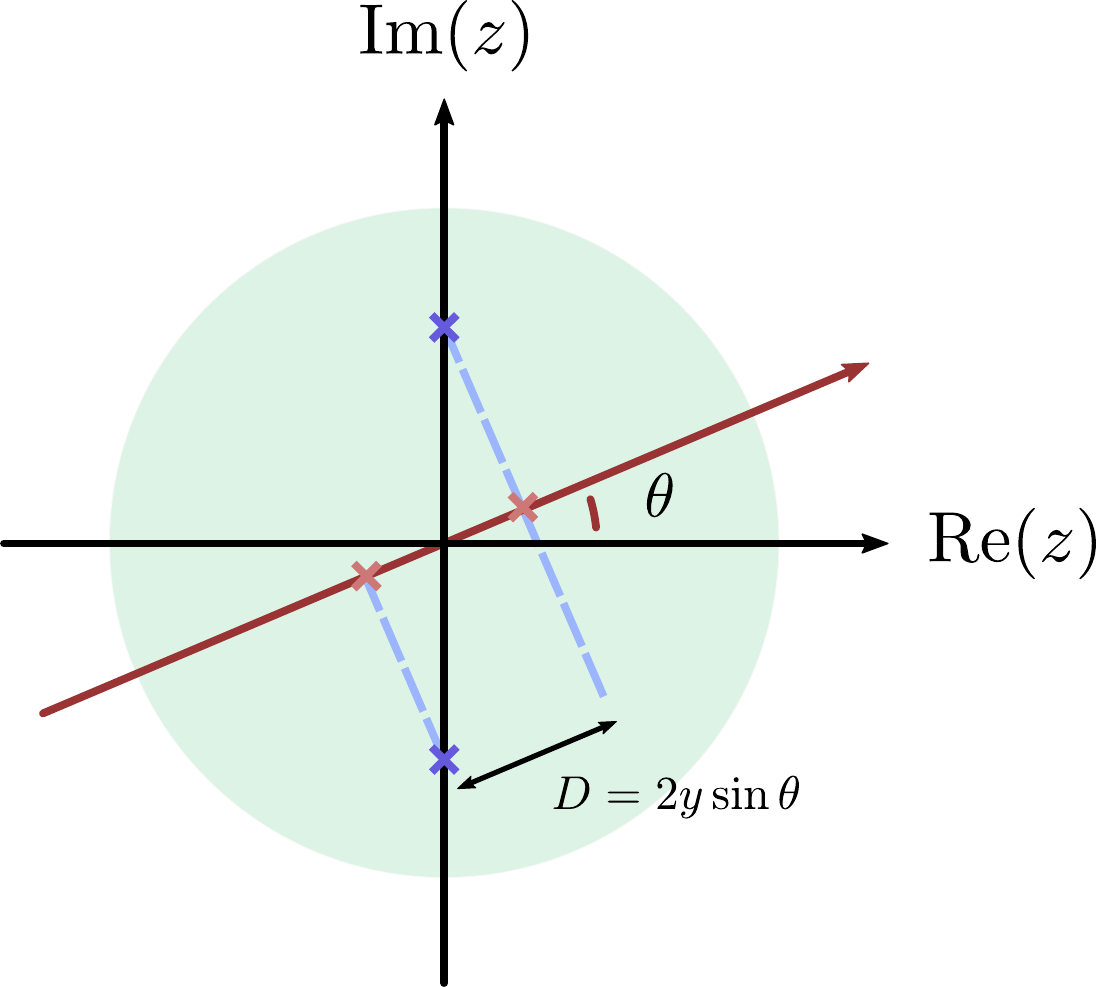}
		\caption{
			Illustration of the projected distance $D$ between a conjugate pair of eigenvalues $(z,z^*)$ (blue) along an axis defined by $\theta$. 
		}
		\label{fig:theta_star_argument}
	\end{figure}
	Here we provide an argument of why $\thetasphi \propto N^{-1/2}$, where $N$ is the matrix size.
	Recall that DSFF $K(|\tau| , \theta)$ of a set of eigenvalues $\{z_m\}$ is equivalent to the SFF $K(|\tau|)$ of the same set of eigenvalues projected onto the axis defined by angle $\theta$.   
	%We will use the convention where $\Dtwo$ denotes the mean level spacing of $$in the complex plane and $\Done$ denotes the mean level spacing 
	%
	Let $(z,z^*)$ be a pair of eigenvalues that are the complex conjugates of each other. Suppose $z=x+iy$ and $z^* = x-iy$ such that their distance in the complex plane is $2y$. 
	The  projected distance between $z$ and $z^*$ along a tilted axis defined by $\theta$ is $2y \sin \theta \approx 2y \theta$ for small $\theta$.
	Now suppose $(z,z^*)$ are separated by the mean level spacing, i.e. $z= x+ i \Delta/2$. 
	The projected distance between $z$ and $z^*$ along the tilted axis is then  $ \theta\Delta$.
	Suppose that $\tildeD=N^{-1}$ is the mean level spacing when $N$ eigenvalues are projected onto the $\theta$-axis.
	For $K(|\tau|, \theta> \thetasone) \approx K(|\tau|, \theta= \pi/4)$ to be true, 
	we must have $\theta \Delta  > \tildeD$, 
	such that the late time plateau is not affected by the degeneracy along the $\theta$-axis. 
	This gives 
	a lower bound $\thetasLB < \thetasphi$, via 
	\begin{align}\label{eq:tildedelta}
	\begin{split}
	\Delta \thetasLB & = O\left( \tildeD \right)
	\\
	\thetasLB & = O \left( \frac{1}{\sqrt{N}} \right)  \; .  
	\end{split}
	\end{align}
	The numerics in Appendix \ref{app:scaling_thetas} suggests that this lower bound is being saturated.
	%

	%%%%%%%%%%%%%%%%%%%%%
	%%%%%%%%%%%%%%%%%%%%%
\end{document}